\documentstyle[10pt,epsfig,dp_delphititle,float,hangcaption,xspace,amssymb,
amsfonts,amsmath,amsthm,cite,graphicx]{dp_delphi}
\makeindex
\def\DpPaperGroup{EP}
\def\DpPaperRef{2000-023}
\def\DpDate{7 February 2000}
\def\DpAuthors{DELPHI Collaboration}
\def\DpSubmit{(Eur. Phys. J. C18(2000)203; Erratum ibid. C25(2002)493)}
\def\DpTitle{{Charged and Identified Particles \\
in the Hadronic Decay of \\
W Bosons and in \boldmath $e^+e^- \rightarrow$q$\bar{\mathbf q}$ from 130 to
200 GeV}}
\def\DpComment{ }
\def\DpEMail{ }

\def\ee{$e\sp{+}e\sp{-}$}
\newcommand{\ppba}{p($\overline{\mathrm{p}}$)}
\newcommand{\Kpm}{$K^{\pm}$}
\newcommand{\pio}{$\pi$}
\newcommand{\idhadrons}{$\pi^+$, K$^+$, K$^0$, p and $\Lambda$}
\newcommand{\Kn}{\mbox{$ {\mathrm K}^{0}$} }
\newcommand{\p}{\mbox{$ {\mathrm p}$} }
\newcommand{\k}{\mbox{${\mathrm K}^{+}$} }
\newcommand{\xip}{$\xi_p$}
\newcommand{\xis}{$\xi^*$}
\newcommand{\Leff}{\Lambda_{\mathrm eff}}

\def\ZP{Z.\ Phys.\ {\bf C}}
\def\PL{Phys.\ Lett.\ {\bf B}}

\def\NP{Nucl.\ Phys.\ {\bf B}}

\def\NIM{Nucl.\ Instr.\ Meth.\ }
\def\Coll{Coll.,\ }
\def\Abreu{DELPHI Coll., P. Abreu {\it et al.,}\ }
\def\etal{{\it et al.,}\ }

\begin{document}
%%%%%%%%%%%%%%%%%%%%%%%%%% They are a problem with Coll.Sty ?
\makeatletter
%\input{dp_system:coll.sty}
% Collapse citation numbers to ranges.  Non-numeric and undefined labels
% are handled.  No sorting is done.  E.g., 1,3,2,3,4,5,foo,1,2,3,?,4,5
% gives 1,3,2-5,foo,1-3,?,4,5
\newcount\@tempcntc
\def\@citex[#1]#2{\if@filesw\immediate\write\@auxout{\string\citation{#2}}\fi
  \@tempcnta\z@\@tempcntb\m@ne\def\@citea{}\@cite{\@for\@citeb:=#2\do
    {\@ifundefined
       {b@\@citeb}{\@citeo\@tempcntb\m@ne\@citea\def\@citea{,}{\bf ?}\@warning
       {Citation `\@citeb' on page \thepage \space undefined}}%
    {\setbox\z@\hbox{\global\@tempcntc0\csname b@\@citeb\endcsname\relax}%
     \ifnum\@tempcntc=\z@ \@citeo\@tempcntb\m@ne
       \@citea\def\@citea{,}\hbox{\csname b@\@citeb\endcsname}%
     \else
      \advance\@tempcntb\@ne
      \ifnum\@tempcntb=\@tempcntc
      \else\advance\@tempcntb\m@ne\@citeo
      \@tempcnta\@tempcntc\@tempcntb\@tempcntc\fi\fi}}\@citeo}{#1}}
\def\@citeo{\ifnum\@tempcnta>\@tempcntb\else\@citea\def\@citea{,}%
  \ifnum\@tempcnta=\@tempcntb\the\@tempcnta\else
   {\advance\@tempcnta\@ne\ifnum\@tempcnta=\@tempcntb \else \def\@citea{--}\fi
    \advance\@tempcnta\m@ne\the\@tempcnta\@citea\the\@tempcntb}\fi\fi}
 
\makeatother
%%%%%%%%%%%%%%%%%%%%%%%%%% ??????????????????????????????????
% Generate the title page
\begin{titlepage}
\pagenumbering{roman}
\CERNpreprint{\DpPaperGroup}{\DpPaperRef} % Reference of the paper
\date{{\small\DpDate}} % Date of the paper
\title{\DpTitle} % Title of the paper
\address{\DpAuthors} % General name of the author(s)
\begin{shortabs} % Start the abstract
\noindent
%   abstract.tex
%
\noindent

Inclusive distributions of charged particles in hadronic W decays
are experimentally investigated using the statistics
collected by the DELPHI experiment at LEP during 1997, 1998 and 1999, 
at centre-of-mass energies from 183 to around 200 GeV. 
The possible effects of interconnection between
the hadronic decays of two Ws are not observed.
\\
Measurements of the average multiplicity for charged and identified
particles in q$\bar{\mathrm q}$  and WW events at centre-of-mass energies 
from 130 to 200 GeV 
and in W decays are presented.
\\
The results on the average multiplicity of identified particles 
and on the position $\xi^*$ of the maximum of the 
$\xi_p = -\mathrm{log} (\frac{2p}{\sqrt{s}})$ 
distribution are compared with predictions
of JETSET and  MLLA calculations.
\end{shortabs}
\vfill
\begin{center}
\DpSubmit \ \\ % Horrible hack to allow to have DpSubmit empty
\DpComment \ \\
\DpEMail \ \\
\end{center}
\vfill
\clearpage
\headsep 10.0pt
\addtolength{\textheight}{10mm}
\addtolength{\footskip}{-5mm}
\begingroup
% Commands to process the author names
%
\newcommand{\DpName}[2]{\hbox{#1$^{\ref{#2}}$},\hfill}
\newcommand{\DpNameTwo}[3]{\hbox{#1$^{\ref{#2},\ref{#3}}$},\hfill}
\newcommand{\DpNameThree}[4]{\hbox{#1$^{\ref{#2},\ref{#3},\ref{#4}}$},\hfill}
\newskip\Bigfill \Bigfill = 0pt plus 1000fill
\newcommand{\DpNameLast}[2]{\hbox{#1$^{\ref{#2}}$}\hspace{\Bigfill}}
%
%\small
\footnotesize
\noindent
\DpName{P.Abreu}{LIP}
\DpName{W.Adam}{VIENNA}
\DpName{T.Adye}{RAL}
\DpName{P.Adzic}{DEMOKRITOS}
\DpName{Z.Albrecht}{KARLSRUHE}
\DpName{T.Alderweireld}{AIM}
\DpName{G.D.Alekseev}{JINR}
\DpName{R.Alemany}{VALENCIA}
\DpName{T.Allmendinger}{KARLSRUHE}
\DpName{P.P.Allport}{LIVERPOOL}
% ...
\normalsize
\endgroup
\titlefoot{Department of Physics and Astronomy, Iowa State
     University, Ames IA 50011-3160, USA
    \label{AMES}}
\titlefoot{Physics Department, Univ. Instelling Antwerpen,
     Universiteitsplein 1, B-2610 Antwerpen, Belgium \\
     \indent~~and IIHE, ULB-VUB,
     Pleinlaan 2, B-1050 Brussels, Belgium \\
     \indent~~and Facult\'e des Sciences,
     Univ. de l'Etat Mons, Av. Maistriau 19, B-7000 Mons, Belgium
    \label{AIM}}
\titlefoot{Physics Laboratory, University of Athens, Solonos Str.
     104, GR-10680 Athens, Greece
    \label{ATHENS}}
\titlefoot{Department of Physics, University of Bergen,
     All\'egaten 55, NO-5007 Bergen, Norway
    \label{BERGEN}}
\titlefoot{Dipartimento di Fisica, Universit\`a di Bologna and INFN,
     Via Irnerio 46, IT-40126 Bologna, Italy
    \label{BOLOGNA}}
\titlefoot{Centro Brasileiro de Pesquisas F\'{\i}sicas, rua Xavier Sigaud 150,
     BR-22290 Rio de Janeiro, Brazil \\
     \indent~~and Depto. de F\'{\i}sica, Pont. Univ. Cat\'olica,
     C.P. 38071 BR-22453 Rio de Janeiro, Brazil \\
     \indent~~and Inst. de F\'{\i}sica, Univ. Estadual do Rio de Janeiro,
     rua S\~{a}o Francisco Xavier 524, Rio de Janeiro, Brazil
    \label{BRASIL}}
\titlefoot{Comenius University, Faculty of Mathematics and Physics,
     Mlynska Dolina, SK-84215 Bratislava, Slovakia
    \label{BRATISLAVA}}
\titlefoot{Coll\`ege de France, Lab. de Physique Corpusculaire, IN2P3-CNRS,
     FR-75231 Paris Cedex 05, France
    \label{CDF}}
\titlefoot{CERN, CH-1211 Geneva 23, Switzerland
    \label{CERN}}
\titlefoot{Institut de Recherches Subatomiques, IN2P3 - CNRS/ULP - BP20,
     FR-67037 Strasbourg Cedex, France
    \label{CRN}}
\titlefoot{Now at DESY-Zeuthen, Platanenallee 6, D-15735 Zeuthen, Germany
    \label{DESY}}
\titlefoot{Institute of Nuclear Physics, N.C.S.R. Demokritos,
     P.O. Box 60228, GR-15310 Athens, Greece
    \label{DEMOKRITOS}}
\titlefoot{FZU, Inst. of Phys. of the C.A.S. High Energy Physics Division,
     Na Slovance 2, CZ-180 40, Praha 8, Czech Republic
    \label{FZU}}
\titlefoot{Dipartimento di Fisica, Universit\`a di Genova and INFN,
     Via Dodecaneso 33, IT-16146 Genova, Italy
    \label{GENOVA}}
\titlefoot{Institut des Sciences Nucl\'eaires, IN2P3-CNRS, Universit\'e
     de Grenoble 1, FR-38026 Grenoble Cedex, France
    \label{GRENOBLE}}
\titlefoot{Helsinki Institute of Physics, HIP,
     P.O. Box 9, FI-00014 Helsinki, Finland
    \label{HELSINKI}}
\titlefoot{Joint Institute for Nuclear Research, Dubna, Head Post
     Office, P.O. Box 79, RU-101 000 Moscow, Russian Federation
    \label{JINR}}
\titlefoot{Institut f\"ur Experimentelle Kernphysik,
     Universit\"at Karlsruhe, Postfach 6980, DE-76128 Karlsruhe,
     Germany
    \label{KARLSRUHE}}
\titlefoot{Institute of Nuclear Physics and University of Mining and Metalurgy,
     Ul. Kawiory 26a, PL-30055 Krakow, Poland
    \label{KRAKOW}}
\titlefoot{Universit\'e de Paris-Sud, Lab. de l'Acc\'el\'erateur
     Lin\'eaire, IN2P3-CNRS, B\^{a}t. 200, FR-91405 Orsay Cedex, France
    \label{LAL}}
\titlefoot{School of Physics and Chemistry, University of Lancaster,
     Lancaster LA1 4YB, UK
    \label{LANCASTER}}
\titlefoot{LIP, IST, FCUL - Av. Elias Garcia, 14-$1^{o}$,
     PT-1000 Lisboa Codex, Portugal
    \label{LIP}}
\titlefoot{Department of Physics, University of Liverpool, P.O.
     Box 147, Liverpool L69 3BX, UK
    \label{LIVERPOOL}}
\titlefoot{LPNHE, IN2P3-CNRS, Univ.~Paris VI et VII, Tour 33 (RdC),
     4 place Jussieu, FR-75252 Paris Cedex 05, France
    \label{LPNHE}}
\titlefoot{Department of Physics, University of Lund,
     S\"olvegatan 14, SE-223 63 Lund, Sweden
    \label{LUND}}
\titlefoot{Universit\'e Claude Bernard de Lyon, IPNL, IN2P3-CNRS,
     FR-69622 Villeurbanne Cedex, France
    \label{LYON}}
\titlefoot{Univ. d'Aix - Marseille II - CPP, IN2P3-CNRS,
     FR-13288 Marseille Cedex 09, France
    \label{MARSEILLE}}
\titlefoot{Dipartimento di Fisica, Universit\`a di Milano and INFN-MILANO,
     Via Celoria 16, IT-20133 Milan, Italy
    \label{MILANO}}
\titlefoot{Dipartimento di Fisica, Univ. di Milano-Bicocca and
     INFN-MILANO, Piazza delle Scienze 2, IT-20126 Milan, Italy
    \label{MILANO2}}
\titlefoot{Niels Bohr Institute, Blegdamsvej 17,
     DK-2100 Copenhagen {\O}, Denmark
    \label{NBI}}
\titlefoot{IPNP of MFF, Charles Univ., Areal MFF,
     V Holesovickach 2, CZ-180 00, Praha 8, Czech Republic
    \label{NC}}
\titlefoot{NIKHEF, Postbus 41882, NL-1009 DB
     Amsterdam, The Netherlands
    \label{NIKHEF}}
\titlefoot{National Technical University, Physics Department,
     Zografou Campus, GR-15773 Athens, Greece
    \label{NTU-ATHENS}}
\titlefoot{Physics Department, University of Oslo, Blindern,
     NO-1000 Oslo 3, Norway
    \label{OSLO}}
\titlefoot{Dpto. Fisica, Univ. Oviedo, Avda. Calvo Sotelo
     s/n, ES-33007 Oviedo, Spain
    \label{OVIEDO}}
\titlefoot{Department of Physics, University of Oxford,
     Keble Road, Oxford OX1 3RH, UK
    \label{OXFORD}}
\titlefoot{Dipartimento di Fisica, Universit\`a di Padova and
     INFN, Via Marzolo 8, IT-35131 Padua, Italy
    \label{PADOVA}}
\titlefoot{Rutherford Appleton Laboratory, Chilton, Didcot
     OX11 OQX, UK
    \label{RAL}}
\titlefoot{Dipartimento di Fisica, Universit\`a di Roma II and
     INFN, Tor Vergata, IT-00173 Rome, Italy
    \label{ROMA2}}
\titlefoot{Dipartimento di Fisica, Universit\`a di Roma III and
     INFN, Via della Vasca Navale 84, IT-00146 Rome, Italy
    \label{ROMA3}}
\titlefoot{DAPNIA/Service de Physique des Particules,
     CEA-Saclay, FR-91191 Gif-sur-Yvette Cedex, France
    \label{SACLAY}}
\titlefoot{Instituto de Fisica de Cantabria (CSIC-UC), Avda.
     los Castros s/n, ES-39006 Santander, Spain
    \label{SANTANDER}}
\titlefoot{Dipartimento di Fisica, Universit\`a degli Studi di Roma
     La Sapienza, Piazzale Aldo Moro 2, IT-00185 Rome, Italy
    \label{SAPIENZA}}
\titlefoot{Inst. for High Energy Physics, Serpukov
     P.O. Box 35, Protvino, (Moscow Region), Russian Federation
    \label{SERPUKHOV}}
\titlefoot{J. Stefan Institute, Jamova 39, SI-1000 Ljubljana, Slovenia
     and Laboratory for Astroparticle Physics,\\
     \indent~~Nova Gorica Polytechnic, Kostanjeviska 16a, SI-5000 Nova Gorica, Slovenia, \\
     \indent~~and Department of Physics, University of Ljubljana,
     SI-1000 Ljubljana, Slovenia
    \label{SLOVENIJA}}
\titlefoot{Fysikum, Stockholm University,
     Box 6730, SE-113 85 Stockholm, Sweden
    \label{STOCKHOLM}}
\titlefoot{Dipartimento di Fisica Sperimentale, Universit\`a di
     Torino and INFN, Via P. Giuria 1, IT-10125 Turin, Italy
    \label{TORINO}}
\titlefoot{Dipartimento di Fisica, Universit\`a di Trieste and
     INFN, Via A. Valerio 2, IT-34127 Trieste, Italy \\
     \indent~~and Istituto di Fisica, Universit\`a di Udine,
     IT-33100 Udine, Italy
    \label{TU}}
\titlefoot{Univ. Federal do Rio de Janeiro, C.P. 68528
     Cidade Univ., Ilha do Fund\~ao
     BR-21945-970 Rio de Janeiro, Brazil
    \label{UFRJ}}
\titlefoot{Department of Radiation Sciences, University of
     Uppsala, P.O. Box 535, SE-751 21 Uppsala, Sweden
    \label{UPPSALA}}
\titlefoot{IFIC, Valencia-CSIC, and D.F.A.M.N., U. de Valencia,
     Avda. Dr. Moliner 50, ES-46100 Burjassot (Valencia), Spain
    \label{VALENCIA}}
\titlefoot{Institut f\"ur Hochenergiephysik, \"Osterr. Akad.
     d. Wissensch., Nikolsdorfergasse 18, AT-1050 Vienna, Austria
    \label{VIENNA}}
\titlefoot{Inst. Nuclear Studies and University of Warsaw, Ul.
     Hoza 69, PL-00681 Warsaw, Poland
    \label{WARSZAWA}}
\titlefoot{Fachbereich Physik, University of Wuppertal, Postfach
     100 127, DE-42097 Wuppertal, Germany
    \label{WUPPERTAL}}
\addtolength{\textheight}{-10mm}
\addtolength{\footskip}{5mm}
\clearpage
\headsep 30.0pt
\end{titlepage}
%%%%%%%%%%%%%%%%%%%%%%%%%
%
% Change for the document body
%%\pagestyle{heading} % for page numbering
\pagenumbering{arabic} % page numbering in number
\setcounter{footnote}{0} %
\large
%\linenumbers %%%CD
%   document.tex
%   9/OCT/2000, 14:15 CET

\section{Introduction}

The study of the production properties of charged and identified
hadrons ($\pi^+$, K$^+$, K$^0$, p and $\Lambda$)\footnote{Unless otherwise
stated antiparticles are included as well.} 
in q$\bar{\mathrm q}$ events at LEP~2 
allows QCD based models to be tested through the comparison with LEP~1 data,
in particular the validity of high-energy extrapolations of Monte Carlo models
tuned at the Z$^0$.

In the case of WW events in which both W bosons decay hadronically, these kind 
of studies,
besides providing checks for QCD-inspired models, are expected to give
insights into possible correlations and/or final state interactions between
the decay products of the two W bosons.

\subsection*{Hadron production in \ee\ and QCD}

The way quarks and gluons transform into hadrons is complex and can not be 
completely described by QCD.  In the Monte Carlo simulations 
(as in~\cite{lund}), the
hadronisation of a q$\bar{\mathrm q}$  
pair is split into 3
phases. In a first phase, gluon emission and parton branching of the
original q$\bar{\mathrm q}$ pair take place. It is believed that this phase
can be 
described by perturbative QCD (most of the calculations have been performed
in leading logarithmic approximation LLA). In a second phase, at a certain
virtuality cut-off scale $Q_0$, where $\alpha_s(Q_0)$ is still small, quarks
and gluons produced in the first phase are clustered in colour singlets and
transform into mesons and baryons. Only phenomenological models, which need
to be tuned to the data, are available to describe this stage of
fragmentation; the models most frequently used in \ee\ annihilations are
based on string and cluster fragmentation. In the third phase, the unstable
states decay into hadrons which can be observed and identified in the
detector. These models account correctly for 
most of the
features of the 
q$\bar{\mathrm q}$ events such as, for instance, the average multiplicity and
inclusive momentum spectra. %; with LEP~2,
%the energy range spanned in $e^+e^-$ 
%interactions is doubled (up to 183 $\mathrm{GeV}$), and it is interesting 
%to check their validity.

A different and purely analytical approach (see e.g.~\cite{ochs} and
references therein) giving quantitative predictions of hadronic spectra are
QCD calculations using the so-called Modified Leading Logarithmic
Approximation (MLLA) under the assumption of Local Parton Hadron Duality
(LPHD)~\cite{dok:mlla,am:mlla}. In this picture multi-hadron 
production is described
by a parton cascade, and the virtuality cut-off $Q_0$ is lowered to values
of the order of 100 MeV, comparable to the hadron masses; it is assumed that
the results obtained for partons are proportional to the corresponding
quantities for hadrons.

The MLLA/LPHD predictions involve three parameters: an effective scale
parameter $\Lambda_{\mathrm{eff}}$, a virtuality cut-off $Q_0$ in the
evolution of the parton cascade and an overall normalisation factor 
$K_{\mathrm{LPHD}}$. 
The momentum spectra of hadrons can be calculated as functions of the
variable $\xi_p=-\ln (\frac{2p}{\sqrt{s}})$,
%$\mbox{$ {\mathrm{\xi_{p}}}$}=-\ln {x_{p}}$ where $x_{p}=2p/\sqrt{s}
%$ ($p$ being the particle momentum and $\sqrt{s}$ the centre-of-mass
%energy): 
with $p$ being the particle's momentum and $\sqrt{s}$ the centre-of-mass
energy: 
\begin{equation}
\frac{d n }{d\xi _{p}}=K_{\mathrm{LPHD}}\cdot f(\xi
_{p},X,\lambda )  \label{eq:evol}
\end{equation}
with 
\begin{eqnarray}
X= \log\frac{\sqrt{s}}{Q_{0}} & ; & \lambda = \log\frac{Q_{0}}{\Lambda_{%
\mathrm{eff}}} \, .
\end{eqnarray}

%The MLLA/LPHD predictions thus involve three parameters: an effective scale
%parameter $\Lambda_\mathrm{eff}$, a virtuality cut-off $Q_0$ in the
%evolution of the parton cascade and an overall normalisation factor $K_%
%\mathrm{LPHD}$. 
Due to uncertainties from higher order corrections $\Leff$ 
cannot be identified with $\Lambda_{\overline{MS}}$.
In equation (\ref{eq:evol}), $n$ is the average multiplicity per bin
of $\xi_p$, and the function $f$ has the form of a
``hump-backed plateau''. It can be approximated by a distorted Gaussian~ 
\cite{dkt-1,bw:mlla}

\begin{equation}
DG(\xi ;N,\overline{\xi },\sigma ,s_{k},k)=\frac{N}{\sigma \sqrt{2\pi }}\exp {%
\left( \frac{1}{8}k+\frac{1}{2}s_{k}\delta -\frac{1}{4}(2+k)\delta ^{2}+%
\frac{1}{6}s_{k}\delta ^{3}+\frac{1}{24}k\delta ^{4}\right) }\ ,
\label{eq:dgauss}
\end{equation}
where $\delta =(\xi -\overline{\xi })/\sigma $, $\bar\xi$ is the mean of the 
distribution, $\sigma$ is the square root of its variance, $s_{k}$ its 
skewness and $k$ its kurtosis. 
For an ordinary Gaussian these
last two parameters vanish. The mean, $\overline{\xi }$, coincides with the
peak of 
the distribution, $\xi^*$, only up to next-to-leading order in $\alpha_s$.

To check the validity of the MLLA/LPHD approach, one can study
the evolution of the position of the maximum, \mbox{$ {\mathrm{\xi^{*}}}$},
as a function of $\sqrt{s}$. In the context of MLLA/LPHD the dependence of %
\mbox{$ {\mathrm{\xi^{*}}}$} on the centre-of-mass energy can be expressed
as~\cite{ochs,dkt-1}:

\begin{equation}
\xi^* = Y \left( \frac{1}{2} +\sqrt{C/Y}-C/Y \right) +F_h (\lambda),
\label{eq:evol2}
\end{equation}
where 
\begin{equation}\label{eq:YC}
Y = {\mathrm log}\frac{\sqrt{s}/2}{\mbox{${\mathrm \Lambda_{\mathrm{eff}}}$}}%
~~,~~ C = \left( \frac{11N_c/3 + 2n_f/(3N_c^2)}{4N_c} \right)^2 \cdot \left( 
\frac{N_c}{11N_c/3 -2n_f/3} \right), 
\end{equation}
with $N_c$ being the number of colours and $n_f$ the number of active quark
flavours in the fragmentation process. The function $F_h (\lambda)$
 depends on the hadron type through 
$\lambda = \log(Q_0/\Leff)$~\cite{ochs}, and it can be approximated as 
%\begin{equation}
$F_h (\lambda) = -1.46\lambda + 0.207\lambda^2$ with an error of $\pm 0.06$.
%\end{equation}

\subsection*{Interference and final state interactions in W decays}

The possible presence of interference due to colour reconnection and
Bose-Einstein correlations (see for 
example~\cite{sjre,gh,were,elre,sjbe,moller} and ~\cite{yellow,ww} for 
reviews) in
hadronic decays of WW pairs %has been discussed on a theoretical basis, 
%in the framework of the measurement of the W mass: 
may provide information on hadron formation at a time scale smaller than 1
fm/$c$. At the same time it can induce a systematic uncertainty on the W
mass measurement in the 4-jet mode~\cite{yellow} comparable with the
expected accuracy of the measurement at LEP~2.

Interconnection can happen due to the fact that the lifetime of the W ($%
\tau_W \simeq \hbar /\Gamma_{W} \simeq 0.1$ fm/$c$) is an order of magnitude
smaller than the typical hadronisation times. The interconnection between
the products of the hadronic decays of different Ws in WW pair events can
occur at several stages: (1)
from colour rearrangement between the quarks coming from the primary
branching,
(2) due to gluon exchanges during the parton cascade,
(3) in the mixing of identical pions or kaons due to Bose-Einstein
correlations. 
The first two are QCD effects. They can mix the two colour singlets and
produce hadrons which cannot be uniquely assigned to either W. The
perturbative effects are colour suppressed and the possible shift is
expected to be only about 5 MeV in the W mass~\cite{sjre}. 

Non-perturbative
effects need model calculations. Several models have been proposed 
(for reviews see for example \cite{yellow} and \cite{sskk}) 
and have already been included in the widely used event 
generators PYTHIA~\cite{lund}, ARIADNE~\cite{ari} and HERWIG\cite{HERWIG}.
In these models the final state quarks after the parton shower can be
rearranged to form colour singlets with probabilities which in some cases
are free parameters. The shift on the W mass in these
models is typically smaller than 50 MeV \cite{wmassnoi}, 
but other observables are affected by colour
rearrangement. Generally these models suggest a small effect on the total
charged particle multiplicity, of the order of $-$1\% 
to $-$2\% \cite{sskk,dejong,ww}. 
Dedicated detailed simulations of the response of the 
DELPHI detector to such models showed that this effect is substantially 
unaffected by the event selection criteria and by the detector performance. 
For identified heavy particles, such as \k\ and \p, 
the effects due to colour reconnection are expected to be 
stronger~\cite{sk99}, but the experimental verification 
is complicated by losses in statistics. The same applies to the 
particle spectrum at low momentum \cite{sskk}.
Bose-Einstein Correlations could also slightly change the multiplicity for 
$(4q)$ events in some models~\cite{moller,fialko}.

The WW events allow a comparison of the characteristics of the W hadronic
decays when both Ws decay in hadronic modes 
%(in the following this shall often be referred to 
(referred to here
as the $(4q)$ mode) with the case in which only one W
decays hadronically ($(2q)$ mode). These characteristics should
be the same in the absence of interference between the hadronic decay
products from different W bosons.

Previous experimental results based on the statistics collected by LEP
experiments during 1997 (see \cite{deafra,watvan,molvan} for reviews) did
not indicate at that level of statistics the presence of interconnection or
correlation effects.

This paper presents measurements of:

\begin{itemize}
\item  the charged particle multiplicities for the q$\bar{\mathrm q}$ events
in the 
data sample collected by the DELPHI experiment at LEP during 1997, 1998 and
1999, at the centre-of-mass energies from 183, 189 and from 
192 to 200~GeV respectively;

\item  the charged particle multiplicity and inclusive distributions for WW
events at 183 and 189~GeV (multiplicity values at 189~GeV are expected to be 
slightly  % gl 189-183 = 0.0106(2q), 0.00412(4q)
higher than that at 183~GeV due to increased phase space. However this effect is
below the precision obtainable with the present data samples); 
%For these energies the main background to WW processes 
%is given by q$\bar{\mathrm q}$ events.
% with a cross section (for an effective centre-of-mass 
%energy larger than 10\% of the maximum annihilation energy) of about 105 pb.

\item  The average multiplicities for identified charged and neutral
particles ($\pi^+$, K$^+$, K$^0$, p and $\Lambda$), and the position 
$\xi^{\ast }$ of 
the maximum of the $\xi _{p}$
distribution for identified particles in q$\bar{\mathrm q}$ events 
from 130 GeV to 189 GeV and in WW events at 189~GeV.
\end{itemize}

\section{Data Sample and Event Preselection at 183 and 189~GeV}

Data corresponding to total luminosities of 157.7~pb$^{-1}$ (54.1 pb$^{-1}$)
 at centre-of-mass energies around 189 (183) GeV 
%were analysed. 
, and 
data taken in 1999 corresponding to total luminosities of 
25.8~pb$^{-1}$, 77.4~pb$^{-1}$ and 83.8~pb$^{-1}$ %and 40.2~pb$^{-1}$ 
%25.84~pb$^{-1}$, 77.42~pb$^{-1}$, 83.81~pb$^{-1}$ and 40.15~pb$^{-1}$ 
%collected by DELPHI 
at centre-of-mass energies around 192~GeV, 196~GeV and 200~GeV %and 202~GeV 
respectively were analysed.
A description of the DELPHI detector can be found in 
reference~\cite{deldet}; its performance is discussed in
reference~\cite{perfo}. 

A preselection of hadronic events was made, requiring at least 6 charged
particles and a total transverse energy of all the particles above 20\%
of the centre-of-mass energy $\sqrt{s}$. In the calculation of the energies 
$E$, all charged particles were assumed to have the pion mass. Charged
particles were required to have momentum $p$ above 100 MeV/$c$ and below 1.5
times the beam energy, a relative error on the momentum measurement $\Delta
p/p < 1$, angle $\theta$ with respect to the beam direction between 20$%
^\circ $ and 160$^\circ$, %a track length of at least 30 cm,
and a distance of closest approach to the interaction point less than 4 cm
in the plane perpendicular to the beam axis (2 cm in the analyses of identified
charged particles) and less than 4/$\sin\theta$~cm along the beam axis (2
cm in the analyses of identified charged particles).

After the event selection charged particles were also required to have 
%then used in the analysis if they had
a track length of at least 30 cm, and in the charged identified particles
analysis a momentum $p > 200$ MeV/$c$.

The influence of the detector on the analysis was studied with the full
DELPHI simulation program, DELSIM~\cite{perfo}; events were generated with 
PYTHIA 5.7, using the JETSET fragmentation with Parton Shower (PS) \cite{lund} 
with parameters tuned to fit LEP~1 data from DELPHI \cite{tuning}.
The initial state for the WW 1999 sample was generated using EXCALIBUR 
version 1.08 \cite{excalibur}. 
The particles were followed through the detailed geometry of DELPHI with
simulated digitizations in each detector. These data were processed with the
same reconstruction and analysis programs as the real data.

To check the ability of the simulation to model the efficiency for the
reconstruction of charged particles, the samples collected at the Z$^0$ pole 
during
1998 and 1997 were used. From these samples, by integrating the distribution
of $\xi_E = -\ln(\frac{2E}{\sqrt{s}})$, where $E$ is the energy of the particle,
corrected bin by bin using the simulation, the average charged particle
multiplicities at the Z$^0$ were measured. The values were found 
to be $20.93 \pm 0.03 (stat)$ and $%
20.60 \pm 0.03 (stat)$ respectively, in satisfactory agreement with the
world average of $21.00 \pm 0.13$ \cite{pdg}. 
%A relative scale systematic uncertainty of 1\% was assumed on the multiplicity.
The ratios of the world average value to the measured multiplicities at the
Z$^0$, respectively $1.0033\pm0.0064$ and $1.0194 \pm 0.0065$ 
from the Z$^0$ data in 1998 and 1997, were
used to correct the measured multiplicities at high energies in the
respective years.

%$

The cross-section for $e^{+}e^{-}\rightarrow 
{\mathrm q}\bar{\mathrm q}(\gamma )$ above the Z$^0$
peak is dominated by radiative q$\bar{\mathrm q}\gamma $ events; the initial
state 
radiated photons (ISR photons) are generally aligned along the beam
direction and not detected. In order to compute the hadronic centre-of-mass
energy, $\sqrt{s^{\prime }}$, the procedure described in 
reference~\cite{sprimen}
was used. The procedure clusters the particles into jets using the DURHAM
algorithm~\cite{durham}, excluding candidate ISR photons and using a $%
y_{cut}=0.002$. The reconstructed jets and additional ISR photons are then
fitted with a three constraint fit (energy and transverse momentum, leaving
free the $z$ component of the missing momentum). The hadronic centre-of-mass
energy, \mbox{$\sqrt{s^\prime}$}, is the invariant mass of the jets using
the fitted jet energies and directions.

%\ref{fig.sprime}.

\section{Analysis of charged particles in 
q\protect\boldmath{$\bar{\mathrm q}$} events}

\subsection{Centre-of-mass energies of 183 and 189 GeV}

Events with $\sqrt{s^{\prime }}/\sqrt{s}$ above 0.9 were used to compute the
multiplicities. A total of 3444 (1297) hadronic events were selected from
the data at 189 (183) GeV, by requiring that the multiplicity for charged
particles was larger than 9, that the total transverse energy of the charged
particles exceeded $0.2\sqrt{s}$, and that the narrow jet 
broadening~\cite{lepII} 
was smaller than 0.065. From the simulation it was calculated that
the expected background coming from WW and Z$^0$Z$^0$ decays was 432+52 (127+21)
events. The contamination from double radiative returns to the Z$^0$, within
10~GeV of the nominal Z$^0$ mass, was estimated by simulation to be below 5\%.
Other contaminations (from Z$^0$ee, We$\nu $, $\gamma \gamma $ interactions and
Bhabhas) are below 2\% 
in total.

The average multiplicity of charged particles with $p > 0.1$~GeV/$c$
measured in the selected events at 189 GeV (183 GeV), after subtraction of
the WW and Z$^0$Z$^0$ backgrounds estimated by simulation, was $24.58 \pm 0.16
(stat) $ ($23.96 \pm 0.23 (stat)$), to be compared to $24.52 \pm 0.05 (stat)$ 
($%
24.30 \pm 0.07 (stat)$) in the q$\bar{\mathrm q}$ PS simulation including
detector effects. 
The dispersion (square root of the variance) of the multiplicity distribution
in the data was
$7.57 \pm 0.11 (stat)$ ($7.00 \pm 0.16 (stat)$), to be compared to the
dispersion from the q$\bar{\mathrm q}$ PS simulation of 
$7.24 \pm 0.03$ ($7.20 \pm 0.05 (stat)$). 
%%(no paragraph here)

Detector
effects and selection biases were corrected for using a q$\bar{\mathrm q}$ 
simulation from PYTHIA with the JETSET fragmentation
tuned by DELPHI without inital state radiation. 
The corrected average charge 
multiplicity was found to be $<n> = 27.37 \pm 0.18 (stat)$ ($<n> = 26.56 \pm
0.26 (stat)$), and 
the dispersion was found to be $D = 8.77 \pm 0.13 (stat)$ ($D =
8.08 \pm 0.19 (stat)$). 

The average multiplicity was computed by integrating the $\xi_{E}$ 
distribution, since the detection
efficiency depends 
mostly on the momentum of the particle, after correcting for
detector effects bin by bin using the simulation. 
The $\xi_E$ distribution was
integrated up to a value of 6.3, and the extrapolation to the region above
this cut was based on the simulation at the generator level. 

After multiplying by the Z$^0$ 
corrections factors from section~2, the following values were obtained:
%The average multiplicity was also computed by integrating the distributions:
%\begin{itemize}
%\item  of the rapidity with respect to the thrust axis $y_{T}=\frac{1}{2}\ln 
%\frac{E+p_{||}}{E-p_{||}}$ ($p_{||}$ is the absolute value of the momentum
%component on the thrust axis);
%
%\item  of $\xi _{E}=-\mathrm{log}(\frac{2E}{\sqrt{s}})$. %\item of $p_T$
%\end{itemize}
%all corrected bin by bin using the simulation. The $\xi_E$ distribution was
%integrated up to a value of 6.3, and the extrapolation to the region above
%this cut was based on the simulation at the generator level. The average
%charge multiplicity of the selected events, including the above corrections,
%is 27.43 and 27.38 (26.59 and 26.53) respectively from the $y_T$ and $\xi_E$
%distribution, consistent with the value from the average observed
%multiplicity.
%
%As a central value for the measurement of the charge multiplicity the result
%of the integration of the $\xi_E$ distribution was taken, since the
%detection efficiency depends mostly on the momentum of the particle, after
%multiplication by the Z correction factors discussed in
%section~2.
%The values
\begin{eqnarray}
\langle n\rangle_{\mathrm{189\,GeV}} = 27.47 \pm 0.18 (stat) \pm 0.30 (syst)
\label{mulv189} \\
D_{\mathrm{189\,GeV}} = 8.77 \pm 0.13 (stat) \pm 0.11 (syst) \\
\langle n\rangle_{\mathrm{183\,GeV}} = 27.05 \pm 0.27 (stat) \pm 0.32 (syst)
\label{mulv183} \\
D_{\mathrm{183\,GeV}} = 8.08 \pm 0.19 (stat) \pm 0.14 (syst)\, .
\end{eqnarray}
%were obtained for the average charge multiplicity and for the dispersion.

These values include the products of the
decays of particles with lifetime $\tau < 10^{-9}$ s.

The systematic errors were obtained by adding in quadrature:

\begin{enumerate}
\item  the propagated uncertainty of the average values in the Z$^0$
correction factors, $\pm 0.18$ ($\pm 0.17$) for the multiplicity.%
% and $\pm 0.09$ ($\pm 0.08$)  for the dispersion.

\item  the effect of the cuts for the reduction of the background. The value
of the cut on the narrow jet broadening was varied from 0.045 to 0.085 in
steps of 0.010, in order to estimate the systematic error associated with
the procedure of removing the contribution from WW events. The new values
for the average charged particle multiplicity and the dispersion were stable
within these variations, and half of the difference between the extreme
values, %the highest shifts with respect to the previous results, 
0.07 and 0.07 (0.06 and 0.12) respectively, were added in quadrature to the
systematic error. The effect of the uncertainty on the WW cross-section was
found to be negligible.

\item  the uncertainty on the modelling of the detector response in the
forward region. %The analysis
%was repeated by using only the tracks with $40^{\circ}<\theta<140^{\circ}$. The
%values obtained for the multiplicity 
%and the dispersion were 26.73$\pm0.24$ and 8.50$\pm0.12$ 
%(26.16$\pm0.37$ and 8.34$\pm$0.18) respectively, 
%and uncertainties of 0.17 and 0.20 
%(0.48 and 0.09) were computed after subtracting in quadrature the difference
%between the statistical errors.
The analysis was repeated by varying the polar angle acceptance of charged
particles from 10-170 degrees to 40-140 degrees, both in the high energy
samples and in the computation of the Z$^0$ correction factors. The spread of
the different values obtained for the multiplicities and for the dispersions
were found to be respectively 0.18 and 0.08 (0.23 and 0.03).
The effect of the variation of other track selection criteria was
found to be negligible, and the same applies to the higher 
centre-of-mass energies.

\item  the systematic errors due to the statistics of the simulated
samples, 0.04 (0.06) for the multiplicity and 0.04 (0.06) for the dispersion.

\item  the uncertainty on the calculation of the efficiency correction
factors in the multiplicity. The values of the multiplicities, before applying
the Z$^0$ correction factors, were also estimated:
\begin{itemize}
\item from the observed multiplicity distribution as 27.37 (26.56);
\item from the integral of the rapidity distribution (with respect to the
thrust axis), $y_{T}=\frac{1}{2}\ln 
\frac{E+p_{||}}{E-p_{||}}$ ($p_{||}$ is the absolute value of the momentum
component on the thrust axis) as 27.43 (26.59).
\end{itemize}
Half of the
differences between the maximum and the minimum values of the
multiplicity calculated from the multiplicity distribution and
from the integration of the $y_{T}$ and $\xi_{E}$ distributions, 0.03 in both
cases, were added in quadrature to the systematic error.

\item  Half of the extrapolated multiplicity in the high-$\xi _{E}$ region,
0.14 (0.12).
\end{enumerate}

As a cross-check, a 
simulated sample based on HERWIG plus DELSIM was also used to unfold the 
data; the results were consistent with those based on PYTHIA
plus DELSIM within the statistical error associated to the size of the
Monte Carlo sample. 

% Figures 1, 2 and their discussion moved to section 8 (Conclusions).

% NEW: 1999 data

\subsection{Centre-of-mass energies of 192 to 200 GeV}

%Data taken in 1999 corresponding to total luminosities of 
%25.8~pb$^{-1}$, 77.4~pb$^{-1}$ and 83.8~pb$^{-1}$ %and 40.2~pb$^{-1}$ 
%%25.84~pb$^{-1}$, 77.42~pb$^{-1}$, 83.81~pb$^{-1}$ and 40.15~pb$^{-1}$ 
%%collected by DELPHI 
%at centre-of-mass energies around 192~GeV, 196~GeV and 200~GeV %and 202~GeV 
%respectively were analysed. 
%
%For each one of the energies the preselection of hadronic events, 
%the study of the influence of the detector on the analysis and the 
%calculation of the hadronic centre-of-mass energy, \mbox{$\sqrt{s^\prime}$},
%were performed separately, following the procedures described in 
%section 2.
%
To check the ability of the simulation to model the efficiency for the
reconstruction of charged particles, the sample collected at the Z$^0$ 
calibration runs of 1999 was used, following the procedure described in
section 2. The average  
charged particle multiplicity at the Z$^0$ was measured to be 
$20.82 \pm 0.03 (stat)$, in satisfactory agreement with the world average.
The ratio of the world average value to the measured multiplicity at the
Z$^0$, $1.0084 \pm 0.0064$, was used to correct the measured multiplicities at
centre-of-mass energies of 192 to 200 GeV.

For each of the energies a separate analysis was performed following the 
procedure described in the previous subsection. 

%Hadronic events were selected by requiring that the total transverse energy of
%the charged particles exceeded $0.2\sqrt{s}$, the multiplicity of charged 
%particles was larger than 9, the effective centre-of-mass energy, 
%\mbox{$\sqrt{s^\prime}$}, exceeded $0.9\sqrt{s}$, and that the narrow jet 
%broadening~\cite{lepII} was smaller than 0.065.

The number of events selected, the number of expected signal and background
events,  
estimated with Monte Carlo simulation, and 
the measured multiplicities and dispersions are listed in
Table~\ref{tab:mult99} for the centre-of-mass energies of 192 to 200~GeV.
The systematic errors were estimated as in 3.1;
a breakdown is shown in the table (the numbering of the sources of 
systematic error corresponds to the one in the previous subsection). 

%Table 1
\begin{table}
\begin{center}
\begin{tabular}{|c | c | c | c |}% c |}
\hline
                  & 192~GeV & 196~GeV & 200~GeV \\ %& 202~GeV \\
\hline      
Selected events   
                  &     526 &    1542 &    1580 \\ %&     760 \\
\hline      
Expected q$\bar{\mathrm q}$ events   
                  &     431 &    1253 &    1277 \\ %&     609 \\
%\hline      
%Expected WW bkg. 
%                  &      76 &     241 &     278 \\ %&     135 \\
%\hline      
%Expected Z$^0$Z$^0$ bkg. 
%                  &       8 &      26 &      29 \\ %&      14 \\
\hline      
Expected background events
                  &       84 &     267 &     307 \\ %&     149 \\
%\hline      
\hline\hline      
$\langle n\rangle$&   27.19 &   27.42 &   27.52 \\ %&   28.09 \\
statistical error &    0.47 &    0.28 &    0.29 \\ %&    0.42 \\
systematic error  &    0.51 &    0.37 &    0.44 \\ %&    0.41 \\
\hline
(syst. 1)  &    0.17 &    0.17 &    0.17 \\ %&    0.41 \\      
(syst. 2)  &    0.35 &    0.08 &    0.18 \\ %&    0.41 \\      
(syst. 3)  &    0.29 &    0.27 &    0.30 \\ %&    0.41 \\      
(syst. 4)  &    0.06 &    0.05 &    0.08 \\ %&    0.41 \\      
(syst. 5)  &    0.03 &    0.01 &    0.01 \\ %&    0.41 \\      
(syst. 6)  &    0.15 &    0.17 &    0.19 \\ %&    0.41 \\      
\hline\hline
$D$               &    8.57 &    8.52 &    8.69 \\ %&    8.75 \\
statistical error &    0.34 &    0.20 &    0.21 \\ %&    0.30 \\
systematic error  &    0.32 &    0.14 &    0.19 \\ %&    0.32 \\
\hline
(syst. 2)  &    0.30 &    0.12 &    0.16 \\ %&    0.41 \\      
(syst. 3)  &    0.09 &    0.06 &    0.05 \\ %&    0.41 \\      
(syst. 4)  &    0.06 &    0.05 &    0.09 \\ %&    0.41 \\      
\hline
\end{tabular}
\end{center}
\caption{Number of events selected, number of events expected for the signal
(q$\bar{\mathrm q}\gamma$) and for the background (WW and ZZ), estimated
from simulation, and the measured multiplicities and dispersions for the three 
different energies.}   
\label{tab:mult99}
\end{table}  

These results were then combined at an average centre-of-mass energy of 
200 GeV, according to the following procedure. First each result was rescaled 
to a centre-of-mass energy of 200 GeV, calculating the scaling factors from 
the simulation. Then a weighted average was computed using the statistical 
error as a weight. The systematic error is taken as the weighted average of 
the systematic errors, increased (in quadrature) by the difference between the 
values obtained when rescaling and when not rescaling to 200 GeV. This gives:
\begin{eqnarray}
%202 \langle n\rangle_{\mathrm{200\,GeV}} = 27.63 \pm 0.17 (stat) \pm 0.43 (syst)
\langle n\rangle_{\mathrm{200\,GeV}} = 27.58 \pm 0.19 (stat) \pm 0.45 (syst)
\label{mulv200} \\
%D_{\mathrm{200\,GeV}} = 8.65 \pm 0.12 (stat) \pm 0.21 (syst) \, .
D_{\mathrm{200\,GeV}} = 8.64 \pm 0.13 (stat) \pm 0.20 (syst) \, .
\end{eqnarray}

As a cross-check, a 
simulated sample based on HERWIG plus DELSIM was also used to unfold the 
data; the results were consistent with those based on PYTHIA
plus DELSIM within the statistical error associated to the size of the
Monte Carlo sample.

\section{Classification of the WW Events and Charged Multiplicity Measurement}

About 4/9 of the WW events are 
WW $\rightarrow {\mathrm q_1\bar{q_2}q_3\bar{q_4}}$ events. At
threshold, their topology is that of two pairs of back-to-back jets, with no
missing energy; the constrained invariant mass of two jet-jet systems is
close to the W mass. Even at 183 and 189 GeV these characteristics allow a
clean selection.

Another 4/9 of the WW events are WW $\rightarrow {\mathrm q_1\bar{q_2}} \, \ell
\bar{\nu}$ events. At threshold, their topology is 2-jets back-to-back, with a
lepton and missing energy opposite to it; the constrained invariant mass of
the jet-jet system and of the lepton-missing energy system equals the W mass.

\subsection{Fully Hadronic Channel 
(WW $\rightarrow {\mathbf q_1\bar{q_2}q_3\bar{q_4}}$)} 
%WW\protect\boldmath{$\rightarrow 4j$})}

Events with both Ws decaying into q$\bar{\mathrm q}$ are characterised by high
multiplicity, large visible energy, and tendency of the particles to be
grouped in 4 jets. The background is dominated by q$\bar{\mathrm q}(\gamma)$
events. 

The events were pre-selected by requiring at least 12 charged particles
(with $p>100$~MeV/$c$), with a total transverse energy (charged plus
neutral) above 20\% of the centre-of-mass energy. To remove the radiative
hadronic events, the effective hadronic centre-of-mass energy $%
\mbox{$\sqrt{s^\prime}$}$, computed as described in section~2,
was required to be above 110~GeV.

The particles in the event were then clustered to 4 jets using the LUCLUS
algorithm~\cite{lund}, and the events were kept if all jets had multiplicity
(charged plus neutral) larger than 3. It was also required that the
separation between the jets ($d_{\mathrm{join}}$ value) be larger than
6~GeV/c.
The combination of these two cuts removed most of the remaining semi-leptonic WW
decays and the 2-jet and 3-jet events of the q$\bar{\mathrm q}\gamma$
background. 

A five constraint fit was applied, imposing energy and momentum conservation
and the equality of two di-jet masses. Of the three fits obtained by
permutation of the jets, the one with the smallest $\chi^2$ was selected.
Events were accepted only if 
\[
D_{\mathrm sel} = \frac{E_{min}\theta_{min}}{E_{max}(E_{max}-E_{min})} >
0.004 \, \mbox{rad GeV}^{-1} 
\]
where $E_{min}$ and $E_{max}$ are respectively the smallest and the largest
fitted jet energy, and $\theta_{min}$ is the smallest angle between the
fitted jet directions. The details of the selection variable $D_{\mathrm sel}$
can be found in~\cite{Wdsel}. The purity and the efficiency of the selected
data sample from the 189 (183) GeV data were estimated using simulation to
be about 76\% and 80\% (75\% and 80\%) respectively. The data sample
consists of 1256 (427) events, where 1255 (422) were expected from the
simulation. The expected background was subtracted bin by bin from the
observed distributions, which were then corrected bin by bin using scaling
factors computed from the simulation
generated using PYTHIA with the JETSET fragmentation tuned by DELPHI 
(EXCALIBUR plus JETSET for the 1999 data)
without initial state radiation.

Finally, the average multiplicity of charged particles 
$\langle n^{(4q)}\rangle$ was
estimated by integrating 
the $\xi_E$ distribution up to a value of 6.3 (and estimated above this
value with simulation)
and multiplying by the Z$^0$ correction factors from section 2.
The following
values were obtained: 
\begin{eqnarray}
\langle n^{(4q)}\rangle_{\mathrm{189\,GeV}} & = & 39.12 \pm 0.33 (stat) \pm 0.36 (syst)
\label{mulha} \\
\langle n^{(4q)}\rangle_{\mathrm{183\,GeV}} & = & 38.11 \pm 0.57 (stat) \pm 0.44 (syst)
\, .  \label{mulha2}
\end{eqnarray}

The systematic errors account for:

\begin{enumerate}
\item  The propagated uncertainty of the average values in the Z$^0$
correction factors, $\pm 0.24$ ($\pm 0.24$). 
%  for the multiplicity and $\pm 0.09$ ($\pm 0.08$)  for the dispersion.
%\item Scale uncertainty 0.38 (0.38).

\item  The spread of the measured values from the reference values by
redoing the analysis varying the selection criteria, 0.05 (0.11).

\item  Modelling of the detector in the forward region. The analysis was
repeated by varying the polar angle acceptance of charged particles from
10-170 degrees to 40-140 degrees, both in the WW samples and in the
computation of the Z$^0$ correction factors. The spreads of the different
measured values were found to be 0.01 (0.08). 
%To investigate a possible dependence of the measured multiplicity on
%the modeling of the detector response in the forward region, the analysis
%was repeated by using only the tracks with $40^{\circ}<\theta<140^{\circ}$. 
%The value obtained for the multiplicity and dispersion
%were $38.34 \pm 0.41$ and $8.43 \pm 0.23$ ($37.38 \pm 0.73$ 
%and $8.77 \pm 0.41$).
%The difference between this multiplicity and the reference ones 
%is $0.02 \pm 0.26$ ($0.41 \pm 0.46$),
%assuming full correlation between this measurement 
%(which contains the same events and subsample of tracks)
%and the reference in \ref{mulha} (\ref{mulha2})
%for the calculation of the  statistical errors.
%Since this source of uncertainty is compatible with its statistical error,
%it is assumed negligible.

\item  Limited statistics in the simulated sample 0.03 (0.05).

\item  Variation of the q$\bar{\mathrm q}\gamma$ cross-sections within 5\%: 0.01
(0.01). 

\item  Calculation of the correction factors. The value of 
$\langle n^{(4q)}\rangle$, before applying the Z$^0$ correction factors, was also estimated:

\begin{itemize}
\item  from the observed multiplicity distribution as 38.96 (37.39);

\item  from the integral of the rapidity distribution (with respect to the
thrust axis) as 39.16 (36.97).

\item  from the integral of the $p_{T}$ distribution (with respect to the
thrust axis) as 38.95 (37.42).
\end{itemize}

Half of the difference between the maximum and the minimum value, 0.11 (0.23),
was added in quadrature to the systematic error.

\item  Uncertainty on the modelling of the 4-jets q$\bar{\mathrm q}$ background, 0.05
(0.14). The uncertainty on the modelling of this background is the sum in
quadrature of %three contributions:
two contributions:

\begin{itemize}
%\item Statistical uncertainty on the q$\bar{\mathrm q}$ multiplicity.
%A relative uncertainty as in
%Eq.~(\ref{mulv189}) (Eq.~(\ref{mulv183}))
%was assumed; this gives a multiplicity error of 0.06 (0.10).

\item  Uncertainty on the modelling of the 4-jet rate. The agreement between
data and simulation was studied in a sample of 4-jet events at the Z$^0$,
selected with the DURHAM algorithm for 
$% 
y_{cut}$ ranging from 0.003 to 0.005. The rate of 4-jet events in the
simulated sample was found to reproduce the data within 10\%. The correction
due to background subtraction was correspondingly varied by 10\%, which
gives an uncertainty of 0.00 (0.01).

\item  Uncertainty on the multiplicity in 4-jet events. The average
multiplicity in 4-jet events selected at the Z$^0$ data in 1998 (1997), with
the DURHAM algorithm 
for a value of $y_{cut}=0.005$, is larger by $0.9\%\pm 0.3\%(stat)$ 
($2.0\%\pm 0.5\%(stat)$) than the corresponding value in the simulation. A
shift by 0.9\% (2.0\%) in the multiplicity for 4-jet events induces a shift
of 0.05 (0.14) on the value in Equation~(\ref{mulha}) (Equation (\ref{mulha2})).
\end{itemize}

\item  Half of the extrapolated multiplicity in the high-$\xi _{E}$ region,
0.23 (0.20).
\end{enumerate}

The presence of interference between the jets coming from the different Ws
could create subtle effects, such as to make the application of the fit
imposing equal masses inadequate. For this reason a different four
constraint fit was performed, leaving the di-jet masses free and imposing
energy-momentum conservation. Of the three possible combinations of the four
jets into WW pairs, the one with minimum mass difference was selected. No $%
\chi^2$ cut was imposed in this case. The average multiplicity obtained was
again fully consistent (within the statistical error) with the one measured
in the standard analysis.

A simulated sample based on HERWIG plus DELSIM was also used to unfold the 
data; the results were consistent with those based on PYTHIA
plus DELSIM within the statistical error associated to the size of the
Monte Carlo sample.

\begin{figure}[t]
%\centerline{DELPHI preliminary}
\begin{center}
\resizebox{0.5\textwidth}{0.44\textheight}{\includegraphics{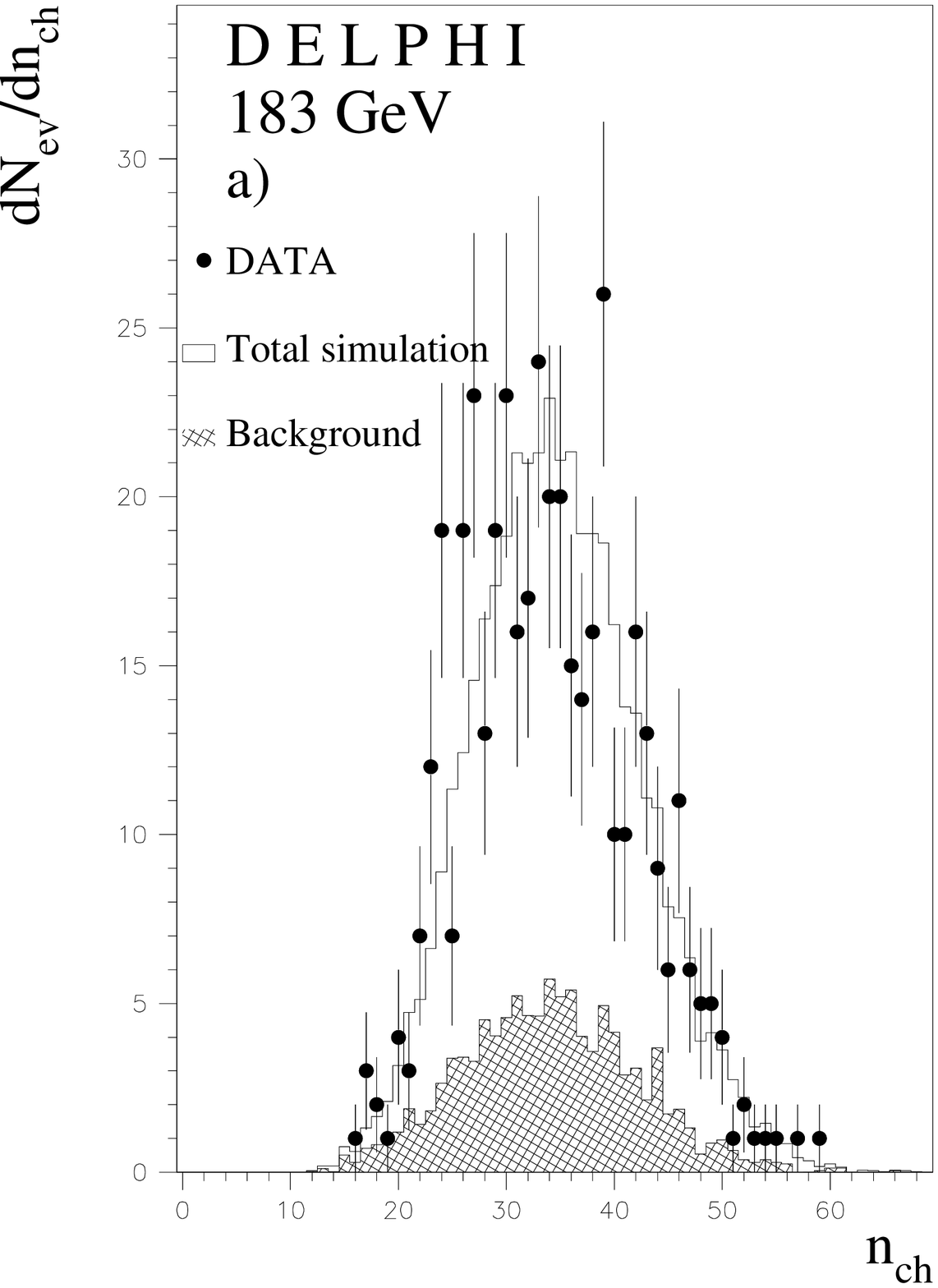}}\resizebox{0.5\textwidth}{0.44\textheight}{\includegraphics{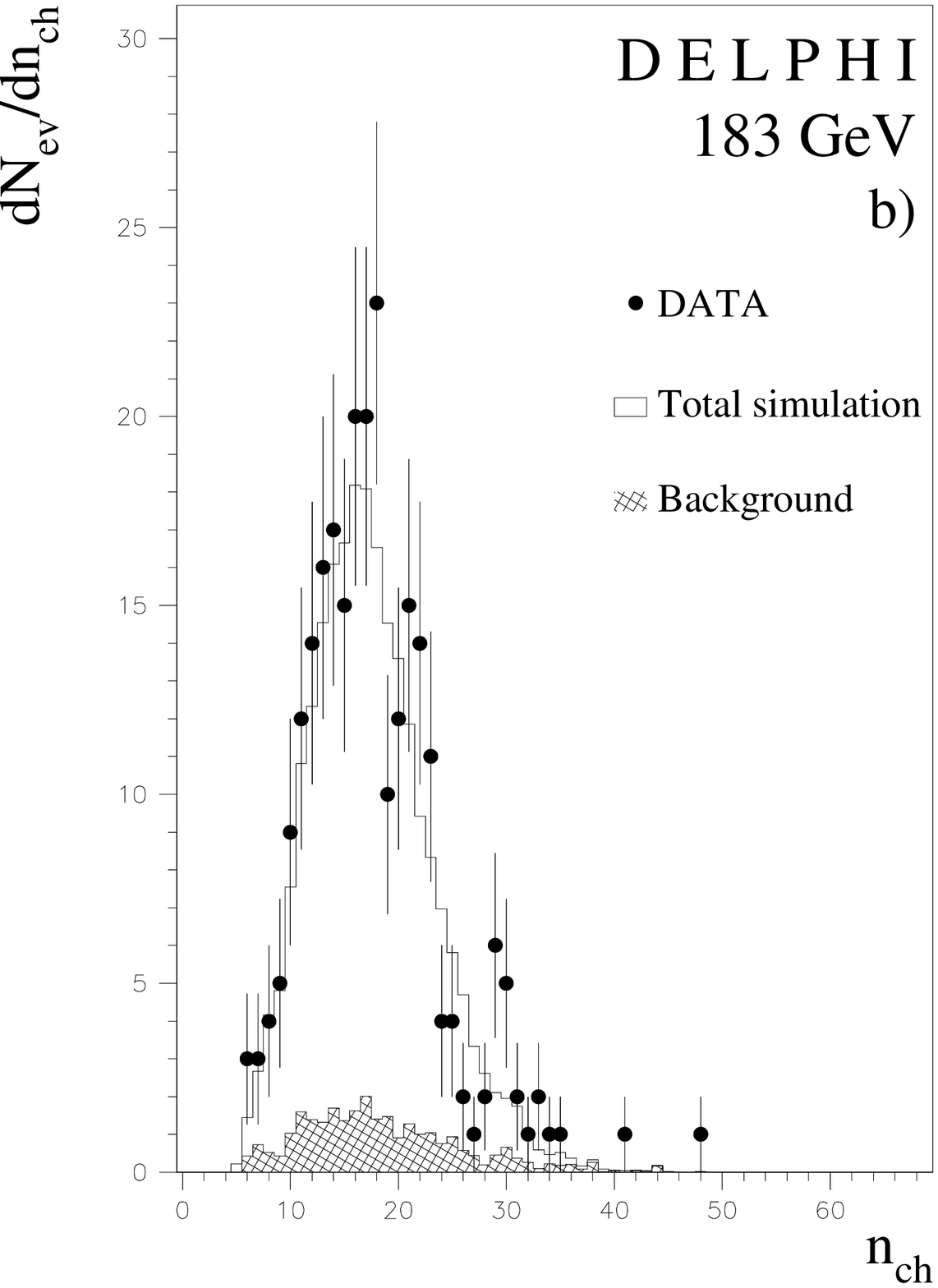}}

\resizebox{0.5\textwidth}{0.44\textheight}{\includegraphics{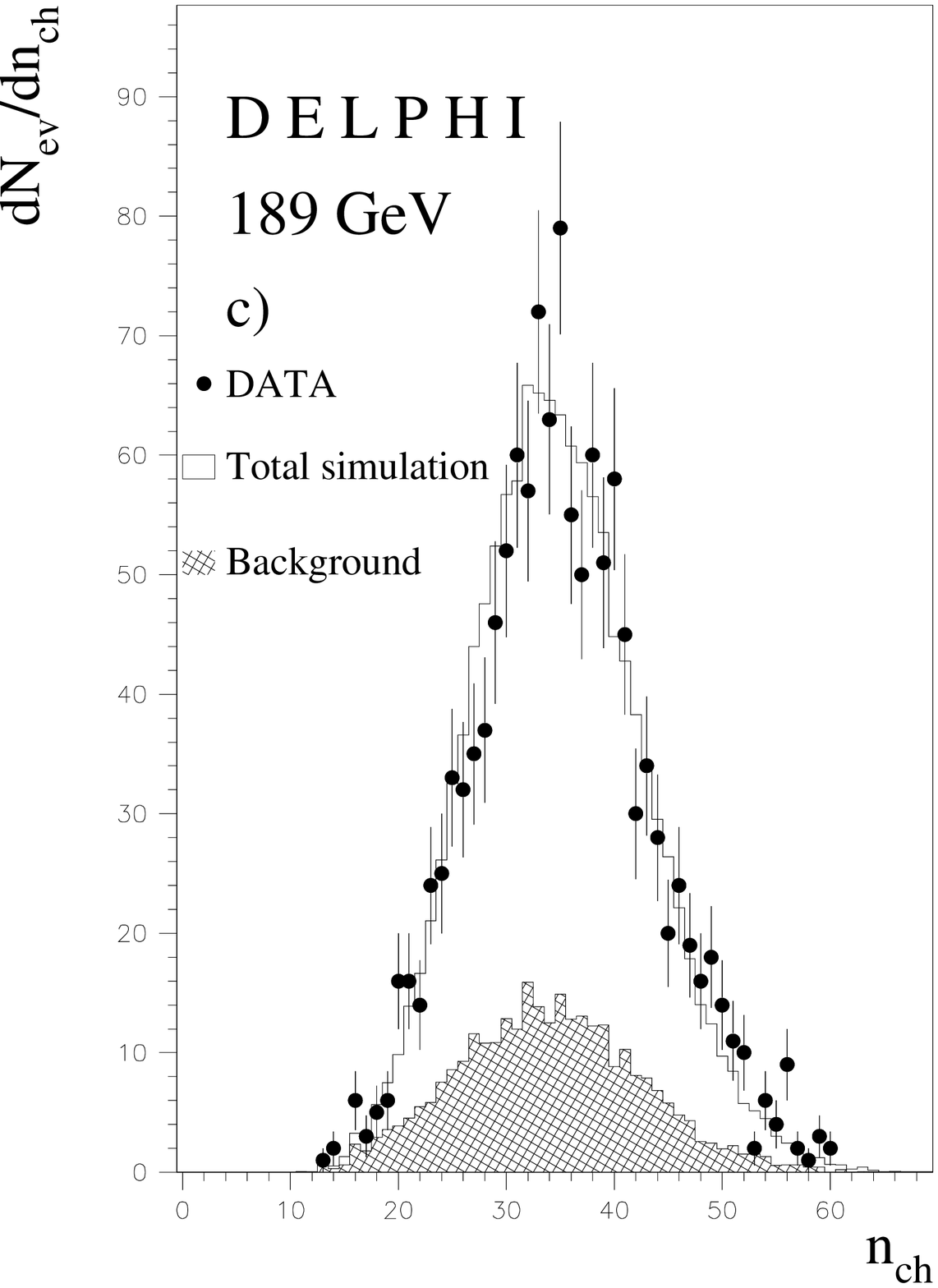}}\resizebox{0.5\textwidth}{0.44\textheight}{\includegraphics{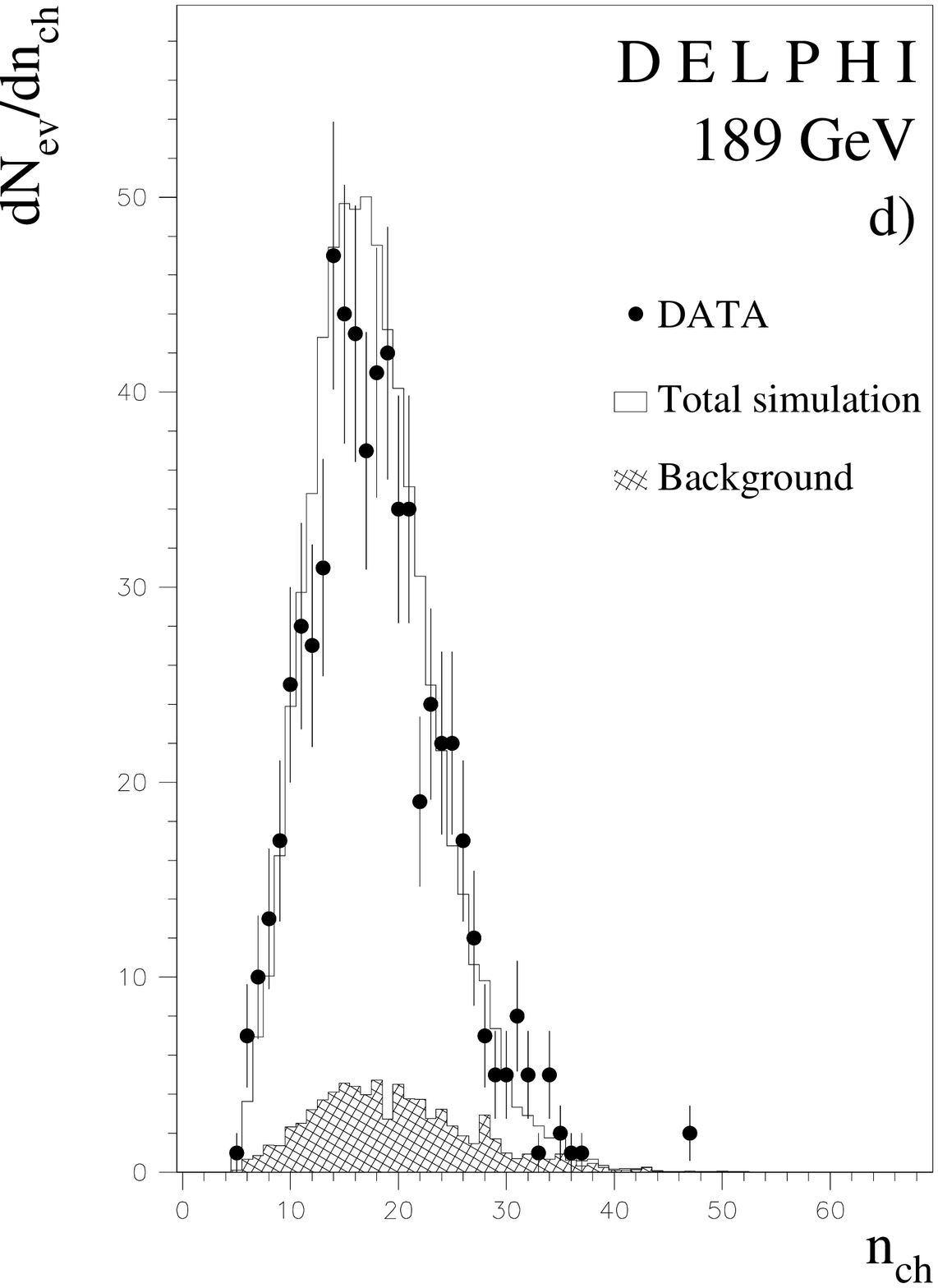}}
\end{center}
\caption[]{
Charged particle multiplicity distributions for (a) the $(4q)$ events
and (b) the $(2q)$ events at 183 GeV, for (c) the $(4q)$ events
and (d) the $(2q)$ events at 189 GeV. 
The error bars in the data represent the statistical errors.
The shaded areas represent the background
contribution; the histograms are the sum of the expected signal
and background.}
\label{wmul}
\end{figure}

The distribution of the observed charged particle multiplicity in $(4q)$ is
shown in Figure~\ref{wmul}c(\ref{wmul}a).

The value of the corrected multiplicity in the low momentum range 0.1 to
1. GeV/$c$, 
where the interconnection effects are expected to be most important, 
was found to be 14.47$\pm0.20$ (13.67$\pm0.34$), where the errors
are statistical only.
%14.12$\pm0.12$ (14.06$\pm0.21$). 

%The dispersion of the
%multiplicity distribution in the data was $6.69 \pm 0.31(stat)$, 
%to be compared to
%the dispersion from the q$\bar{\mathrm q}$ PS simulation of 
%$6.39 \pm 0.04(stat)$.
After correcting for detector effects, the dispersion was found to be: 
\begin{eqnarray}
D^{(4q)}_{\mathrm{189\,GeV}} & = & 8.72 \pm 0.23 (stat) \pm 0.11 (syst) \\
D^{(4q)}_{\mathrm{183\,GeV}} & = & 8.53 \pm 0.39 (stat) \pm 0.16 (syst) \, .
\end{eqnarray}
In the systematic error:

\begin{enumerate}
%\item 0.08 (0.09) accounts for the scale uncertainty;

\item  0.10 (0.15) accounts for the spreads of the measured values from the
reference value when varying the event selection criteria;

\item  0.03 (0.01) is due to the modelling of the detector in the forward
region. The dispersions were also measured using only charged particles with
polar angle between 40 and 140 degrees and the differences with respect to
the reference value were considered in the systematic error;

\item  0.03 (0.05) from the limited simulation statistics.
\end{enumerate}

\subsection{Mixed Hadronic and Leptonic Final States 
(WW\protect\boldmath{$\rightarrow {\mathrm q_1\bar{q_2}}l\protect\nu$})}
%$\rightarrow 2jl\protect\nu$})}

Events in which one W decays into lepton plus neutrino and the other one
into quark and antiquark are characterised by two hadronic jets, one energetic
isolated 
charged lepton, and missing momentum resulting from the neutrino. The main
backgrounds to these events are radiative q$\bar{\mathrm q}$ production and
four-fermion final states containing two quarks and two oppositely charged
leptons of the same flavour.

Events were selected by requiring seven or more charged particles, with a
total energy (charged plus neutral) above $0.2\sqrt{s}$ and a missing
momentum larger than $0.1\sqrt{s}$. Events in the q$\bar{\mathrm q}\gamma$ final
state with ISR photons at small polar angles, which would be lost inside the
beam pipe, were suppressed by requiring the polar angle of the missing
momentum vector to satisfy $|\cos\theta_{miss}|<0.94$.

Including the missing momentum as an additional massless neutral particle
(the candidate neutrino), the particles in the event were clustered to 4 jets
using the DURHAM algorithm. 
%The distribution of the value for the jet resolution parameter 
%$y_{\mbox{cut}}$needed to differentiate 3 to 4 jets (di-jet separation) 
%is shown in Figure~\ref{yc34}.
The jet for which the fractional jet energy carried by the highest momentum
charged particle was greatest was considered as the ``lepton jet''. The most
energetic charged particle in the lepton jet was taken as the lepton
candidate, and the event was rejected if its momentum was smaller than
10~GeV/$c$. %or greater than 65~GeV/$c$.
%The ``neutrino jet'' was considered the jet clustered around the missing
The neutrino was taken to correspond to the missing
momentum. The event was discarded if the invariant mass of the event
(excluding the lepton candidates) was smaller than 20~GeV/$c^2$ or larger
than 110~GeV/$c^2$.

At this point three alternative topologies were considered:

\begin{itemize}
\item  Muon sample: if the lepton candidate was tagged as a muon and its
isolation angle, with respect to other charged particles above 1~GeV/$c$,
was above 10$^{\circ }$, the event was accepted either if the lepton
momentum was greater than 20~GeV/$c$, or if it was greater than 10~GeV/$c$
and the value of the $y_{3\rightarrow 4}^{cut}$ parameter required by
the DURHAM algorithm to force the event from a 3-jet to a 4-jet
configuration was greater than 0.003.

\item  Electron sample: if the lepton candidate had associated
electromagnetic energy deposited in the calorimeters larger than 20~GeV,  and
an isolation angle (defined as above) greater than 10$^\circ$, the event was
accepted if the required value of $y_{3\rightarrow 4}^{cut}$ was
greater than 0.003.

\item  Inclusive sample: the events were also accepted if the momentum of
the lepton candidate was larger than 20~GeV/$c$ and greater than half of 
its associated energy,
the missing momentum was larger than $0.1\sqrt{s}$, the required value of $%
y_{3\rightarrow 4}^{cut}$ was greater than 0.003, and no other
charged particle above 1~GeV/$c$ existed in the lepton jet.
\end{itemize}

The purity and the efficiency of the selected data sample from the 189 (183)
GeV data were estimated using simulation to be about 89\% and 54\% (88\% and
55\%) respectively. The data sample consists of 633 (256) events, where 689
(235) were expected from the simulation. The expected background was
subtracted bin by bin from the observed distributions, which were then
corrected bin by bin using scaling factors computed from the simulation
generated using PYTHIA with the JETSET fragmentation tuned by DELPHI 
(EXCALIBUR plus JETSET for the 1999 data)
without initial state radiation.

By integrating the $\xi_E$ distribution up to a value of 6.3 (and estimated
above this value from simulation at generator level),
and multiplying by the
Z$^0$ correction factors from section 2, the following values were obtained
for the charged multiplicity for 
one W decaying hadronically in a WW event with mixed hadronic and leptonic
final states: 
\begin{eqnarray}
\langle n^{(2q)}\rangle_{\mathrm{189\,GeV}} & = & 19.49 \pm 0.31 (stat) \pm 0.27 (syst)
\label{mulle} \\
\langle n^{(2q)}\rangle_{\mathrm{183\,GeV}} & = & 19.78 \pm 0.49 (stat) \pm 0.43 (syst)
\, .  \label{mulle2}
\end{eqnarray}

In the systematic error:

\begin{enumerate}
%\item 0.20 (0.20) accounts for scale uncertainty, 

\item  0.12 (0.12) accounts for the propagated uncertainty of the 
average values in the Z$^0$ correction factors;

\item  0.16 (0.16) accounts for the spreads from the reference values when
changing the event selection criteria;

\item  0.05 (0.02) accounts for modelling of the detector in the forward
region. The analysis was repeated by varying the polar angle acceptance of
charged particles from 10-170 degrees to 40-140 degrees, both in the WW
samples and in the computation of the Z$^0$ correction factors. The spreads of
the different measured values were found to be 0.05 (0.02); 
%To investigate a possible dependence of the measured multiplicity on
%the modelling of the detector response in the forward region, the analysis
%was repeated by using only the tracks with $40^{\circ}<\theta<140^{\circ}$. 
%The values obtained for the multiplicity 
%and the dispersion were $19.14\pm0.37$ and $6.45\pm0.20$ 
%($20.02\pm0.62$ and $6.83\pm0.34$) respectively. 
%The difference between these values and the reference ones 
%are $0.47 \pm 0.20$ ($0.05 \pm 0.35$).
%The uncertainties of 0.43 and 0.00 were 
%obtained by subtracting in square the statistical error.

\item  0.03 (0.05) accounts for limited statistics in the simulated samples;

\item  the variation of the q$\bar{\mathrm q}(\gamma )$ cross-sections within 5\%,
gives a negligible contribution to the systematic error;

\item  0.12 (0.36) accounts for the uncertainty on the correction factors.
The value of $\langle n^{(2q)}\rangle$, before applying the Z$^0$ correction factors, was
also estimated:

\begin{itemize}
\item  from the observed multiplicity distribution as 19.53 (19.60);

\item  from the integral of the rapidity distribution (with respect to the
thrust axis) as 19.30 (18.88).

\item  from the integral of the $p_{T}$ distribution (with respect to the
thrust axis) as 19.39 (19.36).
\end{itemize}

Half of the difference between the maximum and the minimum value, 0.12 (0.36),
was included in the systematic error.

\item  Half of the extrapolated multiplicity in the high-$\xi _{E}$ region,
0.12 (0.10).
\end{enumerate}

A simulated sample based on HERWIG plus DELSIM was also used to unfold the 
data; the results were consistent with those based on PYTHIA
plus DELSIM within the statistical error associated to the size of the
Monte Carlo sample.

The distribution of the observed charged particle multiplicity in $(2q)$ is
shown in Figure~\ref{wmul}d(\ref{wmul}b).

The value of the corrected multiplicity in the low momentum range 0.1 to
1. GeV/$c$ 
was found to be 7.29$\pm0.19$ (7.15$\pm0.28$) (where the errors are 
statistical only). 
%7.31$\pm0.11$ (7.59$\pm0.17$).

After correcting for detector effects, the dispersions were found to be: 
\begin{eqnarray}
D^{(2q)}_{\mathrm{189\,GeV}} & = & 6.49 \pm 0.21 (stat) \pm 0.43 (syst) \\
D^{(2q)}_{\mathrm{183\,GeV}} & = & 6.51 \pm 0.33 (stat) \pm 0.25 (syst) \, .
\end{eqnarray}
In the systematic error:

\begin{enumerate}
%\item 0.06 (0.06) accounts for the scale uncertainty;

\item  0.14 (0.08) accounts for the variation of the cuts;

\item  0.41 (0.23) accounts for the modelling of the detector in the forward
region. The dispersions were also measured using only charged particles with
polar angle between 40 and 140 degrees and the differences with respect to
the reference value were considered in the systematic error;

\item  0.03 (0.05) is due to the limited simulation statistics.
\end{enumerate}

\section{Analysis of Interconnection Effects from Charged Particle
Multiplicity and Inclusive Distributions}

Most models predict that, in case of colour reconnection, the ratio between
the multiplicity in $(4q)$ events and twice the multiplicity in $(2q)$
events would be smaller than 1; the difference is expected to be at the percent
level. It was measured: 
\begin{eqnarray}
\left(\frac{\langle n^{(4q)}\rangle}{2\langle n^{(2q)}\rangle}\right)_{\mathrm{189\,GeV}} & = & 1.004
\pm 0.018 (stat) \pm 0.014 (syst) \\
\left(\frac{\langle n^{(4q)}\rangle}{2\langle n^{(2q)}\rangle}\right)_{\mathrm{183\,GeV}} & = & 0.963
\pm 0.028 (stat) \pm 0.015 (syst) \, .
\end{eqnarray}
In the calculation of the systematic error on the ratio, the correlations
between the sources of systematic error were taken into account.
%When taking the systematic errors as uncorrelated, 
If the systematic errors are taken as uncorrelated, 
except for the errors on the Z$^0$ correction factors and the modelling of the
detector in the forward region, for which full correlation is assumed, a
compatible value of $\pm 0.014$ ($\pm 0.022$) is obtained for the
systematic error.
% when correlations are explicitly computed for modelling of
%the detector in the forward region, %; 0.025$\pm0.014$=0.020 
%(0.007$\pm0.020$=negligible)
%change in the q$\bar{\mathrm q}$ cross section and computation of multiplicity using
%the multiplicity, $y$ and $p_T$ distributions. %$0.007(0.013)$

% total .990 +- .015 +- .011
Using as weights the inverse of the sum in quadrature of the statistical and
systematic  
errors, one obtains a weighted average of
\begin{equation}
\left(\frac{\langle n^{(4q)}\rangle}{2\langle n^{(2q)}\rangle}\right)  =  0.990
\pm 0.015 (stat) \pm 0.011 (syst) .
\end{equation}

\begin{figure}[ph]
%\centerline{DELPHI preliminary}
\begin{center}
%Only statistical errors
\resizebox{\textwidth}{0.88\textheight}{\includegraphics{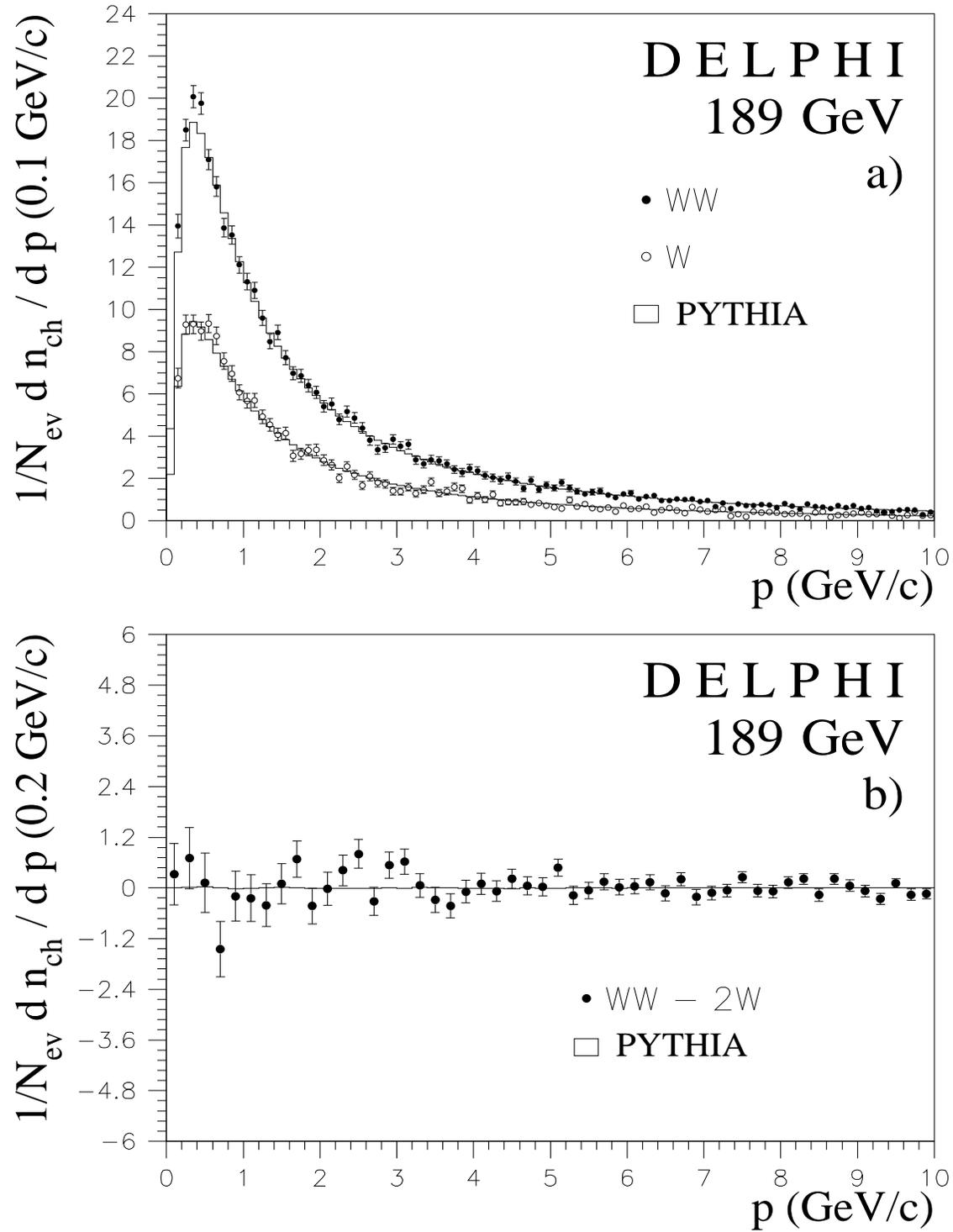}}
\end{center}
\caption[]{\label{xp1} (a)
Corrected momentum 
distributions of charged particles for $(4q)$ events (closed circles) and 
$(2q)$ events (open circles), compared to simulation
without colour reconnection, at 189 GeV.
The error bars in the data represent the statistical errors.
The difference between $(4q)$ and twice $(2q)$ is shown in (b).}
\end{figure}

\begin{figure}[ph]
%\centerline{DELPHI preliminary}
\begin{center}
%Only statistical errors
\resizebox{\textwidth}{0.88\textheight}{\includegraphics{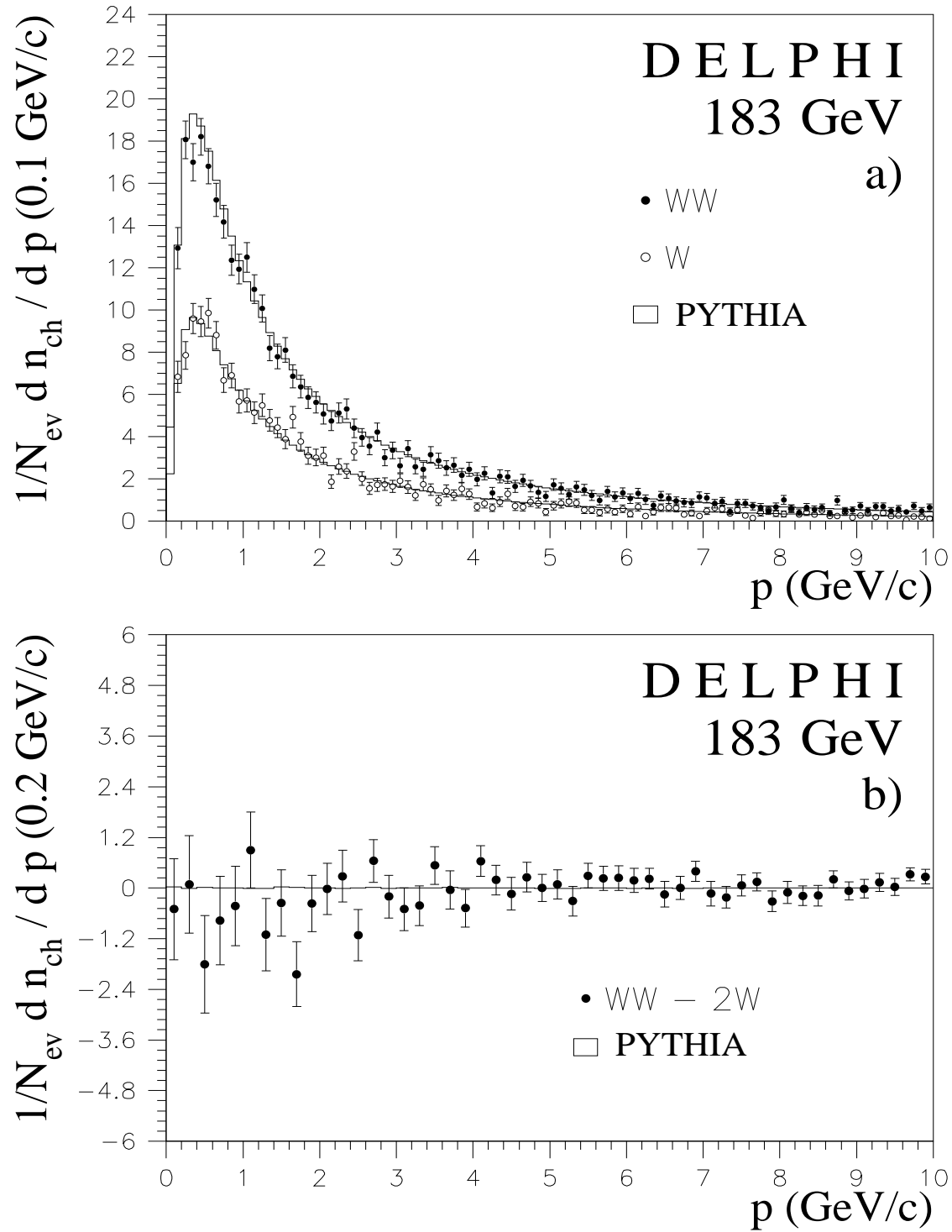}}
\end{center}
\caption[]{\label{xp2} (a)
Corrected momentum 
distributions of charged particles for $(4q)$ events (closed circles) and 
$(2q)$ events (open circles), compared to simulation
without colour reconnection, at 183 GeV.
The difference between $(4q)$ and twice $(2q)$ is shown in (b).}
\end{figure}

\begin{figure}[ph]
%\centerline{DELPHI preliminary}
\begin{center}
%\mbox{\epsfxsize7.0cm\epsfig{file=ww_2w_y_189.eps}}
%\mbox{\epsfxsize7.0cm\epsfig{file=ww_2w_y_183.eps}}
%Only statistical errors
\resizebox{\textwidth}{0.88\textheight}{\includegraphics{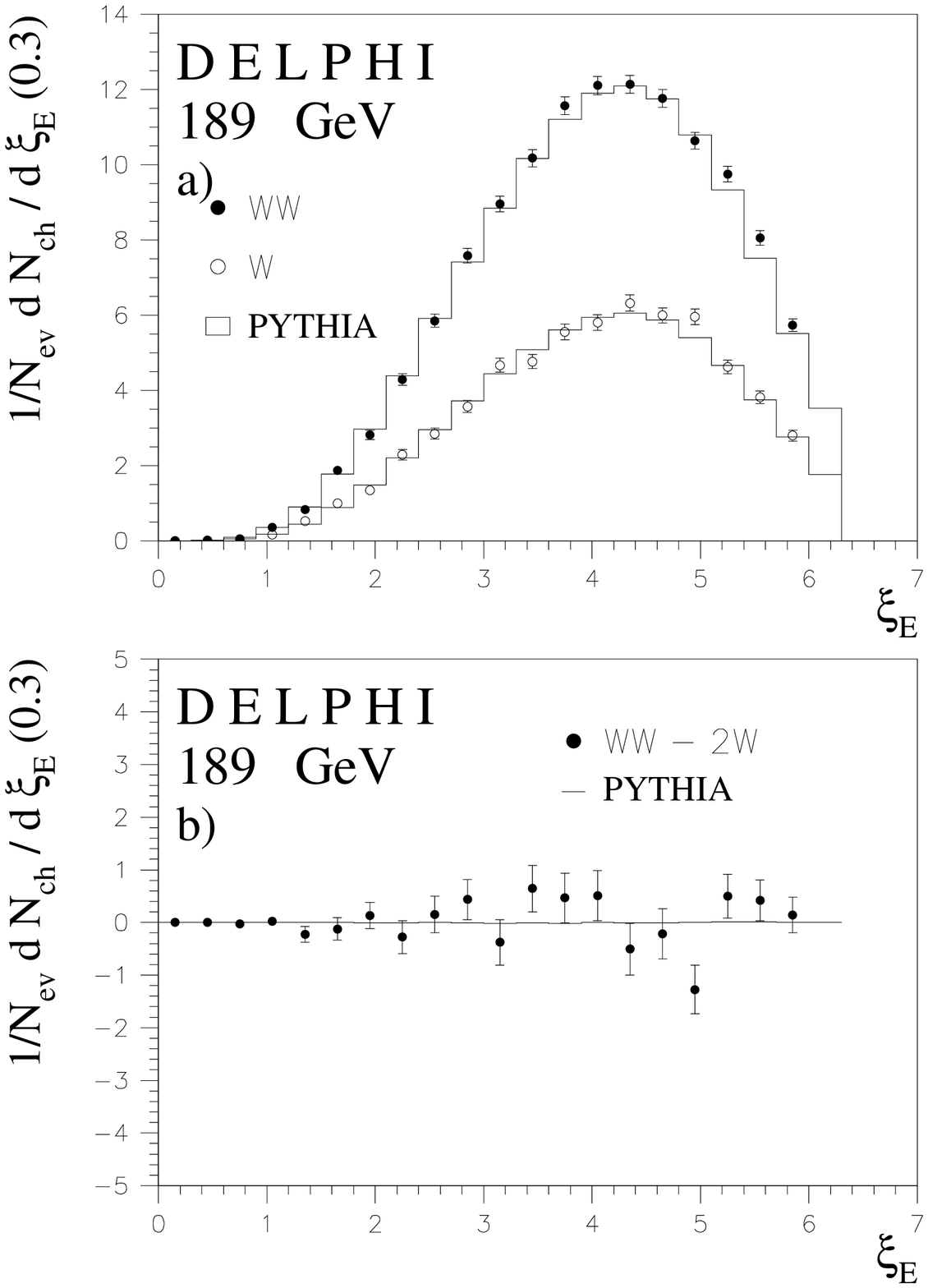}}
\end{center}
\caption[]{\label{xe1} (a)
Corrected 
%rapidity and 
$\xi_E$
distributions of charged particles for $(4q)$ events (closed circles) and 
$(2q)$ events (open circles), at 189~GeV, compared to simulation
without colour reconnection.
The error bars in the data represent the statistical errors.
The difference between $(4q)$ and twice $(2q)$ is shown in (b).}
\end{figure}

\begin{figure}[ph]
%\centerline{DELPHI preliminary}
\begin{center}
%\mbox{\epsfxsize7.0cm\epsfig{file=ww_2w_y_189.eps}}
%\mbox{\epsfxsize7.0cm\epsfig{file=ww_2w_y_183.eps}}
%Only statistical errors
\resizebox{\textwidth}{0.88\textheight}{\includegraphics{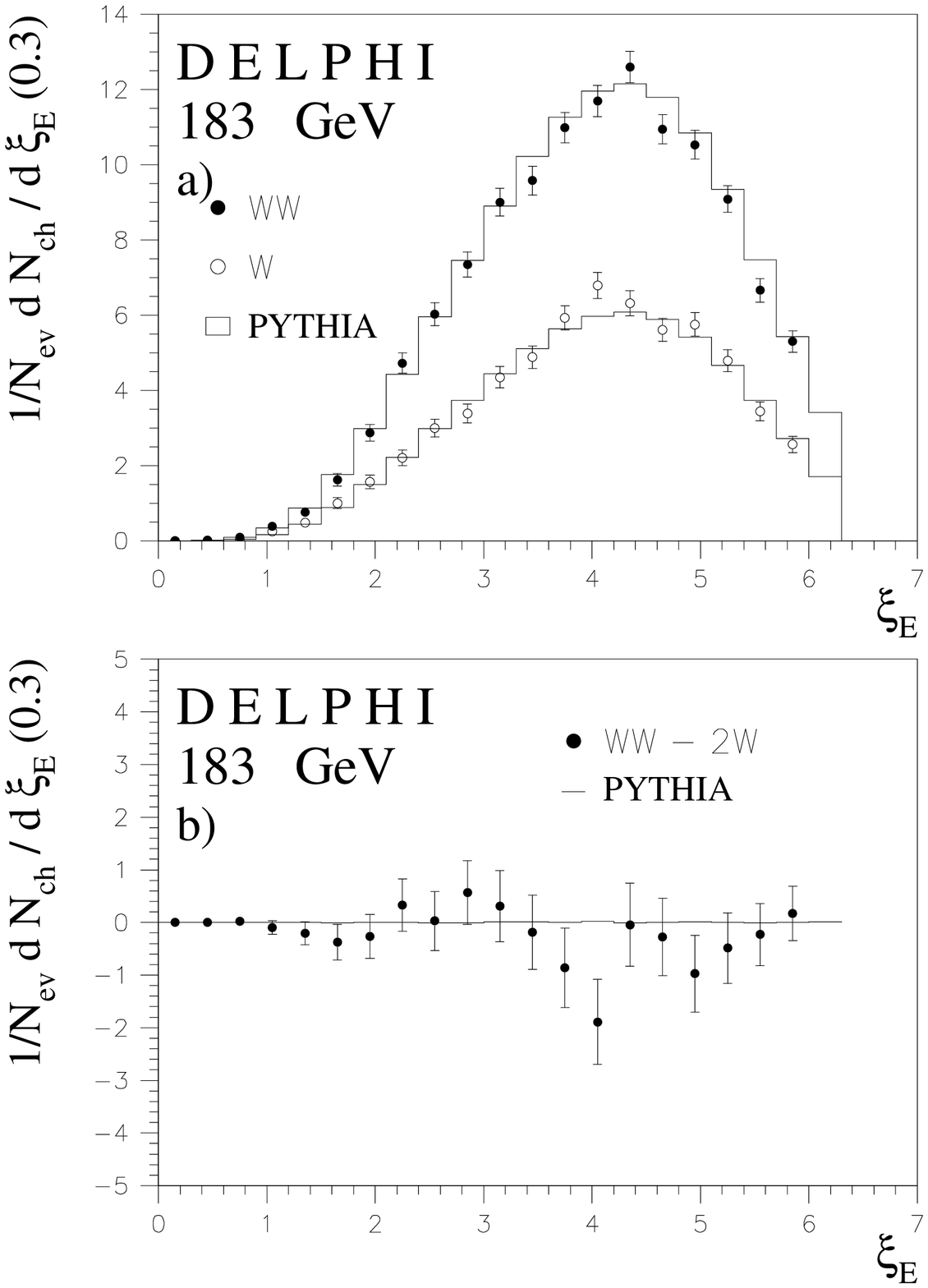}}
\end{center}
\caption[]{\label{xe2} (a)
Corrected 
%rapidity and 
$\xi_E$
distributions of charged particles for $(4q)$ events (closed circles) and 
$(2q)$ events (open circles), at 183~GeV, compared to simulation
without colour reconnection.
The error bars in the data represent the statistical errors.
The difference between $(4q)$ and twice $(2q)$ is shown in (b).}
\end{figure}

In the presence of interconnection, the deficit of multiplicity is expected to
be 
concentrated in the region of low momentum. The corrected momentum
distributions in the $(4q)$ and in the $(2q)$ cases are shown in 
Figures~\ref{xp1} and~\ref{xp2} (Figures~\ref{xe1} and~\ref{xe2} show 
the distributions in terms of the  
%(Figures~\ref{xe1} and \ref{xe2} show the distributions in terms of the
$\xi_E$ variable). The systematic error in the momentum region
between 0.1 and 1 GeV$/c$ was explicitely recomputed.
We measure: 
\begin{eqnarray}
\left.\frac{\langle n^{(4q)}\rangle}{2\langle n^{(2q)}\rangle}\right|_{\mathrm{189\,GeV}}^{0.1 < p < 1
\, {GeV}/c} & = & 0.992 \pm 0.029 (stat) \pm 0.016 (syst) \\
\left.\frac{\langle n^{(4q)}\rangle}{2\langle n^{(2q)}\rangle}
\right|_{\mathrm{183\,GeV}}^{0.1 < p < 1 \, {GeV}/c} & = & 
0.956 \pm 0.044 (stat) \pm 0.022 (syst) \, .
\end{eqnarray}
% total .980 +- .024 +- .011
Using as weights the inverse of the sum in quadrature of the statistical and
systematic  
errors, one obtains a weighted average of
\begin{equation}
\left.\frac{\langle n^{(4q)}\rangle}{2\langle n^{(2q)}\rangle}
\right|^{0.1 < p < 1 \, {GeV}/c}   =  0.981\pm 0.024 (stat) \pm 0.013 (syst) .
\end{equation}

Other inclusive distributions (rapidity and $p_T$ for example) taking into
account the orientation of particles could display a larger sensitivity with
respect to interconnection effects. Special care should be taken, since the
definition of the thrust axis could introduce a bias between $(2q)$ and $%
(4q) $ events. To estimate the effect of this bias, the following procedure
was used. First the distributions of rapidity and $p_T$ were computed as in
the case of the momentum for the $(2q)$ and the $(4q)$. Then, a set of $(4q)$%
-like events was constructed by mixing pairs of $(2q)$ events
 (constructed $(4q)$ sample). To construct this sample
the hadronically decaying Ws in $(2q)$ events were boosted back to 
their rest frame.
For each real 
$(4q)$ event, each W was then replaced by a W from a $(2q)$, 
boosted forward in such a
way that the directions of the jet axis were the same. 

\begin{figure}[ph]
%\centerline{DELPHI preliminary}
\begin{center}
% Statistical + systematics from diff of 2sl-ms
\resizebox{\textwidth}{0.85\textheight}{\includegraphics{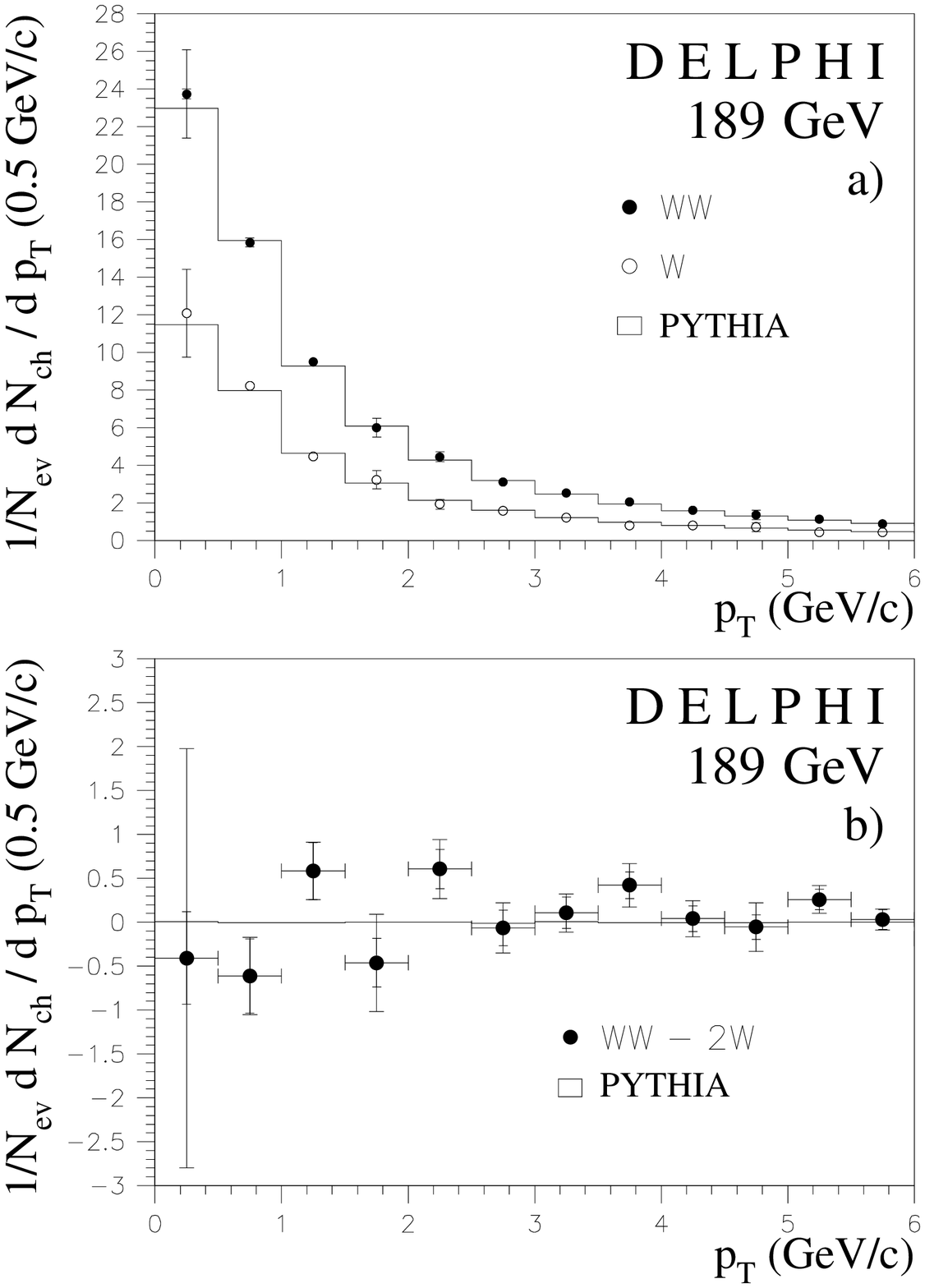}}
\end{center}
\caption[]{\label{xp3} (a)
Corrected $p_T$ 
distributions of charged particles for $(4q)$ events (closed circles) and 
$(2q)$ events (open circles), compared to simulation
without colour reconnection, at 189 GeV.
The internal error bars in the data represent the statistical error and the
external error bars represent the sum in quadrature 
of the statistical and systematic errors.
The difference between $(4q)$ and twice $(2q)$ is shown in (b).}
\end{figure}

\begin{figure}[ph]
%\centerline{DELPHI preliminary}
\begin{center}
% Statistical + systematics from diff of 2sl-ms
\resizebox{\textwidth}{0.85\textheight}{\includegraphics{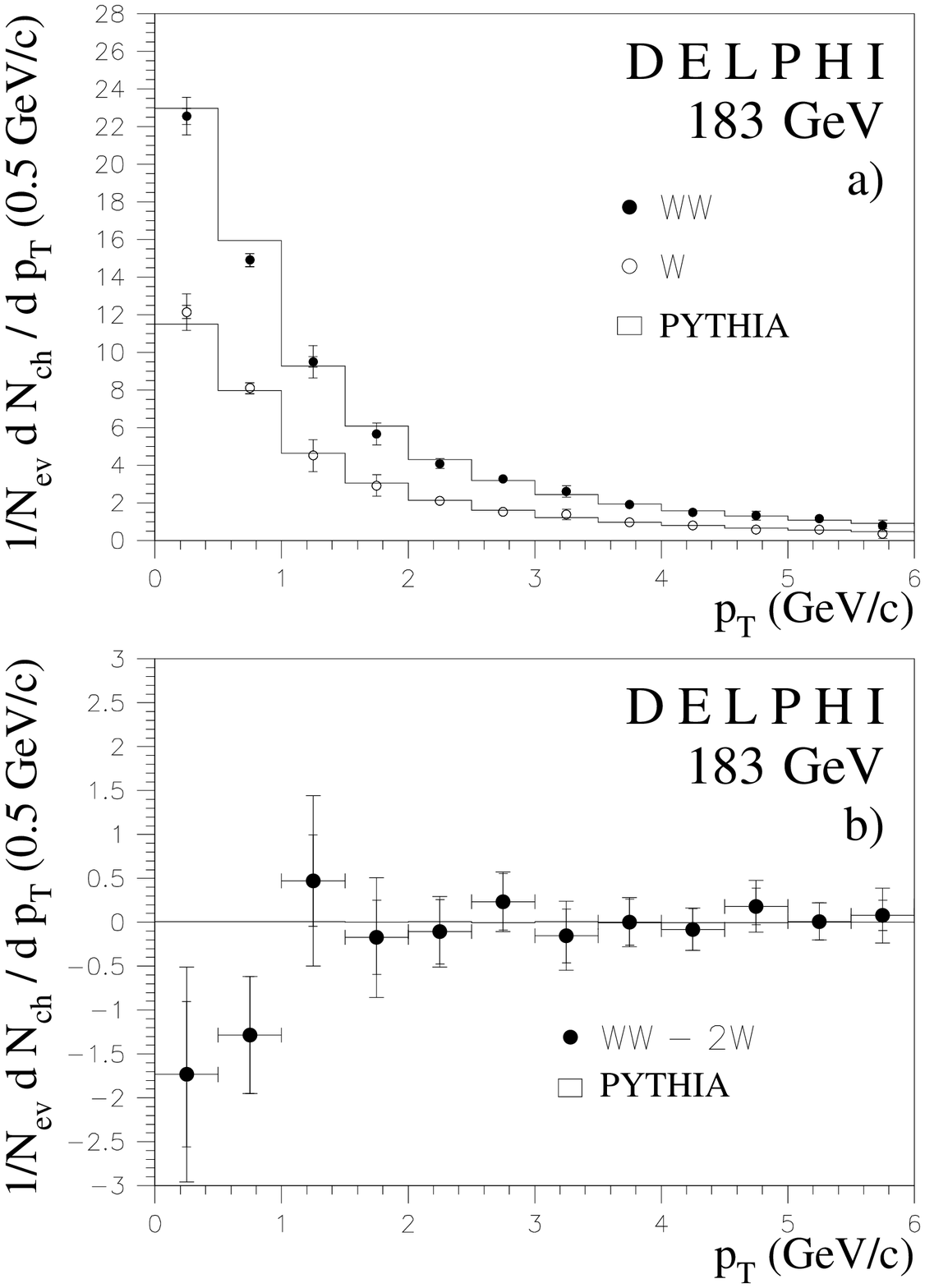}}
\end{center}
\caption[]{\label{xp4} (a)
Corrected $p_T$ 
distributions of charged particles for $(4q)$ events (closed circles) and 
$(2q)$ events (open circles), compared to simulation
without colour reconnection, at 183 GeV.
The internal error bars in the data represent the statistical error and the
external error bars represent the sum in quadrature 
of the statistical and systematic errors.
The difference between $(4q)$ and twice $(2q)$ is shown in (b).}
\end{figure}

If there would be no bias from the definition of
the thrust axis (or if the correction for detector effects could correctly 
account for
the bias), there would be no difference between the $y_T$ and $p_T$
distributions in the constructed $(4q)$ sample and twice the $(2q)$ sample. This
difference has thus been taken as an estimator of the systematic error from 
the bias, and added in quadrature to the $(4q)$ distribution, to twice the 
$(2q)$ distribution, and to their difference. The distributions of transverse
momentum with respect to the thrust axis are shown in Figures \ref{xp3} and 
\ref{xp4}. 
It can be seen that the difference between the $(4q)$ distribution 
and twice the $(2q)$ at 183~GeV is concentrated in the low-$p_T$
region; however, part of the difference is due to the fact that the
$(2q)$ distribution at low $p_T$ lies above the simulation. 
The rapidity distribution, instead, does not display any particular feature.

The dispersion in $(4q)$ events is consistent at both energies with 
$\sqrt{2}$ times the dispersion in $(2q)$ events
(in the calculation of the systematic error on the ratio, the correlations
between the sources of systematic error were taken into account): 
\begin{eqnarray*}
\left(\frac{D^{(4q)}}{\sqrt{2}D^{(2q)}}\right)_{\mathrm{189\,GeV}} & = &
0.95 \pm 0.04 \mathrm{(stat)}\pm 0.07 \mathrm{(syst)} \\
\left(\frac{D^{(4q)}}{\sqrt{2}D^{(2q)}}\right)_{\mathrm{183\,GeV}} & = &
0.93 \pm 0.06 \mathrm{(stat)}\pm 0.04 \mathrm{(syst)} \, . \\
\end{eqnarray*}

Using as weights the inverse of the sum in quadrature of the statistical and
systematic  
errors, one obtains a weighted average of
\begin{equation}
\left(\frac{D^{(4q)}}{\sqrt{2}D^{(2q)}}\right)  =  0.94
\pm 0.03 (stat) \pm 0.03 (syst) .
\end{equation}

In conclusion, no depletion of the multiplicity was observed in fully
hadronic WW events with respect to twice the semileptonic events, at this
statistical precision; a possible depletion at the percent level can however not
be excluded. %and it is consistent with a statistical fluctuation, 
%with a significance less than two standard deviations at 183 GeV
%and less than one standard deviation at 189 GeV. 

\section{Identified Particles from \protect\boldmath$e^+e^-\rightarrow
{\mathrm q}\bar{\mathrm q}$}

This section describes the results obtained for $\pi^+$, K$^+$, K$^0$, p 
and $\Lambda$
with data recorded by DELPHI at LEP~2. After the description of the 
event selection for
identified particles at energies up to 189 GeV,
the additional criteria for hadron identification are described.

\subsection{\protect\boldmath Event selection at 130 and 136 GeV}

After the hadronic preselection described in Section~2, events with 
$\sqrt{s^\prime}>0.85\sqrt{s}$
%a 
%reconstructed
%effective centre-of-mass energy, \mbox{$\sqrt{s^\prime}$} , of at least 85\%
%of the nominal centre-of-mass energy 
were used for further analysis. 
Data recorded at these two energies were combined and are referred to
as the 133 GeV sample. The total of $\sim 12~\mathrm{pb^{-1}}$ recorded by
DELPHI yields $1387$ events while $1406$ are expected from simulation.

\subsection{\protect\boldmath Event selection at 161 and 172 GeV}

\label{sec:evsel161}

Selected events at 161 GeV
were required to have a minimum of 8 and a maximum of 40 charged particles,
$\sqrt{s^\prime}>0.85\sqrt{s}$
and a
visible energy of at least 50\% 
of $\sqrt{s}$. A cut was imposed on the polar
angle $\theta $ of the thrust axis to select events well within the
acceptance of the detector. % Finally, WW events were partially removed. 
It was found that a selection based on the 
narrow jet broadening, \mbox{$B_N$}, is effective in removing the WW
events and minimises the
bias introduced on the remaining event sample. At threshold 
about 30~WW events are expected. Selecting events with $\mbox{$B_N$}\leq 0.12$
reduces this background by 50\%.

Using this selection 342 events are expected from simulation, while 357 were
selected from the data, with an estimated remaining WW background of 15
events.

%\subsection{\protect\boldmath Event selection at 172 \gev}

At 172 GeV,
in addition to the criteria described above, events were required to have
at most 38 charged particles and $\mbox{$B_N$} \leq 0.1$. This leads to
267 selected events, with 264 expected from simulation, out of which 36 are 
WW background.

\subsection{\protect\boldmath Event selection at 183 and 189 GeV}

The event selection for charged identified particles at 189 (183) GeV
follows very closely the procedure already described in sections~2 and~3. Events
with $\mbox{$\sqrt{s^\prime}$}/\sqrt{s}$ above $0.9$ and more
than $9 (8)$ charged particles with $p\geq 200$~MeV$/c$ were used. WW
background was suppressed by demanding $\mbox{$B_N$} \leq 0.1 (0.08)$. A total
of $3617 (1122)$ events were selected with an expected background from WW and
Z$^0$Z$^0$ of 789 (146) events.

\subsection{Selection of charged particles for identification}

%$

A further selection was applied to the charged particle sample to obtain
\emph{well identifiable} particles. Two different momentum regions were
considered, above and below 0.7 GeV/$c$, which correspond respectively to the
separation of samples identified solely by the ionization loss in the Time
Projection Chamber (TPC) and by the 
\v{C}erenkov detectors (RICH) and TPC together. Below 
0.7 GeV/$c$ tighter cuts were applied, namely to eliminate secondary protons.
There had to be at least 30 wire hits in the TPC associated with the track and
the measured track length had be to larger than 100 cm. In addition 
it was required at least
two associated VD layer hits in $r\phi$ and an impact parameter
in the $r\phi$ plane of less then $0.1$~cm. If there were less than two
associated VD layers in $z$, the corresponding impact parameter had to be less
than $1$ cm, else less then $0.1$ cm. Particles above $0.7$ GeV/c were required
to have a measured track length bigger than 30 cm and good RICH quality,
i.e.~presence of primary ionization in the veto regions. Only particles which
were well contained in the barrel region of DELPHI~($\mid\cos(\theta)\mid\leq
0.7$) were accepted.

\subsection{Analysis}

For an efficient identification of charged particles over the full momentum
region, information from the ionization loss in the TPC (``dE/dx'') and
information from the DELPHI RICH detectors were combined, using dedicated
%RICFIX, RICALI and RPROCO 
software packages~\cite{perfo}. 
%which are described
%in~\cite{rproco,ricali}. 
%RICFIX 
One package fine-tunes the Monte Carlo simulation concerning detector related
effects (such as slight fluctuations in pressures and refractive indices,
background arising from photon feedback, crosstalk between the MWPC readout
strips, $\delta $-rays, track ionization photoelectrons, etc.),
%RPROCO
and another package
derives identification likelihoods from the specific energy loss, the number of
reconstructed photons and the mean reconstructed \v{C}erenkov angles
respectively. The likelihoods are then multiplied and rescaled to one. From
these,  a set of ``tags'' which indicate the likelihood for a
particular mass hypothesis ($\pi$, K, and p) are derived. Throughout this
analysis leptons were not separated from pions. Their contribution to
the pion sample was subtracted using simulation.

A matrix inversion formalism was used to calculate the
true particle rates in the detector from the tagged rates.
The $3\times 3$ efficiency matrix is defined by 
\begin{equation}  \label{def:effmat}
{\mathcal{E}}_{i}^{j}=\frac{\mathit{Number~of~type~i~hadrons~tagged~as~type~j~hadrons}}{\mathit{%
Number~of~type~i~hadrons}}\, ,
\end{equation}
where type $i,j$ can be either of \pio,\Kpm,\ppba .
It establishes the connection between the true particles in the RICH/TPC and
the tagged ones: 
\begin{equation}  \label{eq:effmat}
\left( 
\begin{array}{c}
N_{\pi}^{meas} \\ 
N_{K}^{meas} \\ 
N_{p}^{meas}
\end{array}
\right) = \mathbf{\mathcal{E}} \left( 
\begin{array}{c}
N_{\pi}^{true} \\ 
N_{K}^{true} \\ 
N_{p}^{true}
\end{array}
\right)
\end{equation}
The inverse of the efficiency matrix works on the three sets of tagged
particles in two ways. First a particle can have multiple tags, meaning that 
the information from the tagging is ambiguous. This is not unlikely because
in this analysis the low statistics of the data samples force rather loose selection criteria to be applied. Secondly a particle can escape
identification. Both effects can be corrected by this method. The average 
identification efficiency is approximately 85\% 
for pions and 60\%
for kaons and protons, whereas the purities are
approximately 85\% 
and 60\% 
respectively. They show a strong momentum dependence.

\Kn\ and $\Lambda$ candidates were 
%detected 
reconstructed
by their decay in flight into 
$\pi^+\pi^-$ and p$\pi^-$ respectively. Secondary decays candidates, $V^0$,
in the selected sample of hadronic events were found by considering all
pairs of oppositely charged particles. The vertex defined by each pair
was determined so that the $\chi^2$ of the hypothesis of a common vertex
was minimised. The particles were then refitted to the common vertex. The
selection criteria were the ``standard'' ones described in \cite{perfo}. The
average detection efficiency from this procedure is about 36\% for the decay \Kn 
$\rightarrow \pi^+\pi^-$ and about 28\% for the decay $\Lambda \rightarrow$ p$\pi^-$ in
multi-hadronic events. The background under the invariant mass peaks was
subtracted separately for each bin of $V^0$ momentum. The background was
estimated from the data by linearly
interpolating two sidebands in invariant mass:

\begin{itemize}
\item  between 0.40 and 0.45~$\mathrm{GeV}/c^{2}$ and between
0.55 and 0.60~$\mathrm{GeV}/c^{2}$ for the \Kn;

\item  between 1.08 and 1.10~$\mathrm{GeV}/c^{2}$ and between
1.14 and 1.18~$\mathrm{GeV}/c^{2}$ for the $\Lambda $.
\end{itemize}

\subsection{Calibration of the efficiency matrix using Z$^0$ data}

\label{sec:effmat}

The Z$^0$ data recorded during each year for calibration were used to tune
the above described matrix before applying it to high energy data. This is made
possible by the fact that studies at the Z$^0$ 
pole~\cite{emile} established that
the exclusive particle spectra are reproduced 
to a very high level of accuracy by the DELPHI-tuned
version of the generators. Therefore deviations of the rates of
tagged particles between data and simulation in this sample can be
interpreted as detector effects.
Comparison of the tag rates allows a validation of the efficiency matrix, which
would be impossible to measure from the data due to the limited LEP 2 statistics.
%This can be compared to the method used at the \Z~pole where the ample
%statistics collected allowed to use very pure samples of charged particles
%originating from $\Lambda$'s and \Kn's to measure the efficiency matrix
%properly.
%Unfortunately this is not possible with the limited data samples at LEP~2.
%The assumption here is that the simulation describes the calibration samples
%correctly. 
The matrix is corrected so that it reproduces the simulated
rates, assuming that the correction factors are linear in the number of tagged
hadrons. The discrepancies
were found to be smaller than 4\%. This is taken into account when
calculating the systematic uncertainties.

%Figure~\ref{fig:effmat183} shows the efficiency matrix from 183~GeV data.
%The variation of identification efficiency and the rate of misidentification
%as a function of particle momentum is clearly visible.

\begin{figure}
\begin{center}
%This figure comes from /afs/cern.ch/user/n/neufeld/phy/idxi/
\centerline{\resizebox{12cm}{17cm}{\includegraphics{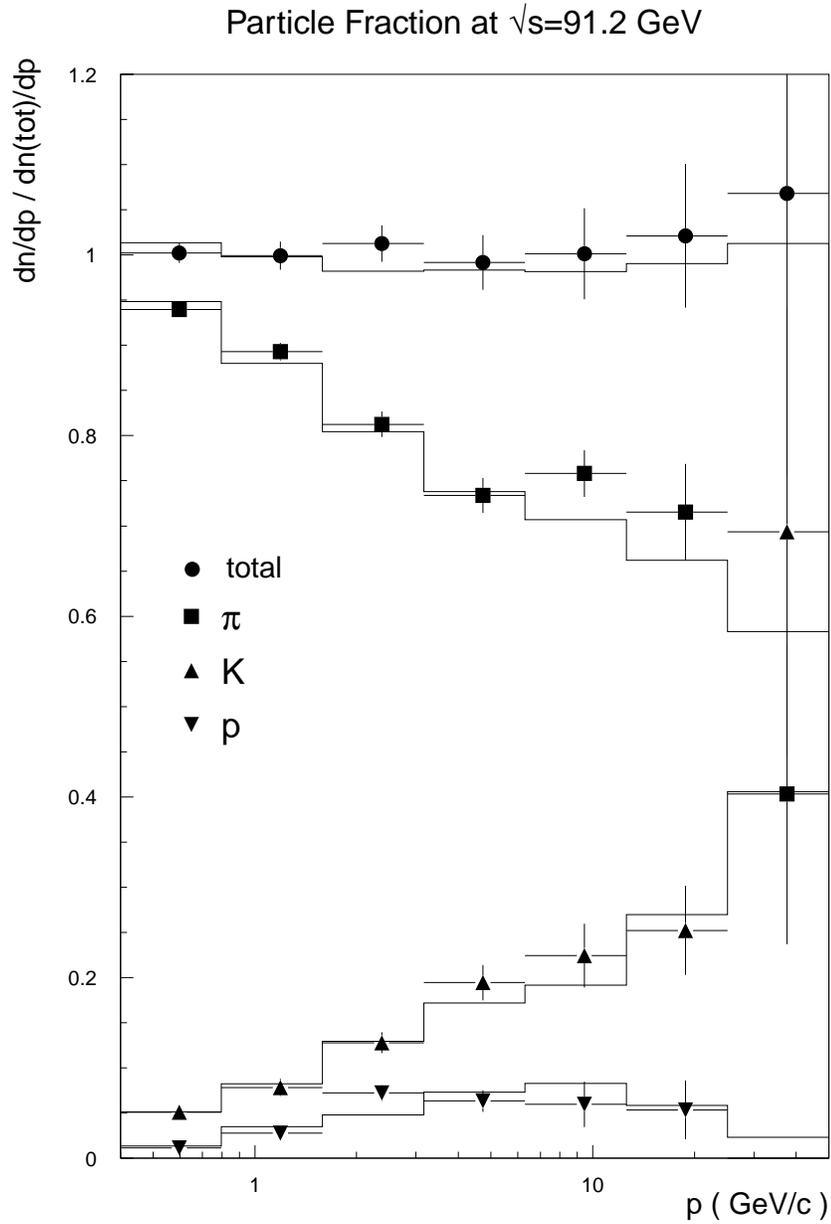}}}
\caption{Fractions of identified particles in radiative Z$^0$
  events from 189 GeV data.  The data have been taken under the same
  conditions as the signal data. Data (points) are in good agreement with the
prediction from JETSET (lines) including full detector simulation. The top
line and points indicate the sum of the fractions after unfolding.  
The error bars represent only the statistical error.}\label{fig:fracradz0189}
\end{center}
\end{figure}

As the high energy events are recorded over a long time period, stability of
the identification devices becomes a major concern. Variations in the
refractive index or the drift velocity in the RICH detectors may significantly
change the performance of the identification. %This variations
%can not be taken into account by the alignment and fixing procedure which
%are done using the calibration sample at the \Z\ pole which
%  is recorded in 2 weeks at the very beginning of the data taking. 
To estimate the effect of these variations on the measurement, the 
radiative returns to the Z$^0$ were used.
Such events were selected among the events passing the
hadronic preselection, by requiring in addition that 
they contained at least 8
charged particles, they had $\sqrt{s^\prime} < 130$
GeV, and a 
total energy transverse to the beam axis of more
than 30~GeV. Figure~\ref{fig:fracradz0189} shows the good agreement for
differential cross-sections for this event sample which may be taken as an
indication of the stability of the detector during the year at the few
percent level.

\subsection{\mbox{$ {\mathrm{\xi_{p}}}$}\ distributions and average
multiplicities}

After background subtraction, the tagged particle fractions were
unfolded using the calibrated matrix. The full covariance matrix was
calculated for the tag rates using multinomial statistics. It was then
propagated to the true rates of identified particles using the unfolding
matrix. 

\begin{figure}[ph]
%The next 4 figures come from /afs/cern.ch/user/n/neufeld/phy/idxi/
\centerline{DELPHI preliminary}
\resizebox{0.5\textwidth}{0.44\textheight}{\includegraphics{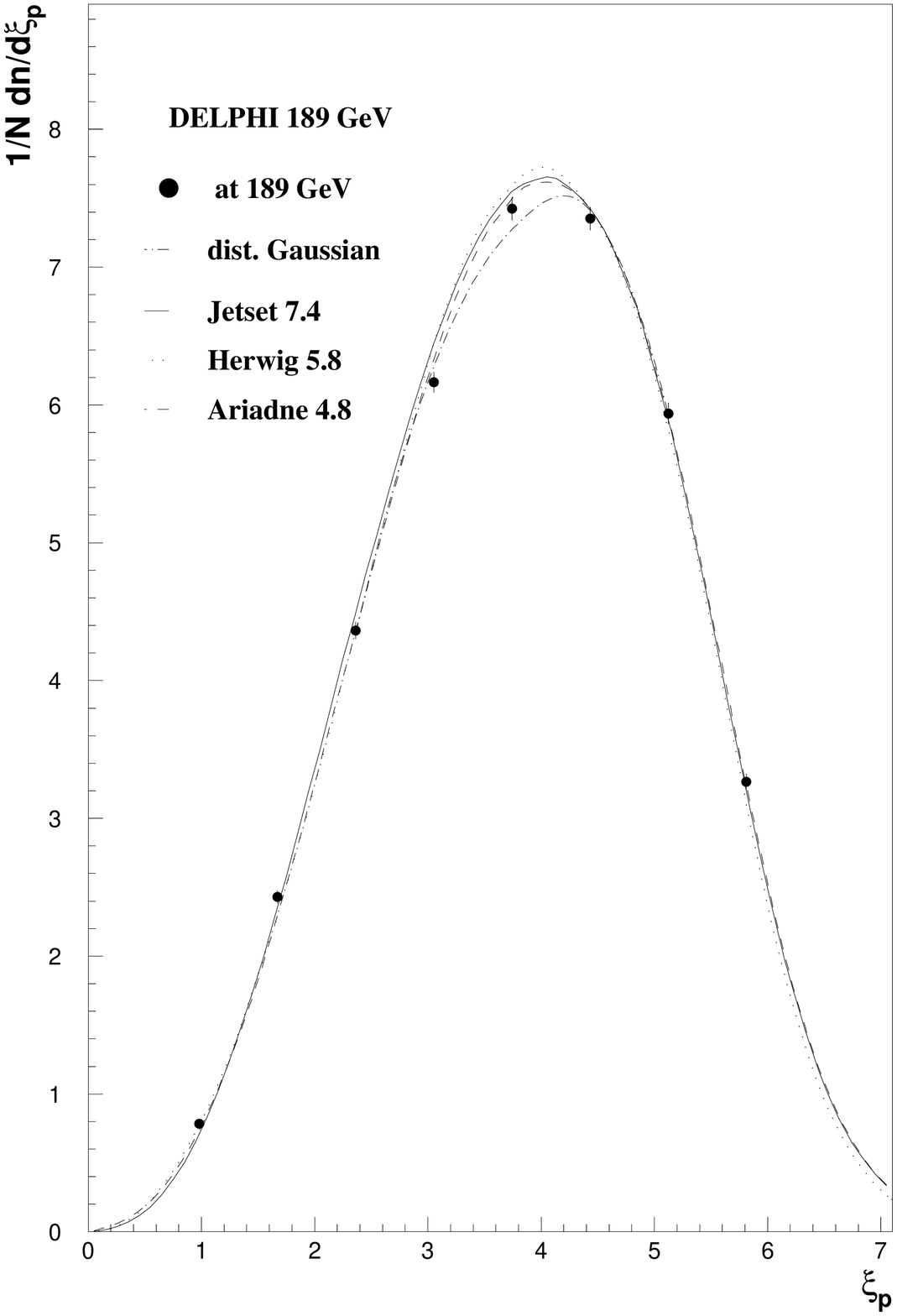}}
\resizebox{0.5\textwidth}{0.44\textheight}{\includegraphics{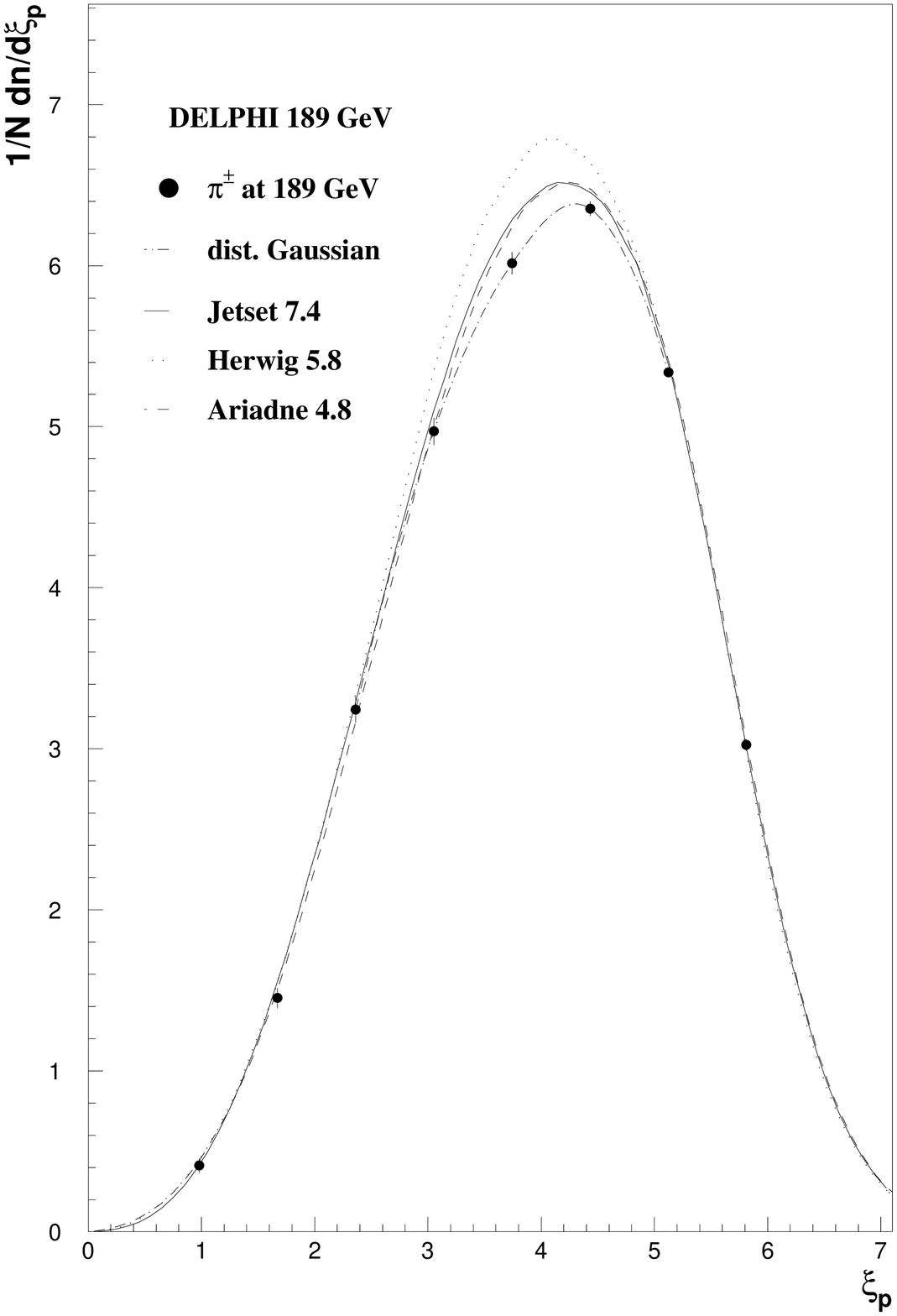}}

\resizebox{0.5\textwidth}{0.44\textheight}{\includegraphics{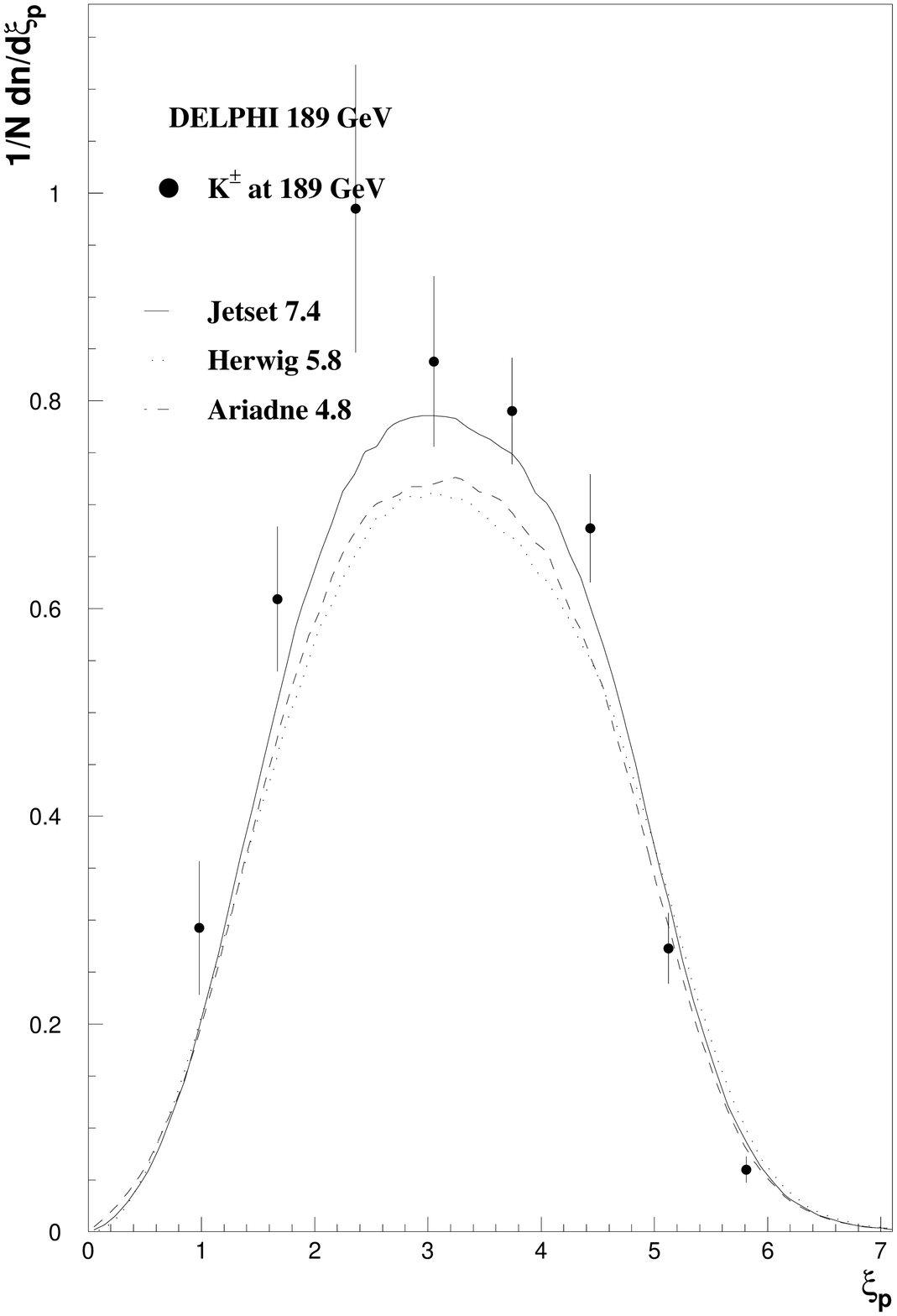}}
\resizebox{0.5\textwidth}{0.44\textheight}{\includegraphics{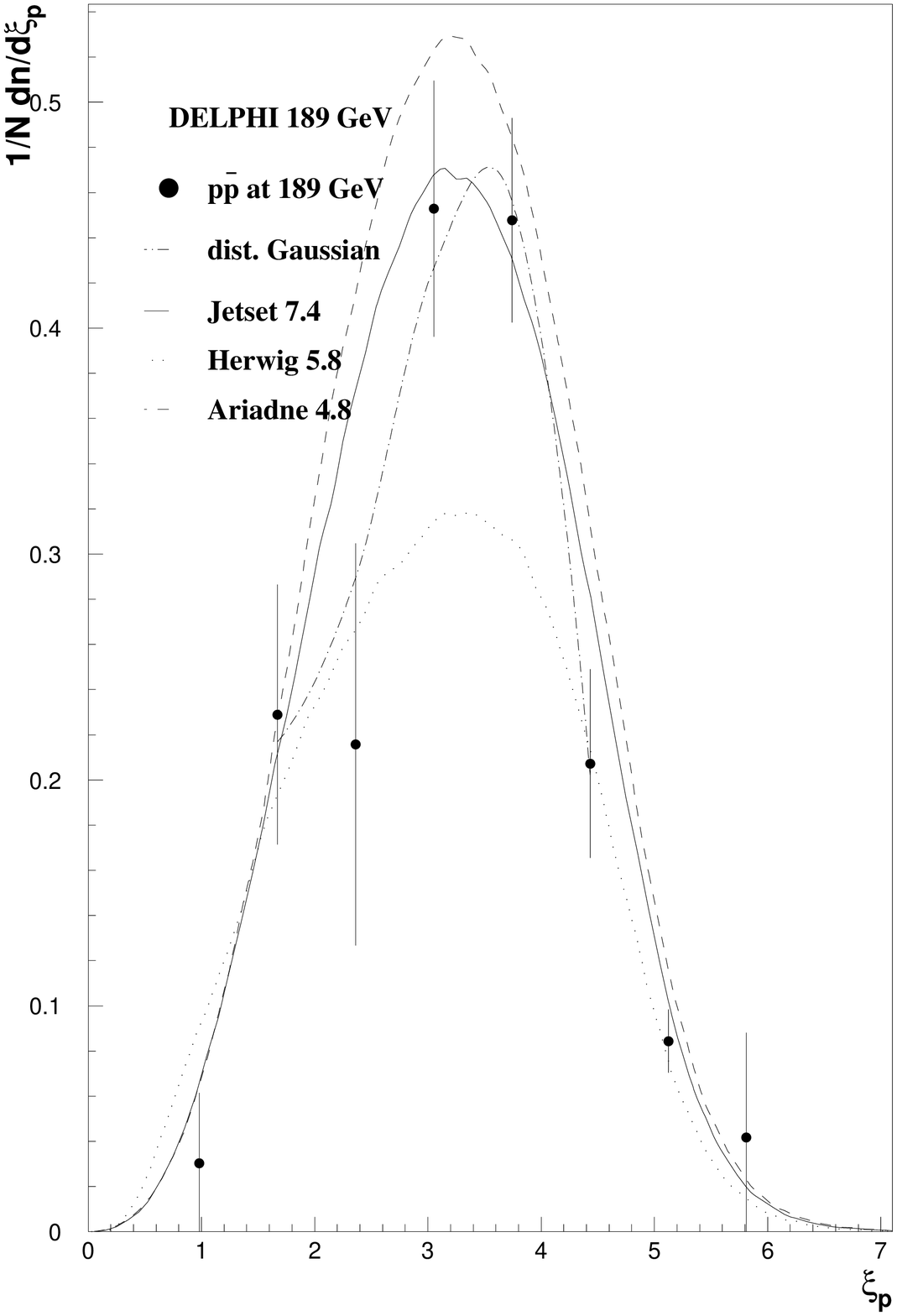}}
\caption{\xip\ distributions (efficiency corrected
and background subtracted) for charged
  particles, pions, kaons and protons in q$\bar{\mathrm q}$ events at
  189~GeV. Data (points) are compared to the prediction from JETSET (solid
line). Only 
  the statistical uncertainties are shown. The dashed dotted line
  shows a fit to equation~(\protect\ref{eq:dgauss}).} 
\label{fig:xiqq}
\end{figure}

\begin{figure}[ph]
%The next 4 figures come from /afs/cern.ch/user/n/neufeld/phy/idxi/
%\centerline{DELPHI preliminary}
\resizebox{0.5\textwidth}{0.44\textheight}{\includegraphics{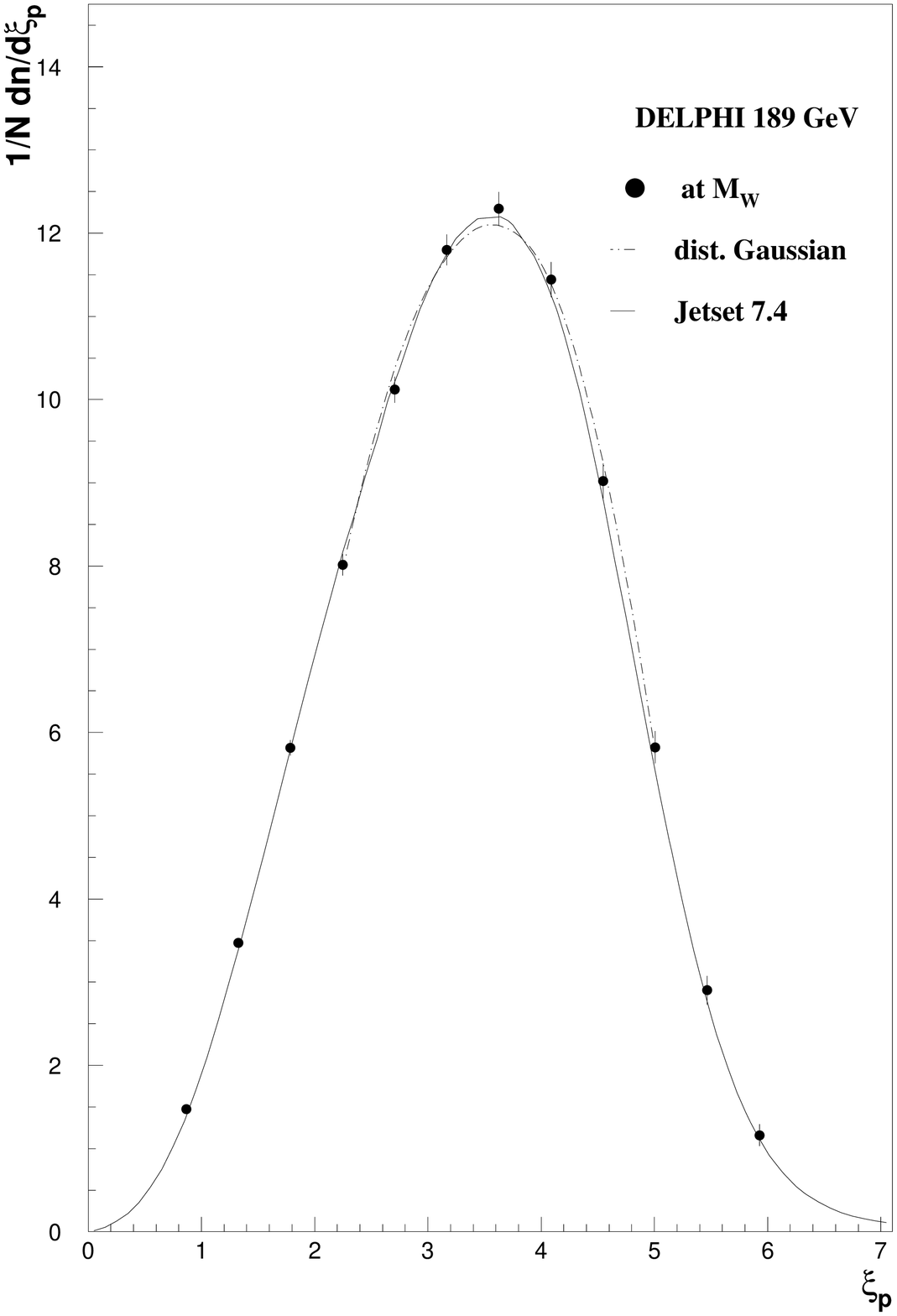}}
\resizebox{0.5\textwidth}{0.44\textheight}{\includegraphics{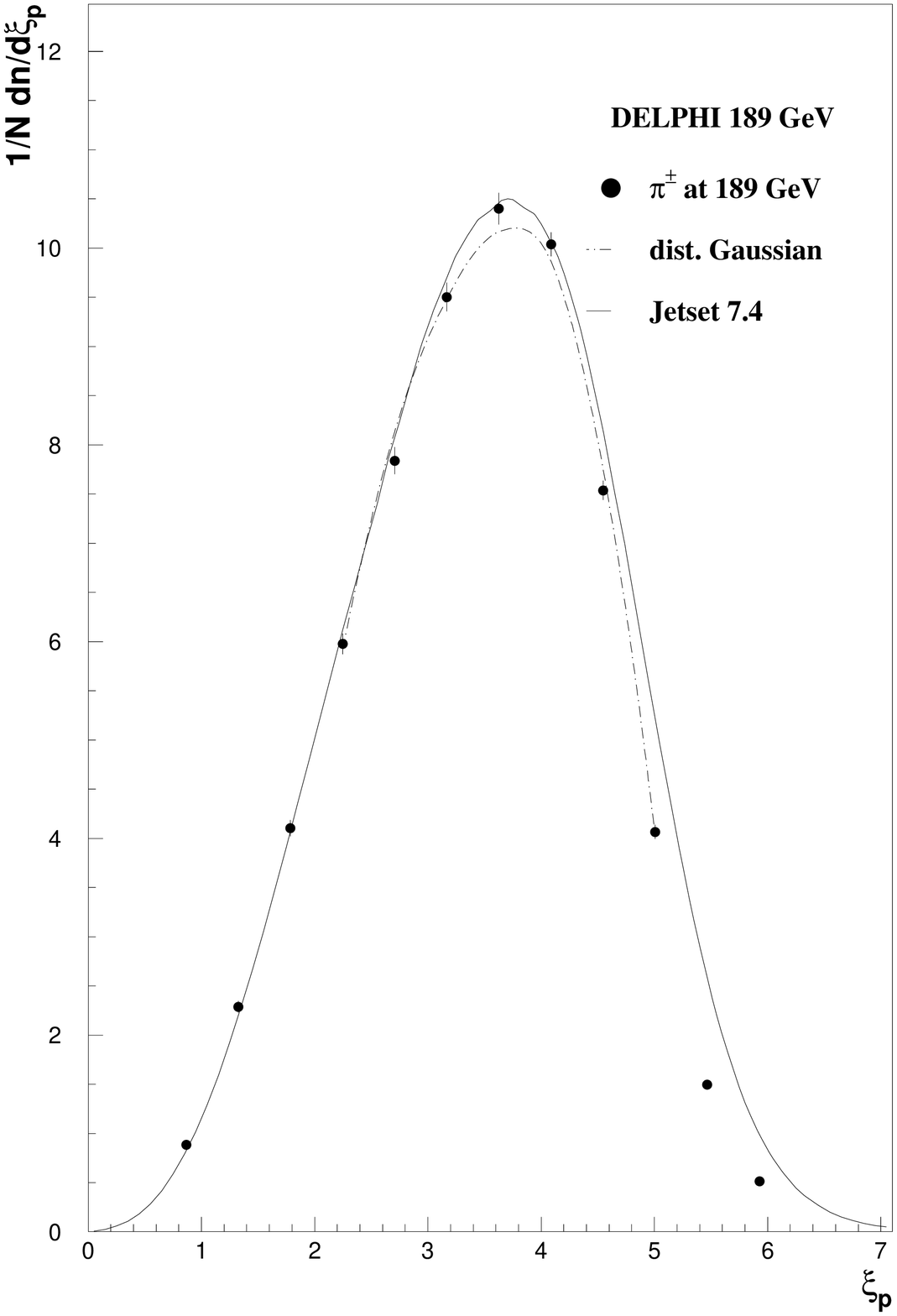}}

\resizebox{0.5\textwidth}{0.44\textheight}{\includegraphics{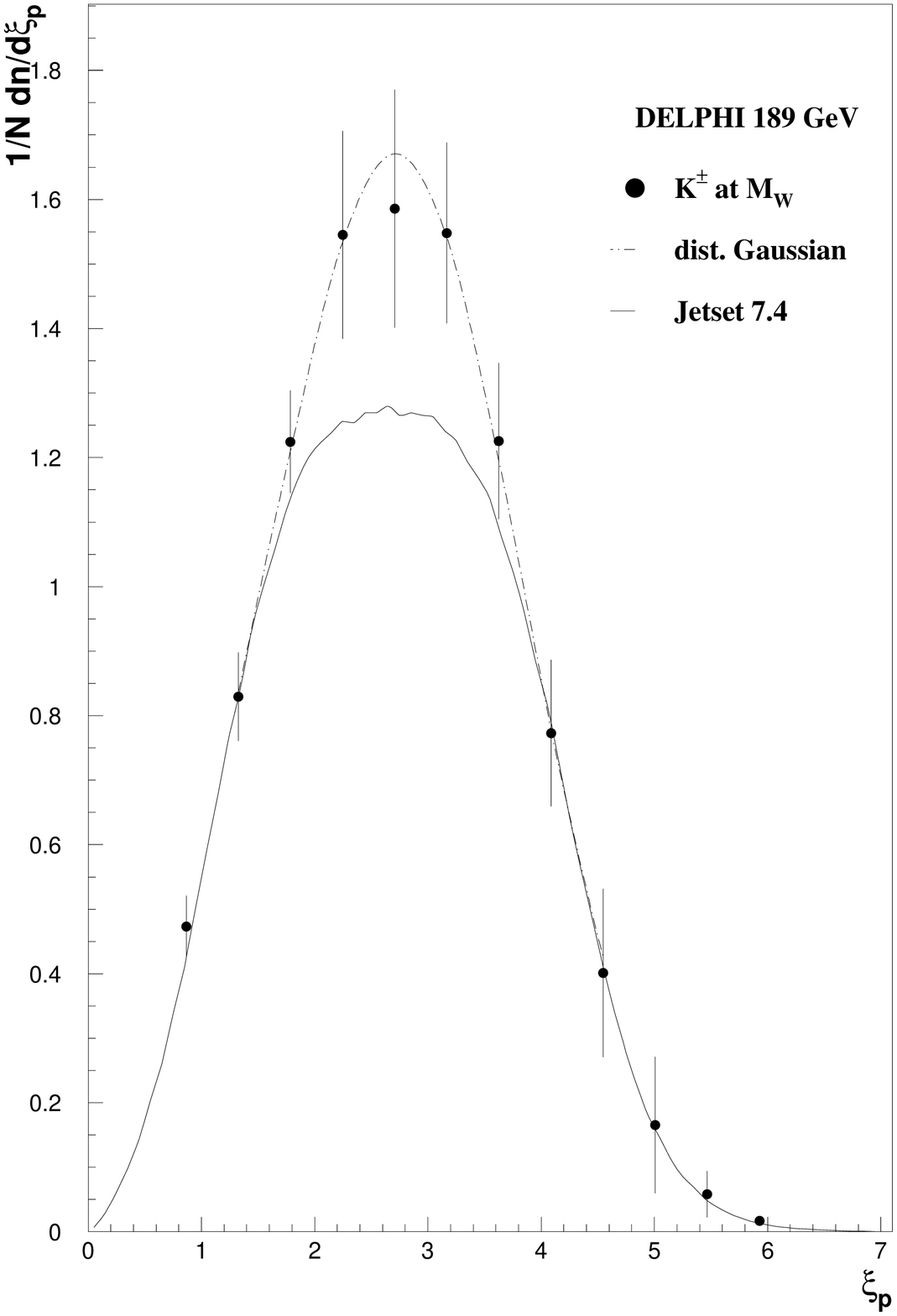}}
\resizebox{0.5\textwidth}{0.44\textheight}{\includegraphics{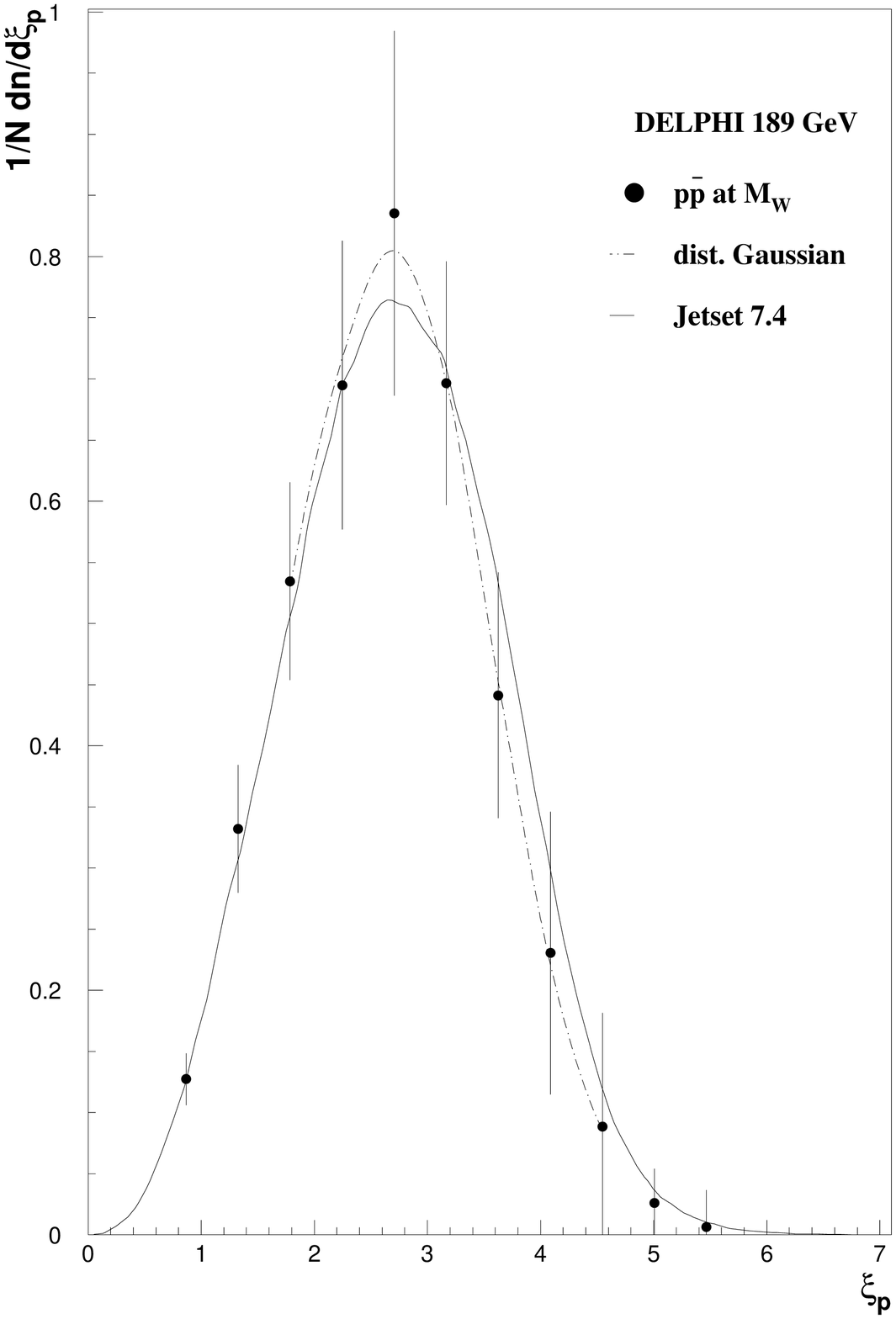}}
\caption{\xip\ distributions (efficiency corrected
and background subtracted) for charged
  particles, pions, kaons and protons in fully hadronic WW events at
  189~GeV. Data (points) which have been boosted back to the W 
  rest-frame, compared to the prediction from JETSET (solid line). Only
  the statistical uncertainties are shown. The dashed dotted line
  shows a fit to equation~(\protect\ref{eq:dgauss}).} 
\label{fig:xiww4q}
\end{figure}

\begin{figure}[ph]
%The next 4 figures come from /afs/cern.ch/user/n/neufeld/phy/idxi/
%\centerline{DELPHI preliminary}
\resizebox{0.5\textwidth}{0.44\textheight}{\includegraphics{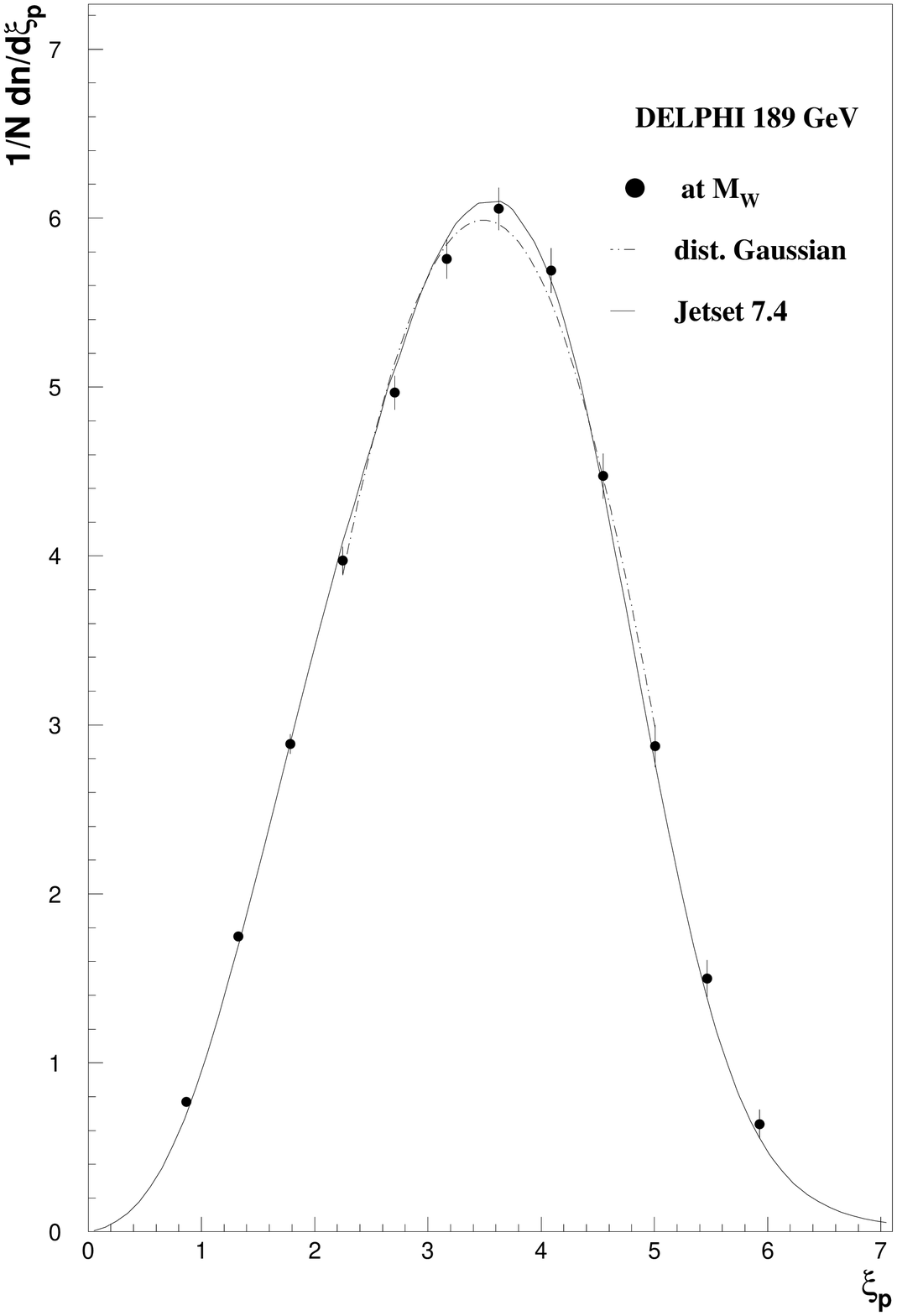}}
\resizebox{0.5\textwidth}{0.44\textheight}{\includegraphics{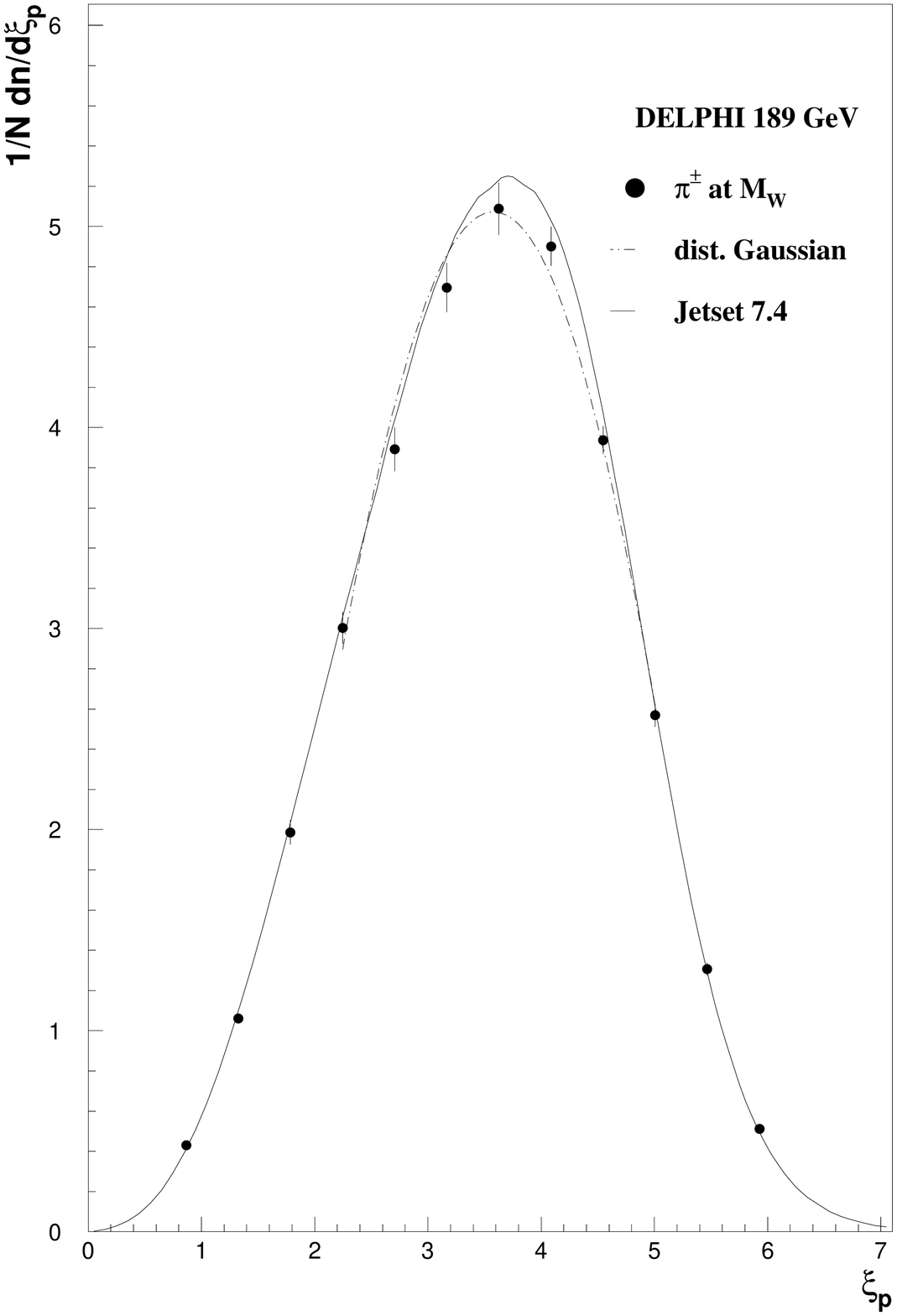}}

\resizebox{0.5\textwidth}{0.44\textheight}{\includegraphics{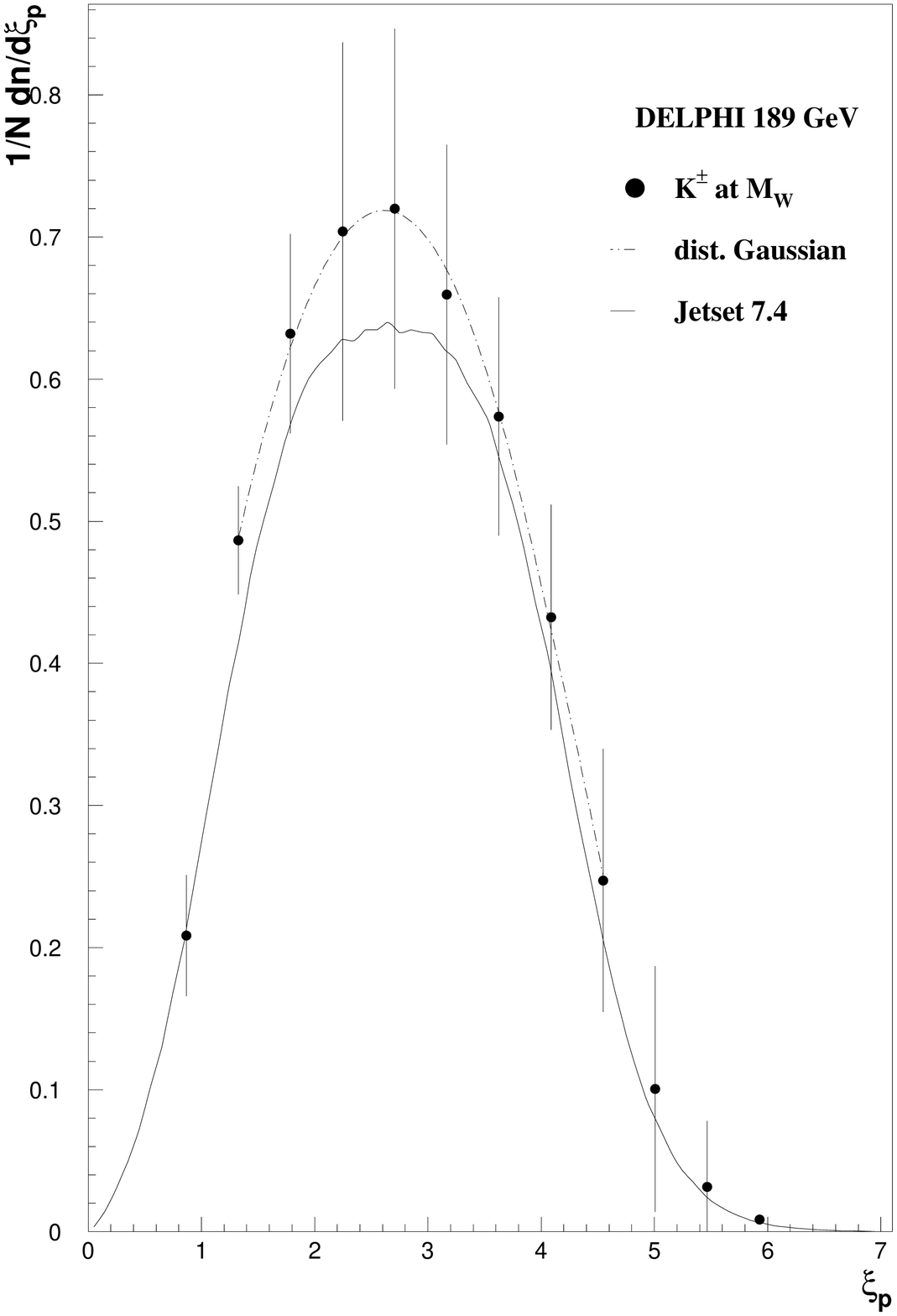}}
\resizebox{0.5\textwidth}{0.44\textheight}{\includegraphics{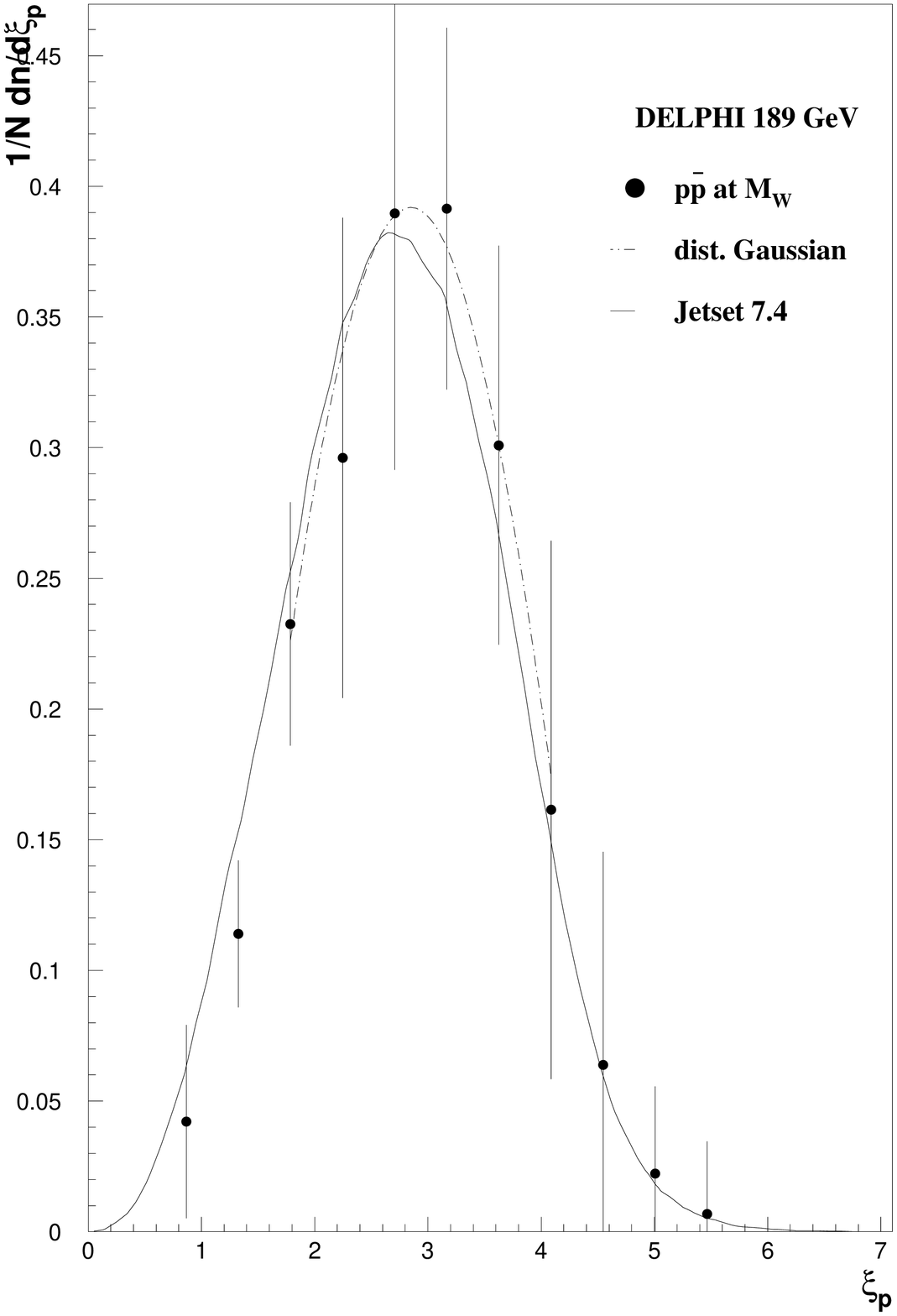}}
\caption{\xip\ distributions (efficiency corrected
and background subtracted) for charged
  particles, pions, kaons and protons in semileptonic WW events at
  189~GeV. Data (points) which have been boosted back to the W 
  rest-frame, compared to the prediction from JETSET (solid line). Only
  the statistical uncertainties are shown. The dashed dotted line
  shows a fit to equation~(\protect\ref{eq:dgauss}).} 
\label{fig:xiww2q}
\end{figure}

\begin{figure}[t]
\resizebox{\textwidth}{0.85\textheight}{\includegraphics{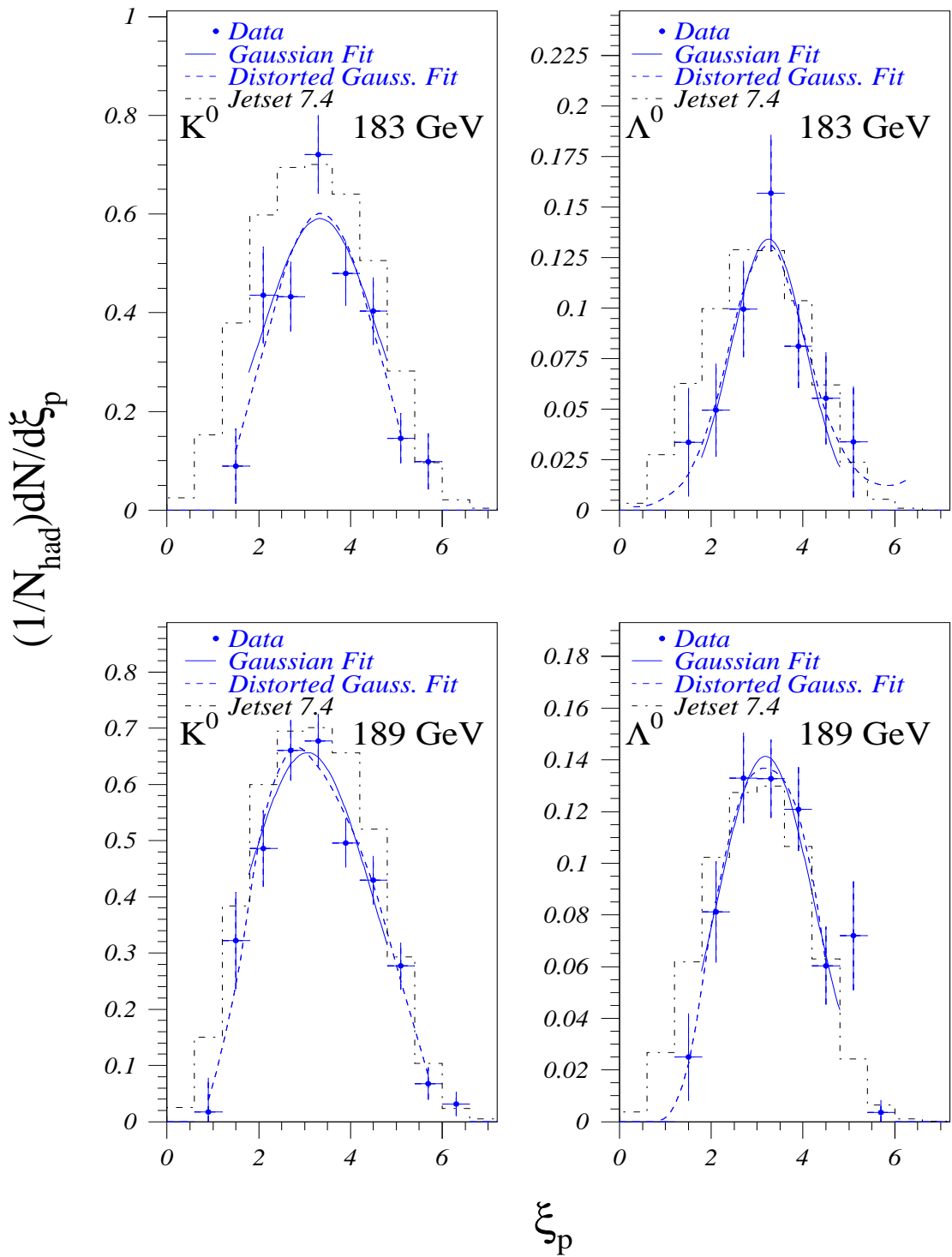}}
\caption{$\xi_p$ distributions (efficiency corrected
and background subtracted) for neutral Kaons
   and neutral Lambdas in $e^+e^-\rightarrow {\mathrm q\bar{q}}$ at 183~GeV
and 189~GeV respectively, for data (points), 
and simulation using JETSET (dashed-dotted histogram). 
   The full curves show the fit of the data to a Gaussian and the dashed ones
to a distorted Gaussian. The error bars represent only the statistical error.} 
  \label{fig:xineut}
\end{figure}

The \mbox{$ {\mathrm{\xi_{p}}}$}\ distribution was corrected bin by bin for
detector acceptance and selection efficiency, using the full detector
simulation. 
As an example, the corrected \mbox{$ {\mathrm{\xi_{p}}}$}\
distributions for charged pions, charged kaons and protons at 189 GeV
are shown in Figures~\ref{fig:xiqq},\ref{fig:xiww4q},\ref{fig:xiww2q} 
for q$\bar{\mathrm q}$, $(4q)$ and $(2q)$ 
events respectively.
%%%%%are shown in figs.~\ref
%{fig:pions}, \ref{fig:kaons}, \ref{fig:protons}, \ref{fig:xineut}
%respectively. 
In the figures the predictions from the DELPHI simulation
% HERWIG 5.8 and ARIADNE~4.8 
as well as a fit to expression~(\ref{eq:dgauss}), if the fit converged, are
also shown. Figure~\ref{fig:xineut} shows the same distributions for K$^0$ and
$\Lambda$ in q$\bar{\mathrm q}$ events at 183 and 189 GeV.
Within the statistics of the data samples analysed, the shapes of the \mbox{$
{\mathrm{\xi_{p}}}$}\ distributions are 
reasonably well described by the generators,
with the exception of the K$^0$ for which 
the agreement is poorer.

The multiplicity of the identified final states per hadronic event was
obtained by integration of the corresponding \mbox{$ {\mathrm{\xi_{p}}}$}\
distributions. The results are shown in table~\ref{tab:multi}. The numbers
given for charged identified hadrons include decay products from particles
with a lifetime $\tau < 10^{-9}$ s. These numbers are compared with the
predictions from PYTHIA 5.7, HERWIG 5.8 and ARIADNE 4.8.

%Table 2
\begin{table}
\begin{center}
\begin{tabular}{| c | c | c c c c |}
\hline
\small$\sqrt{s}$~[GeV]   & Particle  & \multicolumn{4}{c |}{$\langle n \rangle$} \\ \cline{3-6}
                  &           & \small PYTHIA 5.7   & \small HERWIG~5.8 & 
\small ARIADNE~4.8  & \normalsize Data  \\
\hline
%%%%%%%%%%%%% 133 GeV: 
% Results for \pi(JS,HW,AR),K+(JS,HW,AR),K0(JS,HW),p(JS,HW,AR),Lambdas(JS,HW)
% $133$      & $<n_{ch}>$  &  $23.83$   & $23.78$  & $23.76$ &
% $24.04\pm 0.25 \pm 0.09 $   \\ 
 $133$           & $\pi^{\pm}$ &  $19.90$   & $19.97$  & $19.64$ & 
$19.84\pm 0.29 \pm 0.44$   \\ 
             & K$^\pm$   &  $2.37$    & $2.32$   & $2.30$  & 
$~2.60\pm 0.26 \pm 0.13 $   \\
            & K$^0$   &  $ 2.40 $   & $2.64$     &         &       
$~2.51\pm 0.21 \pm 0.11 $ \\ 
           & p$\bar{\mathrm p}$      &  $1.11$    & $0.89$   & $1.29$  &
$~1.56 \pm 0.25 \pm 0.09$   \\
            & $\Lambda$    & $ 0.34 $         & 0.49     &         &
$~0.50\pm 0.07\pm 0.03 $ \\
\hline
%%%%%%%%%%%%% 161 GeV: 
% Results for \pi(JS,HW,AR),K+(JS,HW,AR),K0(JS,HW),p(JS,HW,AR)
% $161 $      & $<n_{ch}>$ &  $25.70$   & $25.44$  & $25.52$ & 
% $ 25.46\pm 0.45 \pm 0.37 $   \\ 
 $161$           & $\pi^{\pm}$  &  $21.24$   & $21.76$  & $21.12$ & 
$20.75 \pm 0.58 \pm  0.50$  \\ 
            & K$^{\pm}$   &  $2.63 $   & $2.44$   & $2.44$  & 
$~2.87 \pm 0.55 \pm 0.25$  \\
            & K$^0$    & $2.56 $      & 2.78     &         &
$~2.65 \pm 0.34\pm 0.14 $ \\
            & p$\bar{\mathrm p}$  &  $1.26$    & $0.96$   & $1.40$  &
$~1.21 \pm 0.48 \pm 0.02$  \\
            & $\Lambda$    &            &          &         &
                           \\          
\hline
%%%%%%%%%%%%% 172 GeV: 
% Results for \pi(JS,HW,AR),K+(JS,HW,AR),p(JS,HW,AR),Lambdas(JS,HW)
% $172$      & $<n_{ch}>$  &  $26.33$   & $26.06$  & $26.19$ & 
% $26.52 \pm 0.53 \pm 0.54$  \\ 
 $172$           & $\pi^{\pm}$ &  $21.77$   & $22.31$  & $21.68$ &
$21.79 \pm 0.68 \pm 0.47 $ \\ 
            & K$^{\pm}$  &  $2.68$    & $2.48$   & $2.50$  &
$~2.09 \pm 0.74 \pm 0.29$   \\
            & K$^0$   &            &          &         & 
                           \\           
            & p$\bar{\mathrm p}$ & $1.30$     & $0.99$   & $1.45$  & 
$~1.78 \pm 0.73 \pm 0.25$  \\
            & $\Lambda$   &            &          &         &
             \\
\hline
%%%%%%%%%%%%% 183 GeV: 
% Results for \pi(JS,HW,AR),K+(JS,HW,AR),K0(JS,HW),p(JS,HW,AR),Lambdas(JS,HW)
% $183$      & $<n_{ch}>$  &  $26.93 $   & $26.64$  & $26.77$ &
% $26.58 \pm 0.24 \pm 0.54$   \\ 
$183$            & $\pi^{\pm}$ &  $22.28$    & $22.82$  & $21.18$ &
$21.79 \pm 0.36 \pm 0.46$   \\
            & K$^{\pm}$   & $2.74$      & $2.53$   & $2.55$ &
$~2.83 \pm 0.37 \pm 0.13 $ \\
            & K$^0$       & $2.66$      & $2.91$   &        &
$~1.81\pm 0.14\pm 0.10 $ \\
            & p$\bar{\mathrm p}$  & $1.33$      & $1.00$   & $1.48$ &
$~1.32 \pm 0.34 \pm  0.17 $   \\
            & $\Lambda$   & $0.39 $     & $0.55$   &      &
$~0.33\pm 0.04\pm 0.03 $ \\
\hline
%%%%%%%%%%%%% 189 GeV: 
% Results for \pi(JS,HW,AR),K+(JS,HW,AR),K0(JS,HW),p(JS,HW,AR),Lambdas(JS,HW)
$189$       & $\pi^{\pm}$   & $22.56$ & $23.10$   & $22.47$         & 
$22.19 \pm 0.24\pm 0.46$ \\      
            & K$^{\pm}$    & $2.77$  & $2.55$    &  $2.57$    & 
$~3.15 \pm 0.21\pm 0.24$       \\ 
            & K$^0$        & $2.67$  & $2.93$    &      &
$~2.10 \pm 0.12\pm 0.10 $ \\
            & p$\bar{\mathrm p}$   & $1.35$  & $1.02$    & $1.50$     & 
$~1.19 \pm 0.17\pm 0.41$ \\ 
            & $\Lambda$    & $0.39$  & $0.56 $    &      &
$~0.40 \pm 0.03\pm 0.03 $ \\

\hline
\end{tabular}
\caption{Average multiplicities of particles \idhadrons~
at 133, 161, 172, 183 and 189 GeV. The first uncertainty is statistical  and the
second is systematic.  
%The values for the inclusive multiplicities have been taken from
% refs.~\cite{mult133,mult161,mult183}.
}\label{tab:multi}

\end{center}
\end{table}
 
The following sources of systematic uncertainties are taken into account.
\begin{enumerate}
\item  Uncertainties due to particle identification.

They are mainly due to the uncertainties in the modelling of the detector
response. In addition for the charged particles 
there are time dependent effects such as variations of
the drift velocity in the RICH detectors.

The unfolding matrix was adjusted using the Z$^0$-calibration data
recorded in the beginning of data taking in 1996, 1997 and 1998, as well as from
the peak period in 1995. This is described in section~\ref{sec:effmat}. For
1997 and 1998 data also radiative return Z$^0$-events were used. These are
better suited as they were recorded under the same conditions as the signal
events. 

Spectra were obtained using the original and the adjusted matrices. The
difference between the results obtained was taken as the corresponding
systematic uncertainty. 

The relative size of this uncertainty, averaged over the centre-of-mass
%energies is $0.0059$, $0.076$ and $0.1355$ for charged pions, kaons
energies is $0.006$, $0.076$ and $0.136$ for charged pions, kaons
and protons respectively. 

A relative systematic uncertainty of 3\%
was estimated for K$^0$~\cite{k0id} and 5\%
for $\Lambda$~\cite{l0id}.

\item  Size of the subtracted WW background (where applicable).

Variation of the selection criteria results in a 10\% 
uncertainty on the fraction of the W contamination in the q$\bar{\mathrm q}$
sample. This corresponds to  a 5\% 
uncertainty in the W cross-section and the size of the W 
background has been varied accordingly. The maximal variation observed in
the distributions has been taken as the corresponding systematic uncertainty.
The relative size of this uncertainty, averaged over the centre-of-mass
%energies is $0.0025$, $0.0099$ and $0.0028$ for charged pions, kaons
energies is $0.003$, $0.010$ and $0.003$ for charged pions, kaons
and protons respectively, and is 0.014 (0.028, negligible) and 0.032 (0.057) 
for K$^0$ and $\Lambda$, respectively, at centre of mass
energies of 189 (183, 161) GeV.

\item  Particles with momenta below 0.2 GeV/c or above 50 GeV/c were not
identified. Their contribution was extrapolated from the simulation.
Half of the extrapolated multiplicity was added in quadrature to the
systematic uncertainty. 
The relative size of this uncertainty, averaged over the centre-of-mass
%energies is $0.0205$, $0.0026$ and $0.0007$ for charged pions, kaons
energies is $0.021$, $0.003$ and $0.001$ for charged pions, kaons
and protons respectively, while for K$^0$ and $\Lambda$ at 133 GeV, 183 GeV
and 189 GeV, is $0.034$, and for K$^0$ at 161 GeV is $0.042$. 
%For the invisible momentum region the predictions from each of the 3
%generators for each particle species are compared. The maximal difference is
%added in quadrature to the uncertainty of each particle type.

\item  For the pions this analysis relies on a subtraction of the lepton
contamination using simulation.

An extra uncertainty of 10\% of the simulation prediction for the total number
of leptons is added.
The relative size of this uncertainty, averaged over the
%centre-of-mass energies is $0.0026$.
centre-of-mass energies is $0.003$.
\end{enumerate}

\begin{figure}[ph]
\resizebox{\textwidth}{0.85\textheight}{\includegraphics{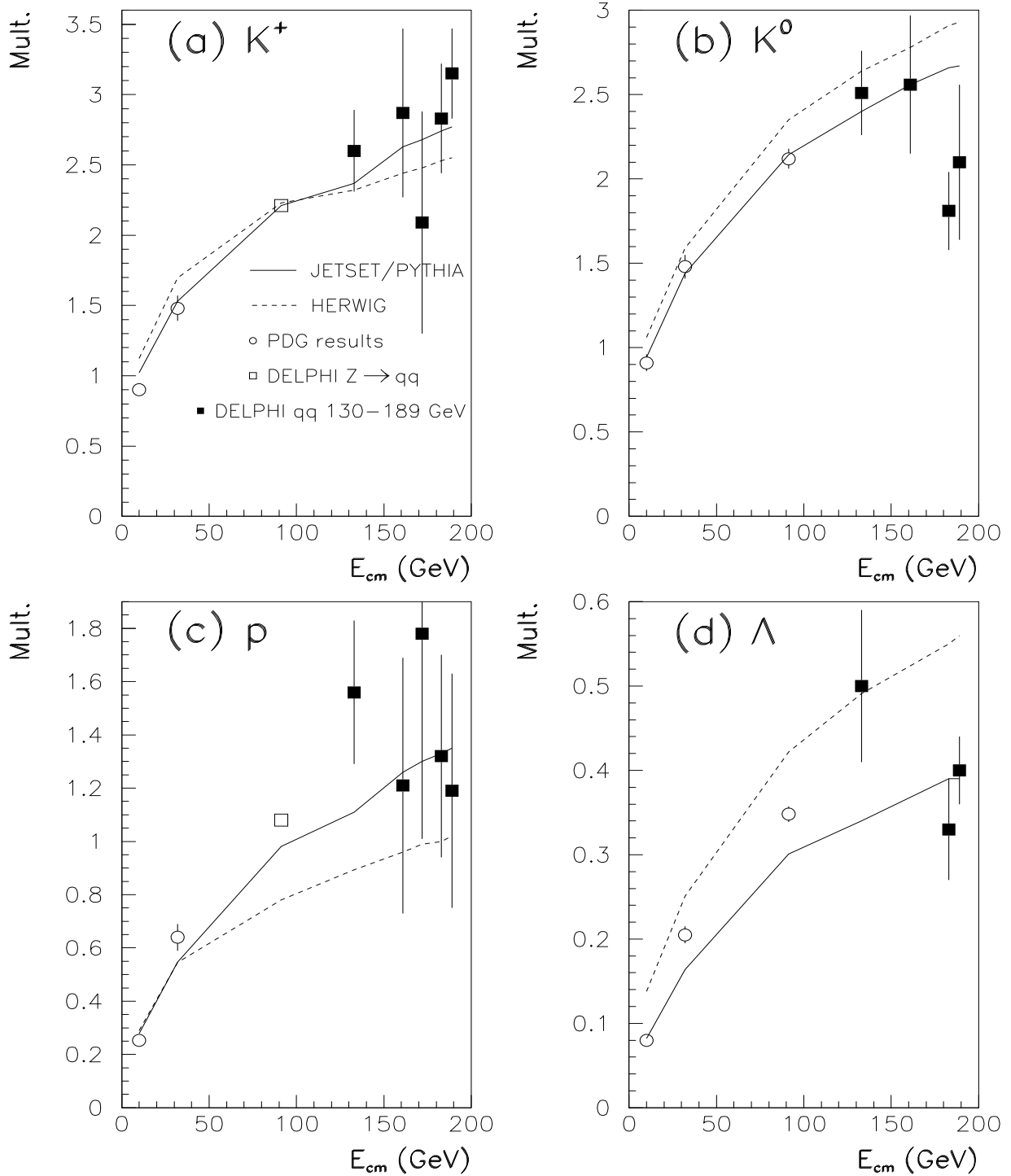}}
  \caption{Average multiplicity of K$^+$ (a), K$^0$ (b), p
          (c) and $\Lambda$ (d) 
           as function of the centre-of-mass energy. Black squares are 
DELPHI high energy data, open squares are measurements from a previous 
DELPHI publication~\cite{emile}, and open bullets are
values taken from the PDG~\cite{pdg}.
Simulations using PYTHIA tuned to DELPHI data (solid line) and HERWIG~5.8
   (dashed line) are superimposed. The error bars represent the sum in quadrature of the statistical and the systematic uncertainties.}\label{fig:mulev}
\end{figure}

In Figure~\ref{fig:mulev} the results are compared to the predictions
from PYTHIA 5.7 and HERWIG~5.8. The 
results shown for energies below 133~GeV (open squares) 
were extracted from reference~\cite{pdg}. 
The models account for the measurements, with the exception of the HERWIG 
predictions for K$^0$ and $\Lambda$ and of the PYTHIA predictions for K$^0$ 
at high energy which are above the measured values.

\subsection{\xis\ and its evolution}

An interesting aspect of the \xip -distribution is
the evolution of its peak position \xis\ with increasing centre-of-mass
energy. It is determined by
fitting a parametrisation of the distribution to the peak region.

%A possible such 
One such 
parametrisation is the distorted
Gaussian in  equation~(\ref{eq:dgauss}).
% The mean, standard deviation,
%kurtosis and skewness have been calculated as functions of
%$\Lambda_{\mathrm{QCD}}$ and constant $O(1)$ term by Fong and
%Webber~\cite{fongwebber}, in the limiting spectrum scenario. The
%higher order terms in $Y$ are neglected. 
%\begin{eqnarray}
%\label{eq:dgpara}
%\overline{\xi}=\frac{1}{2}Y(1+\frac{\rho}{24}\sqrt{\frac{48}{\beta
%    Y}})+O(1) \nonumber \\
%\sigma=\sqrt{\frac{1}{3}Y}(\frac{1}{48}\beta Y)^{\frac{1}{4}}
%       (1 - \frac{\beta}{64}\sqrt{\frac{48}{\beta Y}}) +
%    O(Y^{-\frac{1}{4}}) \nonumber \\
%s=-\frac{\rho}{16}\sqrt{\frac{3}{Y}}(\frac{48}{\beta Y})^{\frac{1}{4}}
%    + O(Y^{-\frac{5}{4}}) \nonumber \\
%k=-\frac{27}{5Y}(\sqrt{\frac{\beta
%    Y}{48}}-\frac{1}{24}\beta)+O(Y^{-\frac{3}{2}}) 
%\end{eqnarray}
%Here $\rho =11+2N_f/N_c^3$ and $\beta = 11-2N_f/N_c$. 
Another parametrisation is a standard Gaussian distribution. 
While being a more crude approximation, it facilitates the analysis in
the case of limited statistics.
 
Since equation~(\ref{eq:dgauss}) is
expected to describe well only the peak region, the fit range has
to be carefully chosen around the peak. 
Table~\ref{tab:xistar} shows the results and the
fit range, with the $\chi^2$ per degree of freedom for the gaussian fit, 
except for
the K$^0$ and $\Lambda$ at 133 GeV where only the distorted gaussian fit
was performed. The errors in the data are the sum in quadrature of the
statistical and systematic uncertainties
(generator values were extracted from samples of one million events generated
with the DELPHI tuned versions of the programs).

%Table 3
\begin{table}
\rotatebox{90}{
\begin{tabular}{| c | c | c c | c c | c c |  c c  | c c c |}
\hline
{\scriptsize $\sqrt{s}$} &   & 
\multicolumn{2}{c|}{\small fit range} & 
\multicolumn{2}{c|}{\small PYTHIA} & \multicolumn{2}{c|}{\small HERWIG} &
\multicolumn{2}{c|}{\small ARIADNE} & \multicolumn{3}{c|}{Data} \\
\hline
     &     &  {\scriptsize Gauss.} & {\scriptsize dist.~G.}  
           &  {\scriptsize Gauss.} & {\scriptsize dist.~G.} 
           &  {\scriptsize Gauss.} & {\scriptsize dist.~G.} 
           &  {\scriptsize Gauss.} & {\scriptsize dist.~G.} 
           &  {\scriptsize Gauss.} & $\chi^2/ndf$ & {\scriptsize dist.~G.}  \\
\hline
133 & $\pi^{\pm}$        &  $[3.3:4.9]$ & $[2.5:5.5]$ 
            & 3.90         & 4.11       
            & 3.90         & 4.12  
            & 4.00         & 4.15 
            & $4.04 \pm 0.13$ & $0.017$ & $4.32 \pm 0.37$ \\  
%    & K$^{\pm}$          &  $[2.1:3.3]$ & $[2.2:3.2]$ 
    & K$^{\pm}$          &  $[2.1:3.3]$ &  
            & 2.85         & 2.99 
            & 2.86         & 2.99  
            & 2.85         & 2.99 
            & $2.90 \pm 0.30$ & $1.39$ &                 \\
    & K$^0$              &              & $[0.6:5.4]$ 
            &              & 2.87  
            &              & 3.15 
            &              & 
            &                 & $1.63$  & $2.86 \pm 0.43$ \\ 
%    & p$\bar{\mathrm p}$ &  $[2.1:3.3]$ & $[2.5:5.5]$ 
    & p$\bar{\mathrm p}$ &  $[2.1:3.3]$ & 
            & 3.04         & 3.09 
            & 3.04         & 3.10  
            & 3.04         & 3.04 
            & $2.74 \pm 0.33$ & $1.86$ &                 \\
    & $\Lambda$          &              & $[0.6:4.8]$ 
            &              &    2.79  
            &              &    2.99 
            &              & 
            &                 & $3.22$ & $2.81 \pm 0.66$ \\
\hline
161 & $\pi^{\pm}$        & $[3.5:5.1]$ & $[2.6:4.5]$
            & 4.09         & 4.17
            & 4.03         & 4.08
            & 4.12         & 4.18
            & $4.11 \pm 0.15$ & $1.10$ & $4.14 \pm 0.48$ \\  
%    & K$^{\pm}$          & $[2.1:3.7]$ & $[2.1:3.7]$
    & K$^{\pm}$          & $[2.1:3.7]$ &
            & 2.99         & 2.98
            & 3.00         & 3.00
            & 3.01         & 2.98
            & $3.16 \pm 0.50$ & $0.57$ &          \\
    & K$^0$              &             &
            &              &  
            &              &
            &              &
            &                 & &                 \\
%    & p$\bar{\mathrm p}$ & $[2.1:3.7]$ & $[2.2:3.1]$
    & p$\bar{\mathrm p}$ & $[2.1:3.7]$ & 
            & 3.01         & 3.14
            & 3.20         &
            & 3.14         & 3.06
            & $3.08 \pm 0.84$ & $1.1$ &                 \\
    & $\Lambda$          &             &           
            &              &  
            &              &
            &              &
            &                 & &                 \\
\hline
172 & $\pi^{\pm}$        & $[3.5:5.1]$ & $[2.6:5.5]$
            & 4.13         & 4.20
            & 4.06         & 4.12
            & 4.16         & 4.21
            & $4.11 \pm 0.14$ & $1.07$ & $4.45 \pm 0.40$ \\  
%    & K$^{\pm}$          & $[2.1:3.7]$ & $[2.1:3.7]$
    & K$^{\pm}$          & $[2.1:3.7]$ & 
            & 3.04         & 3.02
            & 3.03         & 3.01
            & 3.05         & 3.01
            & $2.98 \pm 0.84$ & $1.76$ &                 \\
    & K$^0$              &             &
            &              &
            &              &
            &              &
            &                 & &                 \\
%    & p$\bar{\mathrm p}$ & $[2.1:3.7]$ &  $[2.2:3.7]$
    & p$\bar{\mathrm p}$ & $[2.1:3.7]$ & 
            & 3.16         & 3.14
            & 3.23         & 3.41
            & 3.18         & 3.12
            & $3.55 \pm 1.10$ & $0.98$ &                 \\
    & $\Lambda$          &             &             
            &              &
            &              &       
            &              &      
            &                 & &                 \\ 
\hline
183 & $\pi^{\pm}$        & $[2.6:4.5]$ & $[3.5:5.1]$ 
            & 4.16         & 4.43  
            & 4.10         & 4.30
            & 4.19         & 4.44 
            & $4.23 \pm 0.13$ & $2.8$ & $4.63 \pm 0.20$ \\  
%    & K$^{\pm}$          & $[2.1:3.7]$ & $[2.1:3.7]$ 
    & K$^{\pm}$          & $[2.1:3.7]$ &
            & 3.08         & 3.18 
            & 3.06         & 3.15
            & 3.10         & 3.23 
            & $3.08\pm 0.29$  & $1.19$  &                 \\
    & K$^0$              & $[1.8:4.8]$ & $[1.2:5.4]$  
            & 3.09         & 3.19        
            &              & 
            &              & 
            & $3.33 \pm 0.63$ & $2.75$ & $3.35\pm 0.63$ \\  
    & p$\bar{\mathrm p}$ & $[2.3:3.7]$ & $[2.1:3.7]$
            & 3.21         & 3.32
            & 3.31         &
            & 3.21         & 3.39 
            & $3.29 \pm 0.18$ & $0.98$ & $3.29 \pm 0.52$ \\
    & $\Lambda$          & $[1.8:4.8]$ & $[0.6:6.6]$ 
            & 3.02         & 3.01
            &              & 
            &              & 
            & $3.25 \pm 0.51$ & $0.87$ & $3.22 \pm 0.51$ \\ 
\hline
189 & $\pi^{\pm}$        & $[2.6:4.5]$ &    
            & $4.22$       &
            & $4.08$       &
            &              &    
            & $4.18 \pm 0.10$ & $2.32$ &                 \\
    & K$^\pm$            & $[2.1:4.7]$ &
            & $3.17$       &
            & $3.08$       &
            &              &
            & $3.10 \pm 0.25$ & $0.26$ &                 \\   
    & K$^0$              & $[1.8:4.8]$ & $[0.6:6.0]$  
            & 3.13         & 3.02        
            &              & 
            &              & 
            & $3.06 \pm 0.48$ & $1.66$ & $2.81 \pm 0.43$ \\  
    & p$\bar{\mathrm p}$ & $[2.1:4.7]$  &
            & 3.11         &
            & 3.20         &
            &              & 
            & $3.18 \pm 0.12$ & $0.01$ &                 \\ 
    & $\Lambda$          & $[1.8:4.8]$ & $[0.6:4.8]$ 
            & 3.02         & 2.98
            &              & 
            &              & 
            & $3.18 \pm 0.48$ & $0.38$ & $3.21 \pm 0.49$ \\ 
\hline
\end{tabular}
}
 \caption{Values of \xis\ for  \idhadrons~in $e^+e^-\rightarrow$ $q\bar q$~at 133, 161, 172, 183 and 189
  GeV. The errors are the sum in quadrature of the statistical and the total 
systematic uncertainties, and the $\chi^2/ndf$ corresponds to the Gaussian fit,
except for K$^0$ and $\Lambda$ at 133 GeV.}
\label{tab:xistar}
\end{table}

The systematic uncertainty has the following contributions 
which were added in quadrature to the statistical uncertainty of the
fit.
%The procedure is the same for the Gaussian and distorted Gaussian fit.

\begin{enumerate}
\item Uncertainty of the background evaluation (above the W threshold).

This source was evaluated as the maximal difference obtained by a variation 
of the WW background cross-section.

The relative size of this uncertainty, averaged over the centre-of-mass
energies is $0.005$, $0.010$, $0.002$, $0.024$ and $0.023$ 
for charged pions, kaons,
protons, K$^0$ and $\Lambda$ respectively. 
\item Uncertainty due to the particle identification.

The analysis was repeated using the calibrated matrices
or changes by 3\% 
(K$^0$) 
or 5\% 
($\Lambda$) in the bin contents of the $\xi_p$
distribution, and the fit redone.
The maximal differences in the position of the peak 
thus obtained were added in quadrature.

The relative size of this uncertainty, averaged over the centre-of-mass
%energies is $0.011$, $0.017$, $0.0019$, $??$ and $??$  
energies is $0.011$, $0.017$, $0.002$, $0.019$ and $0.017$  
for charged pions, kaons,
protons, K$^0$ and $\Lambda$ respectively. 
\item Stability of the fit and dependence on the fit range.

To estimate this effect, which arises from the combination of
the limited statistics, the resulting need to choose a coarse binning,
and the choice of the fit range, systematic
shifts have been imposed on the data by variation within the
statistical uncertainty. 
One standard deviation has been added to the values left of the
peak and  one standard deviation has been subtracted from the values to its
right and vice versa.  The maximum variation is taken as the contribution to
the systematic uncertainty. 

The relative size of this uncertainty, averaged over the centre-of-mass
energies is $0.026$, $0.112$, $0.161$, $0.163$ and $0.147$
 for charged pions, kaons, protons, K$^0$ and $\Lambda$ respectively. 
\end{enumerate}

\begin{figure}[ph]
%PTA%NO%This figure comes from /afs/cern.ch/user/n/neufeld/phy/idxi/
%\resizebox{!}{0.9\textheight}{\rotatebox{90}{\includegraphics{xistar99_bw.eps}}}
\resizebox{!}{0.9\textheight}{\rotatebox{90}{\includegraphics{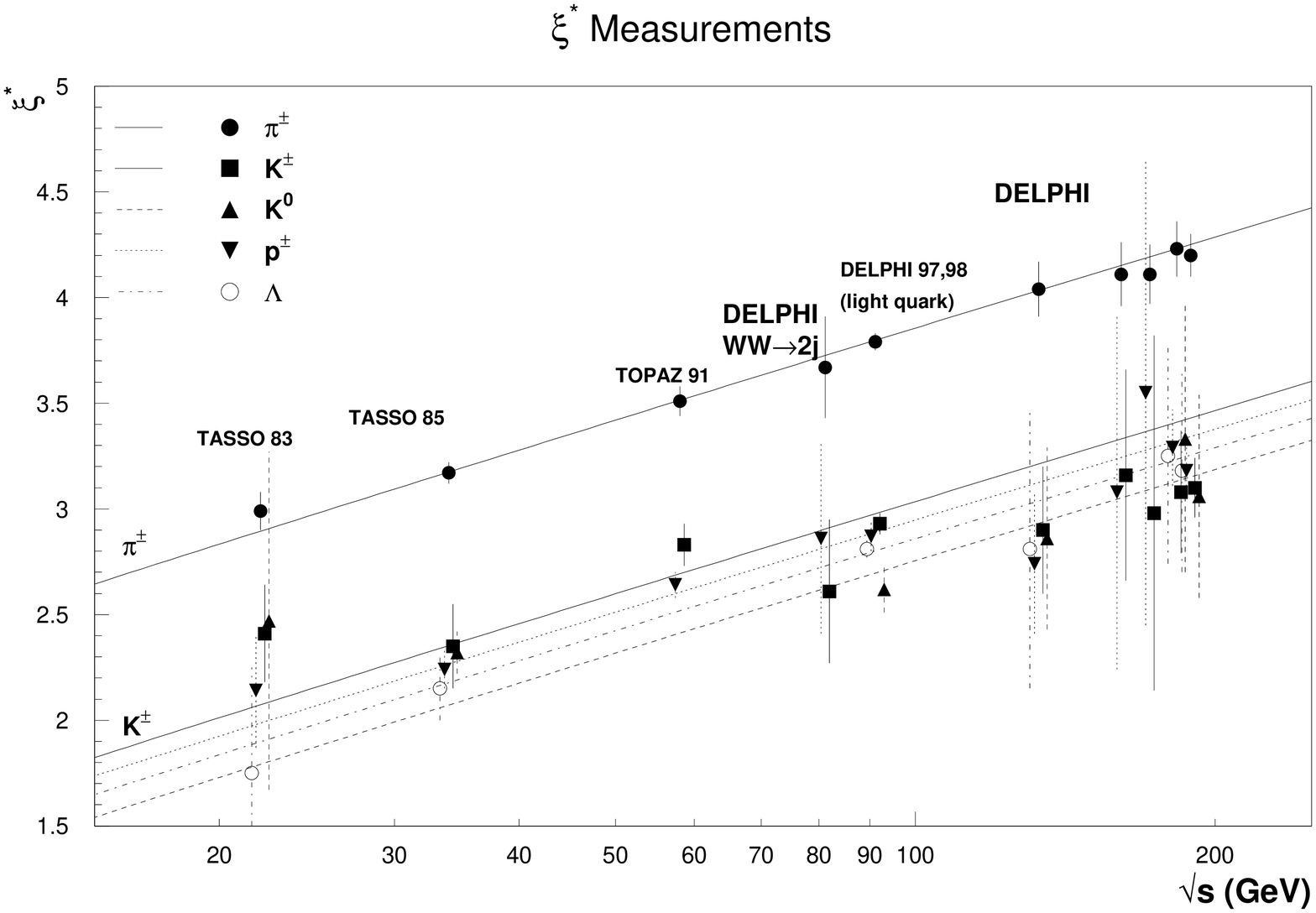}}}
\caption{Evolution of $\xi^*$ in $e^+e^-\rightarrow 
{\mathrm q}\bar{\mathrm q}$ with increasing 
centre-of-mass
  energy. This measurement is compared with previous measurements from
  DELPHI from references~\protect\cite{emile} and lower energy
  experiments~\protect\cite{brummer}. The lines are fits to
  equation~(\protect\ref{eq:evol2}) for the hadrons listed in
  the figure. The error bars represent the sum in quadrature of the statistical and the systematic uncertainties.
  For better legibility the data points
  have been shifted from the nominal energy.}\label{fig:xistar}
\end{figure}

Figure~\ref{fig:xistar} shows the $\xi^*$ values from the Gaussian fits
as a function of the centre-of-mass energy. 
%In MLLA this was
%predicted~\cite{ochs,brummer} to be
%\begin{equation}
%\xi^* = Y \left( \frac{1}{2} +\sqrt{C/Y}-C/Y \right) +F_h (\lambda),
%\label{eq:xistar}
%\end{equation}
%$ (dummy dollar here for Xemacs colour bug)
% 
%Next para. repeated just below!
%The comparison with equation~(\ref{eq:xistar}) and previous measurements
%taken from~\cite{brummer} is
%shown in Figure~\ref{fig:xistar}. 
%The fit of the data points to expression~(\ref{eq:dgauss}) 
%was used to extract the peak position of 
%the $\xi_p$ distribution, $\xi^*$. 
%
%In Figure~\ref{fig:xistar} the evolution
%of $\xi^*$ with the centre-of-mass energy is presented. 
The data up to centre-of-mass energies of
91~$\mathrm{GeV}$ were taken from previous measurements~\cite{brummer}. 
The fits to expression~(\ref{eq:evol2}), with $Y$ and
$C=0.351$ %for nf=5 and Nc=3
defined as in equation~(\ref{eq:YC}) and 
$F_h$ and $\Leff$ 
taken as free parameters, were done separately for each particle type and
are superimposed 
on the data points (lines).
Figure~\ref{fig:xistar} shows that 
(within the statistics of the data samples analysed) the fitted
functions follow well the data points. This suggests that 
MLLA+LPHD gives a good description
of the observed particle spectra. 
From table~\ref{tab:xistar} it is shown that there is fair agreement 
between
the data and the predictions from the generators (PYTHIA 5.7, HERWIG~5.8, and
ARIADNE~4.8). 

\section{Identified hadrons in WW events}

%A different event selection was used for the
%analysis of identified hadron production at 189 \gev .
The selections of $(4q)$ and $(2q)$ WW events 
in the analysis of identified hadrons in WW events at 189~GeV are similar to 
the
procedures described in~\cite{Wxsec}. A feed forward
neural network, trained with back-propagation on PYTHIA 
simulated events,
was used to improve the separation of $(4q)$ WW events
%$W^+W^- \rightarrow q \overline{q} q \overline{q}$ 
from 2-fermion (mainly
Z$^{0}/\gamma  \rightarrow {\mathrm q} \overline{\mathrm q} (\gamma)$) and
4-fermion background (mainly Z$^0$Z$^0$ events).
The network 
%is based on the JETNET package, uses the back-propagation
%algorithm and consists of three layers with 13 input, 7 hidden and one
%output node. 
input variables, the training procedure and its performance 
are described in detail in~\cite{Wxsec}. 
%The samples for training and testing the feed forward net consist of
%PYTHIA simulation and real data. The
% training is performed with 3500 events from signal and
%$Z^{0}$/$\gamma$ simulation. Afterwards the network output is calculated for
%other independent samples of WW, $Z^{0}$/$\gamma$ and ZZ MC and all real
%data as test samples.
The purities and the selection efficiencies were determined from
the simulation to be respectively 73\% 
and 83\%
in the $(4q)$ sample and 89\%
and 75\%
in the $(2q)$ sample.

\subsection{\xip\ distributions for identified particles at \boldmath$M_W$}

The mass of the W boson is the
relevant energy scale which enters in the \xip\ variable;
it is thus appropriate to compute the \xip\ distributions and their peak 
positions from W decays after a boost to the rest frame of the W. 
The boosting procedure relies on the simulation for
the part of the spectrum which was not accessible before
the boosting, and thus introduces a certain model dependence. This does
not affect significantly the peak region, which is used to extract \xis.
Figure~\ref{fig:xiww2q} shows the \xip\ distributions obtained from
semi-leptonic W events.

Table~\ref{tab:xisww} shows the values of \xis\ for the semi-leptonic W 
data at 189 GeV and for the WW simulation. 
The \xis\ values were
obtained by a Gaussian fit in the indicated region. 
The systematic uncertainties were estimated in the same way as described
in the previous section.
The relative uncertainty due to particle identification is 
$0.03$, $0.05$ and $0.05$ for charged pions, kaons and protons
respectively. The relative uncertainty due to stability evaluation of
the fit is $0.06$, $0.12$ and $0.15$ for charged pions, kaons and protons.
%Table 4
\begin{table}
\begin{center}
\begin{tabular}{|c | c | c | c | c |}
\hline
                   & fit range     &  WW simulation  &  data
                   & $\chi^2/\mathit{n.d.f.}$\\
\hline      
$\pi^{\pm}$        & $[2.7:4.5]$   & $3.66$ & $3.65\pm 0.02\pm 0.24$ &
                   $1.20/2$ \\ 
K$^{\pm}$          & $[1.10:3.86]$ & $2.70$ & $2.61\pm 0.09\pm 0.33$ &
                   $0.02$\\
p$\bar{\mathrm p}$ & $[1.10:3.86]$ & $2.73$ & $2.86\pm 0.11\pm 0.44$ &
                   $0.08$ \\
\hline
\end{tabular}
\end{center}
\caption{Values for \xis\ for $\pi^{\pm}$ , K$^{\pm}$\ and protons from a
        Gaussian fit to the semi-leptonic W data at 189 GeV. The fit
        was made to the 
        spectra in the W rest frame. 
        The first error is statistical and the second reflects the total
systematic uncertainties. Also indicated is the
        $\chi^2/\mathit{n.d.f}$. }   
\label{tab:xisww}
\end{table}  

Within the limited statistics of this sample the data are in good
agreement with the generator prediction as well as with the prediction
from MLLA (equation (\ref{eq:evol2})).

\subsection{Average multiplicity}
\label{anamul}

 After the event selection the analysis proceeds along the same
lines as described for the q$\bar{\mathrm q}$ analysis. Fully corrected  \xip
-distributions are obtained and afterwards integrated to obtain the
multiplicity. The results for the average multiplicity are 
shown in table~\ref{tbmultip}
and compared to the predictions from PYTHIA 5.7 and HERWIG~5.8, with the
systematic errors estimated as in subsection~6.7.
The relative size of the systematic uncertainty due to extrapolation
into the unseen region
is $0.023$ ($0.024$), $0.0023$ ($0.0022$) and $0.0011$ ($0.0011$) for charged
pions, kaons and protons from fully hadronic (semi-leptonic) $W$
decays.
The relative size of the systematic uncertainty due to particle identification
is $0.0052$ ($0.0093$), $0.026$ ($0.076$) and $0.088$ ($0.59$) for charged
pions, kaons and protons from fully hadronic (semi-leptonic) $W$
decays.
Finally, the relative size of the systematic uncertainty due to
uncertainty in the estimate of the subtracted background 
is $0.0013$ ($0.0003$), $0.0027$ ($0.0004$) and $0.0022$ ($0.0011$) for charged
pions, kaons and protons from fully hadronic (semi-leptonic) $W$
decays.

%Table 5
\begin{table}
\begin{center} 
\begin{tabular}{|l||c|c|c|c|} \hline
 event & Particle  &    \multicolumn{3}{|c|}{$\langle n\rangle$} \\ \cline{3-5} 
 type  &           &    (Data)              & PYTHIA 5.7 & HERWIG 5.8 \\ \hline

$\mathrm{WW} \rightarrow q \bar q q \bar q$  
     & $\pi^{\pm}$        &$31.65\pm 0.48\pm 0.76$ & $31.74$ & $31.55$ \\ 
     & K$^{\pm}$          &$~4.38\pm 0.42\pm 0.12$ & $~4.05$ & $~3.63$ \\ 
     & p$\bar{\mathrm p}$ &$~1.82\pm 0.29\pm 0.16$ & $~1.89$ & $~1.66$ \\ \hline
$\mathrm{WW} \rightarrow q \bar ql\nu$
     & $\pi^{\pm}$        &$15.51\pm 0.38\pm 0.40$ & $15.90$ & $16.08$ \\
     & K$^{\pm}$          &$~2.23\pm 0.32\pm 0.17$ & $~2.02$ & $~1.82$ \\ 
     & p$\bar{\mathrm p}$ &$~0.94\pm 0.23\pm 0.06$ & $~0.95$ & $~0.83$ \\ \hline
\end{tabular}
\end{center}
\caption{Average multiplicity for 
$\pi^\pm$, 
K$^\pm$ and protons
%$\pi^{\pm}$, 
%\Kpm\ and protons
  for WW events at 189 GeV. 
In the data the first uncertainty is statistical and the second is systematic.}
\label{tbmultip}
\end{table}

The ratios of the average multiplicities in (4q) events to twice the
multiplicities in (2q) events for different hadron species, 
conservatively assuming
uncorrelated errors, for the full
momentum range and for momentum between 0.2 and 1.25 GeV/c are shown in 
table~\ref{tbratmult}, and are compatible with unity. The systematic errors
for the restricted momentum range were assumed to be proportional to the sum
in quadrature of the systematic errors in the full momentum range, excluding
the contributions from the extrapolation.

%Table 6
\begin{table}[!ht]
\begin{center}  
\begin{tabular}{|c|c|c|} \hline
 Particle  & \multicolumn{2}{|c|}{$\langle n\rangle(4q)/(2\cdot\langle n\rangle(2q))$} \\ \cline{2-3} 
           & All $p$ & 0.2 GeV/c $< p <$ 1.25 GeV/c \\ \hline
 $\pi^{\pm}$    & $1.02\pm 0.03\pm 0.04$   & $1.03 \pm 0.03 \pm 0.01$ \\ 
 K$^{\pm}$      & $0.98\pm 0.17\pm 0.08$   & $0.96 \pm 0.38 \pm 0.08$ \\ 
 p$\bar{\mathrm p}$ &$0.97\pm 0.28\pm 0.11$& $0.72 \pm 0.57 \pm 0.08$ \\ \hline 
\end{tabular}
\end{center}
\caption{Ratio of average multiplicities in (4q) events to twice the values
in (2q) events for 
$\pi^\pm$, 
K$^\pm$ and protons
%$\pi^{\pm}$, 
%\Kpm\ and protons
  from WW events at 189 GeV, for different momentum ranges.
The first uncertainty is statistical and the second is systematic.}
\label{tbratmult}
\end{table}

\section{Summary and Discussion}

The mean charged particle multiplicities $\langle n\rangle$
and dispersions $D$ in q$\bar{\mathrm q}$ events at the different
centre-of-mass energies were measured to be:
\begin{eqnarray}
\langle n\rangle_{\mathrm{183\,GeV}} & = &  27.05 \pm 0.27 (stat) \pm 0.32 (syst)\\
D_{\mathrm{183\,GeV}}   & = &  ~8.08 \pm 0.19 (stat) \pm 0.14 (syst)\\
\langle n\rangle_{\mathrm{189\,GeV}} & = &  27.47 \pm 0.18 (stat) \pm 0.30 (syst)\\
D_{\mathrm{189\,GeV}}   & = &  ~8.77 \pm 0.13 (stat) \pm 0.11 (syst)\\
%202 \langle n\rangle_{\mathrm{200\,GeV}} & = &  27.63 \pm 0.17 (stat) \pm 0.43 (syst)\\
\langle n\rangle_{\mathrm{200\,GeV}} & = &  27.58 \pm 0.19 (stat) \pm 0.45 (syst)\\
%202 D_{\mathrm{200\,GeV}}   & = &  ~8.65 \pm 0.12 (stat) \pm 0.21 (syst) \, .
D_{\mathrm{200\,GeV}}   & = &  ~8.64 \pm 0.13 (stat) \pm 0.20 (syst) \, .
\end{eqnarray}

\begin{figure}[th]
%\centerline{DELPHI preliminary}
\resizebox{\textwidth}{14cm}{\includegraphics{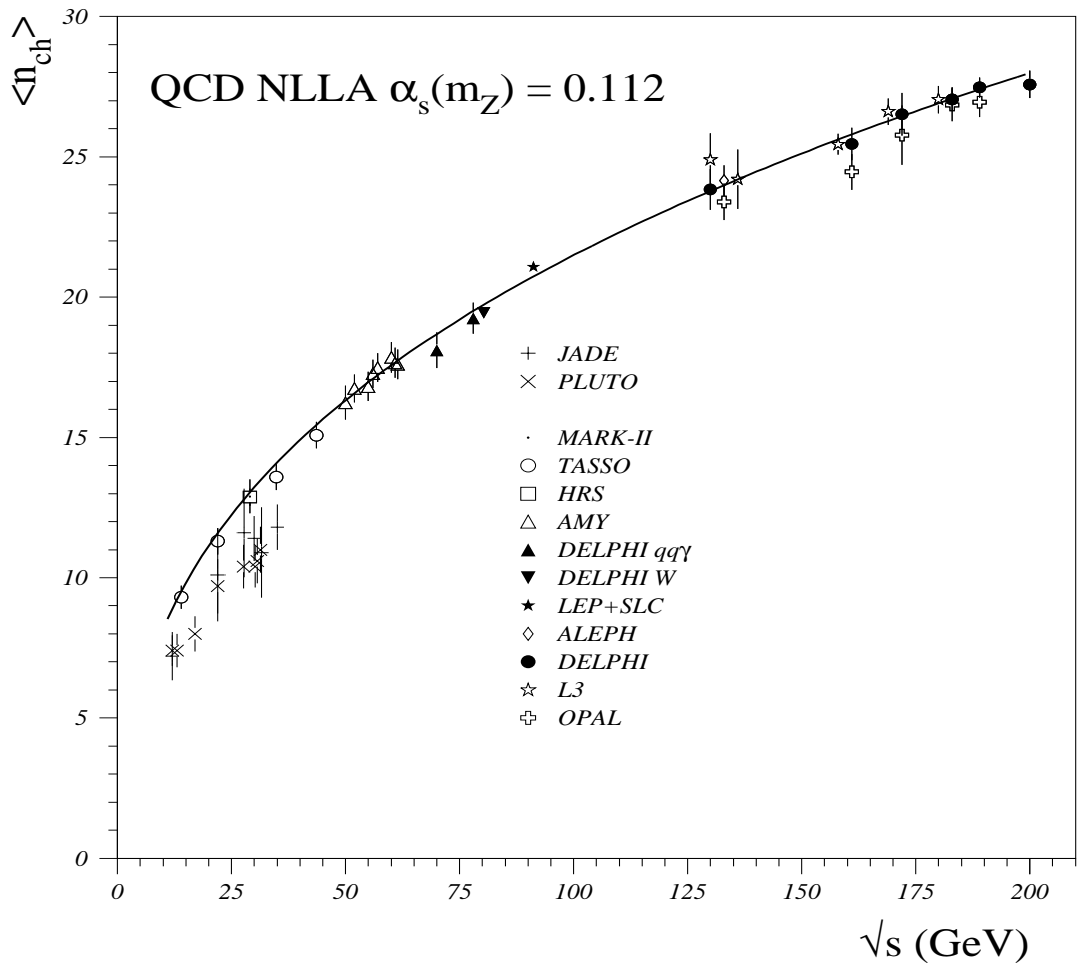}}
\caption[]{Measured average charged particle multiplicity
in $e^+e^- \rightarrow {\mathrm q}\bar{\mathrm q}$ events as a function
of centre-of-mass energy $\sqrt{s}$. DELPHI high energy results
are compared with other experimental results
and with a fit to a prediction from
QCD in Next to Leading Order.  The error bars represent the sum in quadrature of the statistical and the systematic uncertainties. Some points are slightly shifted on the abscissa
for clarity.

The average charged multiplicity in W decays
is also shown at an energy corresponding to the W mass.
The measurements have been corrected for the different
proportions of $b\bar{b}$ and $c\bar{c}$ events at the various energies.}
\label{mulea}

\end{figure}

Figure~\ref{mulea} shows the value of the average charged particle
multiplicity in $e^+e^- \rightarrow {\mathrm q}\bar{\mathrm q}$ events at 183,
189 and 200~GeV 
compared with lower energy points from JADE~\cite{jade}, PLUTO~\cite{pateta},
MARK~II~\cite{markII}, TASSO~\cite{tasso}, HRS~\cite{HRS}, and 
AMY~\cite{amy}, with DELPHI results in q$\bar{\mathrm{q}}\gamma$ 
events at the Z$^0$~\cite{qgjet},
with the world average at the Z$^0$~\cite{pdg}, and with LEP
results at high 
energy~\cite{aleph133,delphi130,delphi172,l3_136,l3_172,l3_183,opal133,opal161,opal189}.
%133, 161 and 172 GeV calculated in~\cite{frascati}
%according to the prescriptions in~\cite{pdg}. 
The points from JADE, PLUTO and MARK II do not include the decay products of short
lived K$^0$ and $\Lambda$.
%A point corresponding to the multiplicity observed by DELPHI
%in W decays is also included at $\sqrt{s} = M_W$ and is discussed later in
%section~8. 
The value at $M_{\mathrm{Z}^0}$ %($M_W$)
has been lowered by 0.20 %(incremented by 0.35), 
to account for the
different proportion of $b\bar{b}$ and $c\bar{c}$ events at the Z$^0$ with
respect to the $e^+e^-\rightarrow\gamma^{*}\rightarrow 
{\mathrm q}\bar{\mathrm q}$ \cite{dea}.
Similarly, the values at 133, 161, 172, 183 and 189~GeV were lowered by
0.15, 0.12, 0.11, 0.08, 0.05 and 0.07 respectively. 
%To the statistical errors 
%on the charged multiplicities in q$\bar{\mathrm q}\gamma$ events at the
%Z~\cite{fuster},
%a systematic error assumed to be $\pm 0.50$ has been added in quadrature.
%$
The QCD prediction for the charged particle multiplicity has been computed as
a function 
of $\alpha_s$ including the resummation of leading (LLA) and next-to-leading
(NLLA) corrections~\cite{webber}:%,sprime}: 
\begin{equation}  \label{truffle}
\langle n\rangle(\sqrt{s}) = a [\alpha_s(\sqrt{s})]^b e^{c/\sqrt{\alpha_s(\sqrt{s})}} %
\left[1+O(\sqrt{\alpha_s(\sqrt{s})})\right] \ ,
\end{equation}
where $s$ is the squared centre-of-mass energy and $a$ is a parameter (not
calculable from perturbation theory) whose value has been fitted from the
data. The constants $b=0.49$ and $c=2.27$ are predicted by 
theory~\cite{webber} and $\alpha_s(\sqrt{s})$ is the strong coupling
constant. The fitted curve to the data 
between 14~GeV and 200~GeV, excluding the results 
from JADE, PLUTO and MARK~II, is plotted in
Figure~\ref{mulea}, %with the curve 
corresponding to $a = 0.045$ and $\alpha_s(m_{\mathrm{Z}})=0.112$.
The multiplicity values 
are consistent with the QCD prediction on the
multiplicity evolution with centre-of-mass energy.

\begin{figure}[th]
%\centerline{DELPHI preliminary}
\resizebox{\textwidth}{!}{\includegraphics{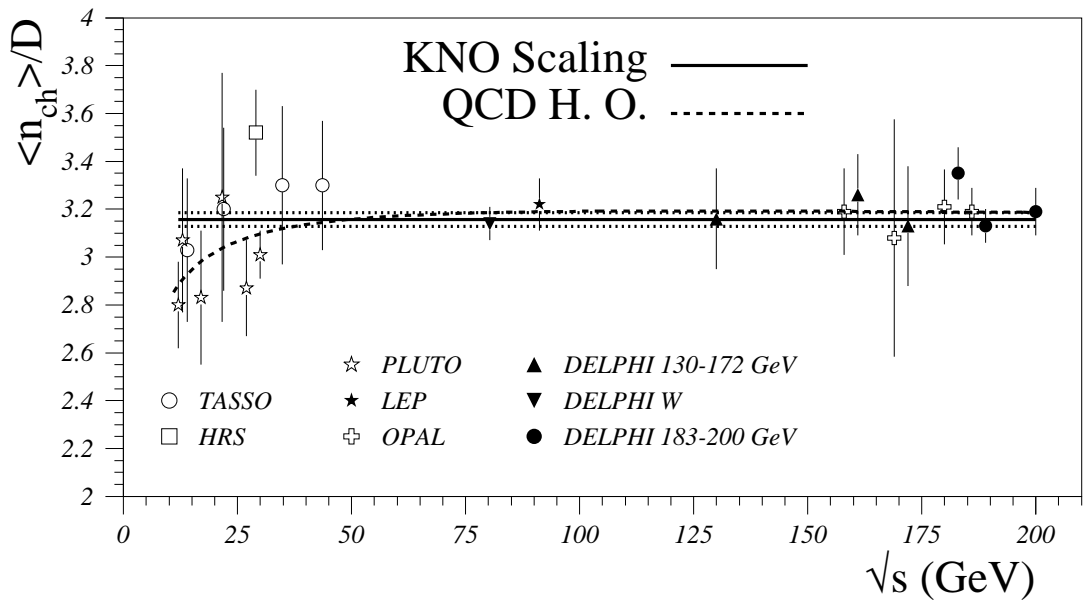}}
\caption[]{Ratio of the average charged particle multiplicity to the dispersion
in $e^+e^-\rightarrow {\mathrm q}\bar{\mathrm q}$ events
at 183, 189 and 200~GeV, compared with lower
energy measurements.  The error bars represent the sum in quadrature of the statistical and the systematic uncertainties.
Some points are slightly shifted on the abscissa for clarity.
The ratio in W decays
is also shown at an energy corresponding to the W mass.
The straight solid and dotted lines represent
the weighted average of the data points and its error.
The dashed line represents the prediction from QCD (see text).}
\label{muleb}
\end{figure}

The ratios of the average multiplicity to the dispersion measured at 183~GeV, 
189~GeV and 200~GeV, $\langle n\rangle/D = 3.35 \pm 0.11$, 
$\langle n\rangle/D = 3.14 \pm 0.07$ and %202 $\langle n\rangle/D = 3.19 \pm 0.10$
$\langle n\rangle/D = 3.19 \pm 0.10$
(where the errors are the sum in quadrature of
the statistical and of the systematic)
respectively, are consistent with the weighted 
average from the measurements at lower centre-of-mass energies 
(3.13 $\pm$ 0.04), as can be seen in Figure~\ref{muleb}. From
Koba-Nielsen-Olesen scaling~\cite{kno} this ratio is predicted to be 
energy-independent. The ratio measured is also consistent with the
predictions of QCD including 1-loop Higher Order terms (H.O.)
\cite{webberkno}.

For WW events the measured multiplicities in the fully hadronic
channel are: 
\begin{eqnarray}
\langle n^{(4q)}\rangle_{\mathrm{189\,GeV}} & = & 39.12 \pm 0.33 (stat) 
\pm 0.36 (syst)\\
\langle n^{(4q)}\rangle_{\mathrm{183\,GeV}} & = & 38.11 \pm 0.57 (stat) 
\pm 0.44 (syst) \, ,
\end{eqnarray}
while for the semileptonic channel they are:
\begin{eqnarray}
\langle n^{(2q)}\rangle_{\mathrm{189\,GeV}} & = & 19.49 \pm 0.31 (stat) 
\pm 0.27 (syst) \\
\langle n^{(2q)}\rangle_{\mathrm{183\,GeV}} & = & 19.78 \pm 0.49 (stat) 
\pm 0.43 (syst) \, .
\end{eqnarray}
The PYTHIA Monte Carlo program
%without colour reconnection effects, 
with parameters tuned to the DELPHI data at LEP~1, predicts
multiplicities of 38.2 and 19.1 for the fully hadronic and
semileptonic events respectively.

A possible depletion of the multiplicity in
fully hadronic WW events with respect to twice the semileptonic events,
as predicted by most colour reconnection models,
is not observed in the full momentum range, in agreement with the results
from other LEP collaborations \cite{otherlep}:
\begin{equation}
\left(\frac{\langle n^{(4q)}\rangle}{2\langle n^{(2q)}\rangle}\right)  =  0.990
\pm 0.015 (stat) \pm 0.011 (syst)\, ,
\end{equation}
%The depletion is at the percent level and concentrated in the
%low-$p$ (and low-$p_T$) region.
%(as predicted by most colour reconnection models).
nor in the momentum range $0.1<p<1.0$~GeV/$c$:
\begin{equation}
\left.\frac{\langle n^{(4q)}\rangle}{2\langle n^{(2q)}\rangle}
\right|^{0.1 < p < 1 \, {GeV}/c}   =  0.981 \pm 0.024 (stat) \pm 0.013 (syst) .
\end{equation}

No significant difference is observed between the dispersion 
in fully hadronic events and $\sqrt{2}$ times the dispersion 
in semileptonic events:
\begin{eqnarray*}
\frac{D^{(4q)}}{\sqrt{2}D^{(2q)}} & = &
0.94 \pm 0.03 \mathrm{(stat)}\pm 0.03 \mathrm{(syst)} \, . \\
\end{eqnarray*}

%If the
%differences in multiplicity are due to a statistical fluctuation, one
%can average the results to
%obtain the best determination of the charged
%particle multiplicity from hadronic decays of the W,
%and of the dispersion of the distribution. The multiplicity in the
%$(4q)$ case has to be divided by a factor of 2, while the dispersion 
%has to be 
%divided by a factor $\sqrt{2}$.
%Considering the
%systematic errors as independent (excluding the scale error), 
%one finally has:
Assuming uncorrelated systematic errors for the two centre-of-mass 
energies, 183 and 189 GeV,
%and using for the
%weights in the averaging the statistical errors and the systematic errors 
%added in quadrature (the errors and the
%dispersions in (4q) divided by $\sqrt{2}$), we obtain:
the inverse of the sums in quadrature of the statistical and the systematic
errors  were used as weights to give the averages:
\begin{eqnarray}
\langle n^{(W)}\rangle  
 & = & 19.44 \pm 0.13 (stat) \pm 0.12 (syst) \\
D^{(W)}  
 & = & ~6.20 \pm 0.11 (stat)\pm 0.06 (syst) \, ,
\end{eqnarray}
where the multiplicities and their errors in (4q) were divided by 2 and the
dispersions and their errors in (4q) were divided by $\sqrt{2}$.

The value of $\langle n^{(W)}\rangle$ is plotted in Figure~\ref{mulea} at an energy value
corresponding to the W mass, with an
increase of 0.35 applied 
to account for the different proportion of events with a $b$ or a $c$ quark.
%in W hadronic decays than in continuum $e^+e^-$ events \cite{dea}.
The measurement lies on the same curve as the neutral current data. 
The value of 
$\langle n^{(W)}\rangle/D^{(W)} = 3.14 \pm 0.07 (stat+syst)$, 
plotted in Figure~\ref{muleb}, is also consistent with the
$e^+e^-\rightarrow\gamma^{*}\rightarrow 
{\mathrm q}\bar{\mathrm q}$ average. 

The production of \idhadrons~ 
%$\pi^{\pm}$ ,K$^{\pm}$\ and protons  
from q$\bar{\mathrm q}$ and WW events at 189 GeV 
was also studied.
% using data taken with the DELPHI detector at LEP.
The results on the average multiplicity of  identified particles 
and on the position $\xi^*$ of the maximum of the 
%and on the position \xis\ of the maximum of the 
$\xi_p = -\mathrm{log} (\frac{2p}{\sqrt{s}})$ 
distribution were compared with predictions
of PYTHIA and with calculations based on MLLA+LPHD 
approximations.
Within their uncertainties the data are in good agreement with the
prediction from the generator as well as with the predictions based on
the analytical calculations in the MLLA framework.  

The ratio of the multiplicities of identified heavy hadrons in (4q) events 
to twice those in (2q) events is compatible with unity, 
both for the full momentum range and momenta between 0.2 and 1.25~GeV/c:
\begin{center}  
\begin{tabular}{ccc}
 Particle  & All $p$ & $0.2$ GeV/c $< p < 1.25$ GeV/c \\
%\pi       &                           & $0.983~\pm~0.014~\pm 0.021$ \\
 K$^{\pm}$          & $0.98\pm 0.17\pm 0.08$ & $0.96 \pm 0.38 \pm 0.08$\,
\phantom{.} \\ 
 p$\bar{\mathrm p}$ & $0.97\pm 0.28\pm 0.11$ & $0.72 \pm 0.57 \pm 0.08$\, . \\ 
\end{tabular}
\end{center}

The evolution for the $\xi^*$ for identified hadrons is in good agreement with
the prediction from perturbative QCD 
(equation~(\ref{eq:evol2})). This underlines the
applicability of MLLA/LPHD for the description of hadron production in 
$e^+e^-$ annihilation over the full LEP energy range.

%         Modified on 04-06-1999 by dimartino
%-------------------------------------------------------------------
\subsection*{Acknowledgements}
\vskip 3 mm
 We are greatly indebted to our technical 
collaborators, to the members of the CERN-SL Division for the excellent 
performance of the LEP collider, and to the funding agencies for their
support in building and operating the DELPHI detector.\\
We acknowledge in particular the support of \\
Austrian Federal Ministry of Science and Traffics, GZ 616.364/2-III/2a/98, \\
FNRS--FWO, Belgium,  \\
FINEP, CNPq, CAPES, FUJB and FAPERJ, Brazil, \\
Czech Ministry of Industry and Trade, GA CR 202/96/0450 and GA AVCR A1010521,\\
Danish Natural Research Council, \\
Commission of the European Communities (DG XII), \\
Direction des Sciences de la Mati$\grave{\mbox{\rm e}}$re, CEA, France, \\
Bundesministerium f$\ddot{\mbox{\rm u}}$r Bildung, Wissenschaft, Forschung 
und Technologie, Germany,\\
General Secretariat for Research and Technology, Greece, \\
National Science Foundation (NWO) and Foundation for Research on Matter (FOM),
The Netherlands, \\
Norwegian Research Council,  \\
State Committee for Scientific Research, Poland, 2P03B06015, 2P03B1116 and
SPUB/P03/178/98, \\
JNICT--Junta Nacional de Investiga\c{c}\~{a}o Cient\'{\i}fica 
e Tecnol$\acute{\mbox{\rm o}}$gica, Portugal, \\
Vedecka grantova agentura MS SR, Slovakia, Nr. 95/5195/134, \\
Ministry of Science and Technology of the Republic of Slovenia, \\
CICYT, Spain, AEN96--1661 and AEN96-1681,  \\
The Swedish Natural Science Research Council,      \\
Particle Physics and Astronomy Research Council, UK, \\
Department of Energy, USA, DE--FG02--94ER40817. \\
%=========================================================================%
\hfill
%=========================================================================%

\newpage

% Figure 1

\newpage

% Figure 2
\newpage

% Figure 3

\newpage

% Figure 4

\newpage

% Figure 5

\newpage

% Figure 6

\newpage

% Figure 7

\newpage

% Figure 8

\newpage

% Figure 9

\newpage

% Figure 10

\newpage

% Figure 11

\newpage

% Figure 12

\newpage

% Figure 13

\newpage 

% Figure 14

\newpage

% Figure 15

% Figure 16

\end{document}